\documentclass[10pt]{iopart}
\usepackage{epsfig,graphics,amssymb}
\def\u{{\bf u}}
\def\r{{\bf r}}
\def\d{{\rm d}}
\def\n{{\bf n}}
\def\F{{\bf F}}
\def\U{{\bf U}}
\def\L{{\bf L}}
\def\G{{\bf G}}
\def\p{\partial}
\def\Re{{\rm Re}}
\def\x{{\bf x}}
\def\r{{\bf r}}
\def\t{{\bf t}}
\def\f{{\bf f}}
\def\g{{\bf g}}
\def\aA{{\bf A}}
\def\B{{\bf B}}
\def\C{{\bf C}}
\def\D{{\bf D}}
\def\M{{\bf M}}
\def\N{{\bf N}}
\def\OO{{\bf O}}

\def\X{{\bf X}}
\def\e{{\bf e}}
\def\k{{\bf k}}

\newcommand{\bsi}{\mbox{\boldmath$\sigma$}}
\newcommand{\bom}{\mbox{\boldmath$\omega$}}

\newcommand{\btau}{\mbox{\boldmath$\tau$}}
\newcommand{\Om}{\mbox{\boldmath$\Omega$}}

\begin{document}

\title{The hydrodynamics of swimming microorganisms}

\author{Eric Lauga$^1$ and Thomas R. Powers$^2$}
\address{$^1$Department of Mechanical and Aerospace Engineering, \\ University of California, San Diego, La Jolla, CA 92093-0411, United States}
\vskip-0.08in
\ead{elauga@ucsd.edu}
\medskip
\address{$^2$Division of Engineering, \\ Brown University, Providence, RI 02912-9104, United States }
\vskip-0.08in
\ead{Thomas\_Powers@brown.edu}

\begin{abstract}

Cell motility in viscous fluids is 
ubiquitous and affects many biological processes, including reproduction, infection, and the marine life ecosystem. Here we review the biophysical and mechanical principles of locomotion at the 
small scales 
relevant to cell swimming  (tens of microns and below). The focus is on the fundamental flow physics phenomena occurring in this inertia-less realm, and the emphasis is on the simple physical picture. We review the basic properties of flows at low Reynolds number, paying special attention to aspects most relevant for swimming, such as resistance matrices for solid bodies, flow singularities, and kinematic requirements for net translation.
Then we review classical theoretical work on cell motility: early calculations of the speed of a swimmer with prescribed stroke, and the application of resistive-force theory and slender-body theory to flagellar locomotion. After reviewing the physical means by which flagella are actuated, we outline areas of active research, including hydrodynamic interactions, biological locomotion in complex fluids, the design of small-scale artificial swimmers, and the optimization of locomotion strategies.

\end{abstract}
\pacs{47.63.-b, 47.63.Gd, 87.17.Jj, 87.18.Ed, 47.63.mf, 47.61.-k, 47.15.G-}
%47.63.-b	Biological fluid dynamics
%47.63.Gd 	Swimming microorganisms
%87.17.Jj	Cell locomotion, chemotaxis
%87.18.Ed	Cell aggregation
%47.63.mf	Low-Reynolds-number motions
%47.61.-k	Micro- and nano- scale flow phenomena
%47.15.G-	Low-Reynolds-number (creeping) flows
\submitto{\RPP}

\date{\today}

\maketitle

\section{Introduction}
Our world is filled with swimming microorganisms: The spermatozoon that fuse with the ovum during fertilization,  the  bacteria that inhabit our guts, 
the protozoa in our ponds, and the algae in the ocean.

The reasons microorganisms move are familiar.
Bacteria such as {\it Escherichia coli}  detect gradients in nutrients and move to regions of higher concentration~\cite{bergbook}. The spermatozoa of many organisms swim to the ovum, sometimes in challenging environments such as tidal pools in the case of sea urchins or cervical mucus in the case of humans~\cite{suarez06}. {\it Paramecium} cells swim to evade predator rotifers.  

What is perhaps less familiar is the fact that the physics governing swimming at the micron scale is different from the physics of swimming at the macroscopic scale. The world of microorganisms is the world of low ``Reynolds number," a world where inertia plays little role and viscous damping is paramount. As we describe below, the Reynolds number $\mathrm{Re}$  is defined as $\mathrm{Re}=\rho U L/\eta$, where $\rho$ is the fluid density, $\eta$ is the viscosity, and  $U$ and $L$ are a characteristic velocity and length scale of the flow, respectively. Swimming strategies employed by larger organisms that operate at high Reynolds number, such as fish, birds, or insects~\cite{childress81,ellington84,vogel96,alexander02,dudley02,vogel03}, do not work at the small scale. For example, any attempt to move by imparting momentum to the fluid, as is done in paddling, will be foiled by the large viscous damping. Therefore microorganisms have evolved propulsion strategies that successfully overcome and exploit drag. The aim of this review is to explain the fundamental physics upon which these strategies rest. 

The study of the physics of locomotion at low Reynolds number has a long history. In 1930, Ludwig~\cite{ludwig} pointed out that a microorganism that waves rigid arms like oars is incapable of net motion. Over the years there have been many classic reviews, from the general perspective of animal locomotion~\cite{gray68},
from the perspective of fluid dynamics at low Reynolds number~\cite{lighthill75,lighthill76,brennen77,purcell77,childress81,yates86,fauci06}, and from the perspective of the biophysics and biology of cell motility~\cite{holwill66,jahn72,blum79,berg00_PT,braybook,bergbook}.
Nevertheless, the number of publications in the field has grown substantially in the past few years. This growth has been spurred in part by new experimental techniques for studying cell motility. Traditionally, motile cells have been passively observed and tracked using light microscopy. This approach has led to crucial insights such as the nature of the chemotaxis strategy of {\it E. coli}~\cite{bergbook}. Advances in visualization techniques, such as the fluorescent staining of flagella~\cite{turner00} in living, swimming bacteria, continue to elucidate the mechanics of motility.  A powerful new contribution is  the ability to measure forces at the scale of single organisms and single motors. For example, it is now possible to  measure the force required to hold 
a swimming spermatozoon \cite{tadir89,tadir90,konig96}, algae \cite{mccord05}
or bacterium~\cite{chattopadhyay06} in an optical trap.
Atomic force microscopy also allows direct measurement of the force exerted by cilia \cite{teff07}. Thus the relation between force and the motion of the flagellum can be directly assessed. These measurements of force allow new approaches to biological questions, such the heterogeneity of motor behaviour in genetically identical bacteria. Measurements of force together with quantitative observation of cell motion motivate the development of detailed hydrodynamic theories that can constrain or rule out models of cell motion.

The goal of this review is to describe the theoretical  framework for locomotion at low Reynolds number.  Our focus  is on analytical results, but our aim is to emphasize physical intuition.  In \S\ref{OverviewSwimming}, we give some examples of how microorganisms swim.
After a brief general review of low-Reynolds number hydrodynamics (\S\ref{fluids}), we outline the fundamental properties of locomotion without inerti (\S\ref{life}). We then discuss the classic contributions of Taylor~\cite{taylor51}, Hancock~\cite{hancock53} and Gray~\cite{gray55}, who  all but started the field more than 50 years ago (\S\ref{classics});  we also outline 
many of the subsequent works that followed. We proceed by introducing the different ways to physically actuate a flagella-based swimmer (\S\ref{actuation}). We then move on to introduce topics of active research. These areas include the role 
of hydrodynamic interactions, such as the interactions between   two swimmers, or between a wall and a swimmer  (\S\ref{interactions}); locomotion in non-Newtonian fluids such as the mucus of the female mammalian reproductive tract (\S\ref{complex});  and the design of artificial swimmers and the optimization of locomotion strategies in an environment at low Reynolds number  (\S\ref{artificial}).  
Our coverage of these topics is motivated by intellectual  curiosity and the desire to understand the fundamental physics of swimming; the relevance of swimming in biological processes such as reproduction or bacterial infection; and the practical desire to build artificial swimmers, pumps, and transporters in microfluidic systems.

Our review is necessarily limited to a small cross-section of current research. There are many closely related aspects of  ``life at low Reynolds number" that we do not address, such as 
nutrient uptake or quorum sensing; instead we focus on flow physics. 
Our hope is to capture some of the current excitement in this research area, which lies at the intersection of physics, mechanics, biology, and applied mathematics, and is driven by clever experiments that shed a new light on the hidden world of microorganisms. Given the interdisciplinary nature of the subject, we have tried to make
the review a self-contained starting point for the interested student or scientist.

\section{Overview of mechanisms of swimming motility}
\label{OverviewSwimming}
\begin{figure}[t]
\begin{center}
 \includegraphics[width=0.9\textwidth]{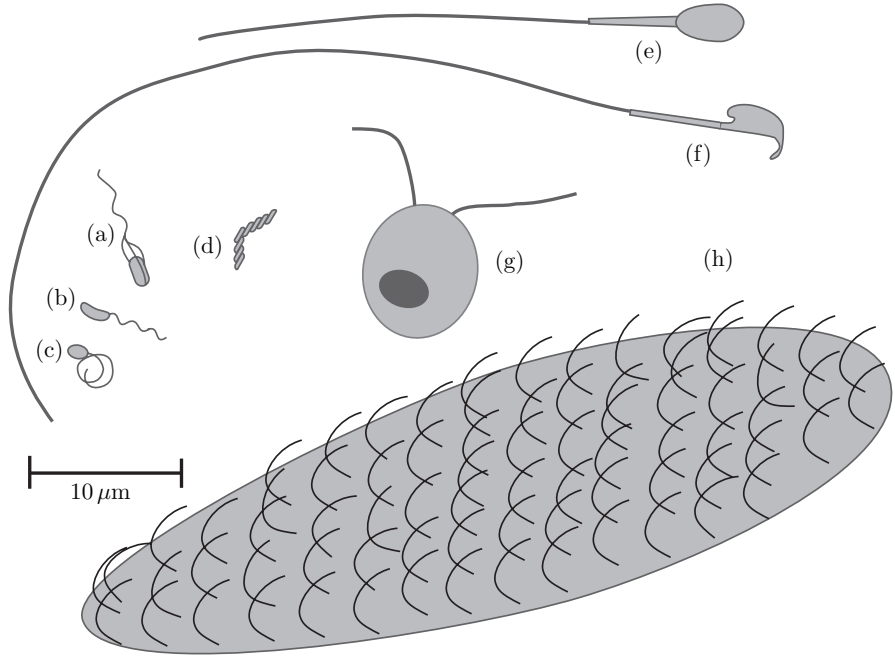}
\end{center}
 \caption{Sketches of microscopic swimmers, to scale. (a) \textit{E. coli}. (b) \textit{C. crescentus}. (c) \textit{R. sphaeroides}, with flagellar filament in the coiled state. (d) \textit{Spiroplasma}, with a single kink separating regions of right-handed and left-handed coiling. (e) Human spermatozoon. (f) Mouse spermatozoon. (g) \textit{Chlamydomonas}. (h) A smallish \textit{Paramecium}. }
\label{swimmerfig}
\end{figure}

In this section we motivate our review with a short overview of mechanisms for swimming motility. We define a ``swimmer" to be a creature or object that moves by changing its body shape in a periodic way. To keep the scope of the article manageable, we do not consider other mechanisms that could reasonably be termed ``swimming," such as the  polymerization of the actin of a host cell by pathogens of the genus \textit{Listeria}~\cite{StevensGalyovStevens2006}, or the gas-vesicle mediated buoyancy of aquatic micoorganisms such as \textit{Cyanobacteria}~\cite{Walsby1994}. 
 
Many microscopic swimmers use one or more appendages for propulsion. The appendage could be a relatively stiff helix that is rotated by a motor embedded in the cell wall, as in the case of \textit{E. coli} (Fig.~\ref{swimmerfig}a), or it could be a flexible filament undergoing whip-like motions due to the action of molecular motors distributed along the length of the filament, as in the sperm of many species~\cite{braybook}  (Figs.~\ref{swimmerfig}e and f).  For example, the organelle of motility in \textit{E. coli} and \textit{Salmonella typhimurium} is the bacterial flagellum, consisting of a rotary motor, a helical filament, and a hook which connects the motor to the filament~
\cite{berg00,berg03,berg2004}. The filament has a diameter of $\approx20$\,nm, and traces out a helix with contour length $\approx10$\,$\mu$m. In the absence of external forces and moments, the helix is left-handed with a pitch $\approx2.5$\,$\mu$m and a helical diameter $\approx0.5$\,$\mu$m~\cite{turner00}. There are usually several flagella per cell. When the motor turns counter-clockwise (when viewed from outside the cell body), the filaments wrap into a bundle that pushes the cell along at speeds of 25--35\,$\mu$m/s (see \S\ref{bundling}) \cite{LoweMeisterBerg1987}. When one or more of the motors reverse, the corresponding filaments leave the bundle and undergo ``polymorphic" transformations in which the handedness of the helix changes; these polymorphic transformations can change the swimming direction of the cell~\cite{turner00}.

There are many variations on these basic elements among swimming bacteria. For example, \textit{Caulobacter crescentus} has a single right-handed helical filament (Fig.~\ref{swimmerfig}b), driven by a rotary motor that can turn in either direction. The motor preferentially turns clockwise, turning the filament in the sense to push the body forward~\cite{KoyasuShirakihara1984}. During counterclockwise rotation the filament pulls the body instead of pushing. 
The motor of the bacterium \textit{Rhodobacter sphaeroides} turns in only one direction but stops from time to time~\cite{ArmitageSchmitt1997}. The flagellar filament forms a compact coil when the motor is stopped (Fig.~\ref{swimmerfig}c), and extends into a helical shape when the motor turns.
Several archaea also use rotating flagella to swim, although far less is known about the archaea compared to bacteria. Archaea such as the various species of \textit{Halobacterium} swim more slowly than bacteria, with typical speeds of  2--3\,$\mu$m/s~\cite{MarwanAlamOesterhelt1984}. Although archaeal flagella also have a structure comprised of motor, hook, and filament, molecular analysis of the constituent proteins shows that archaeal and bacterial flagella are unrelated (see~\cite{JarrellMcBride2008} and references therein).

There are also bacteria that swim with no external flagellar filaments. The flagella of spirochetes lie in the thin periplasmic space between the inner and outer cell membranes~\cite{Canale-Parola1984}. The flagellar motors are embedded in the cell wall at both poles of the elongated body of the spirochete, and the flagellar filaments emerge from the motor and wrap around the body. Depending on the species, there may be one or many filaments emerging from each end of the body. In some cases, such as the Lyme disease spirochete \textit{Borrelia burgdorferi}, the body of the spirochete is observed to deform as it swims, and it is thought that the rotation of the periplasmic flagella causes this deformation which in turn leads to propulsion~\cite{Berg1976,GoldsteinCharonKreiling1994}. The deformation can be helical or planar.   These bacteria swim faster in gel-like viscous environments than bacteria with external flagella~\cite{BergTurner1979,Nakamura_etal2006}. 
Other spirochetes, such as \textit{Treponema primitia}, do not change shape at all as they swim, and it is thought that motility develops due to rotation of the outer membrane and cytoplasmic membrane in opposite senses~\cite{Berg1976,Murphy_etal2008}. 
Finally, we mention the case of \textit{Spiroplasma}, helically shaped bacteria with no flagella (Fig.~\ref{swimmerfig}d). These cells swim via the propagation of pairs of kinks along the length of the body~\cite{ShaevitzLeeFletcher2005}. Instead of periplasmic flagella, the kinks are thought to be generated by contraction of the cytoskeleton~\cite{Berg2002,TrachtenbergGilad2001,Wolgemuth2003}.

Eukaryotic flagella and cilia are much larger than bacterial flagella, with a typical diameter of $\approx$ 200\,nm, and with an intricate internal structure~\cite{braybook}. The most common structure has nine microtubule doublets spaced around the circumference and running along the length of a flagellum or cilium, with two microtubules along the center.  Molecular  motors (dynein) between the doublets slide them back and forth, leading to bending deformations that propagate along the flagellum. There is a vast diversity in the beat pattern and length of eukaryotic flagella and cilia. For example, the sperm of many organisms consists of a head containing the genetic material propelled by a filament with a planar or even helical beat pattern, depending on the species~\cite{WernerSimmons2008}. The length of the flagellum is $12$\,$\mu$m in some Hymenoptera~\cite{Quicke1992}, $\approx 20$\,$\mu$m for hippos, $\approx40$\,$\mu$m for humans~{\cite{suarez06}} (Fig.~\ref{swimmerfig}e), $\approx80$\,$\mu$m for mice (Fig.~\ref{swimmerfig}f), and can be 1\,mm~\cite{HatsumiWakahami1986} or even several cm long in some fruit flies~\cite{JolyBressacLachaise1995} (although in the last case the flagella are rolled up into pellets and offered to the female via a ``pea-shooter" effect~\cite{JolyBressacLachaise1995}).
 
Many organisms have multiple flagella. \textit{Chlamydomonas reinhardtii} is an algae with two flagella that can exhibit both ciliary and flagellar beat patterns (Fig.~\ref{swimmerfig}g). In the ciliary case, each flagellum has an asymmetric beat pattern~\cite{braybook}. In the power stroke, each flagellum extends and bends at the base, sweeping back like the arms of a person doing the breaststroke. On the recovery stroke, the flagellum folds, leading as we shall see below to less drag. When exposed to bright light, the alga swims in reverse, with its two flagella extended and propagating bending waves away from the cell body as in the case of sperm cells described above~\cite{Mitchell2000}.  \textit{Paramecium} is another classic example of a ciliated microorganism. Its surface is covered by thousands of cilia that beat in a coordinated manner~\cite{Gibbons1981}, propelling the cell at speeds of $\approx500$\,$\mu$m/s (Fig.~\ref{swimmerfig}h). Arrays of beating cilia are also found lining the airway where they sweep mucus and foreign particles up toward the nasal passage~\cite{sleigh88}.

\section{Flows at low Reynolds number}
\label{fluids}

%%%%%%%%%%%%%%%%%%%%%%%%%%%%%%%%
\subsection{General properties}

We first briefly discuss the general properties of flow at low Reynolds numbers. For more detail we refer to the classic monographs by Happel and Brenner~\cite{happel}, Kim and Karilla~\cite{kimbook}, and Leal~\cite{leal}; a nice introduction is also offered by Hinch~\cite{hinch_lowRe}, as well as a more formal treatment by Pozrikidis \cite{poz}.

To solve for the force distribution on a 
organism, we need to solve for the flow field $\u$ and pressure $p$ in the surrounding fluid. For an incompressible Newtonian fluid with density $\rho$ and viscosity $\eta$, 
the flow satisfies the Navier-Stokes equations
\begin{equation}\label{NS}
\rho \left(\frac{\p}{\p t} + \u \cdot \nabla\right)\u = -\nabla p + \eta \nabla^2 \u, \quad 
\nabla \cdot \u = 0,
\end{equation}
with 
boundary conditions 
appropriate to the problem at hand. The Navier-Stokes 
equations are
a pointwise statement of momentum conservation.
Once $\u$ and $p$ are known, the stress tensor is given by 
$\bsi=-p{\bf 1} + \eta [\nabla\u + (\nabla\u)^T]$, and the force $\F$ and torque $\L$ acting on the body are found by integrating along its surface $S$
\begin{equation}
\F(t)=\int\!\!\!\int_S \bsi\cdot\n\,\d S
,\quad 
\L(t)=\int\!\!\!\int_S\x\times(\bsi\cdot\n)\,\d S.
\end{equation}

The Reynolds number is a dimensionless quantity which qualitatively captures the characteristics of the flow regime obtained by solving Eq.~(\ref{NS}), 
and it has 
several different physical interpretations.  Consider a steady flow with typical velocity $U$ around a body of size $L$. The Reynolds number $\Re$ is classically defined as the ratio of the typical inertial terms  
in the Navier-Stokes equation, $\sim \rho \u \cdot \nabla \u$, to the viscous forces per unit volume, $\sim \eta \nabla^2 \u$.  
Thus, $\Re={\rho L U}/{\eta}$. A  low-Reynolds number flow is one for which viscous forces dominate in the fluid. 

A second interpretation can be given as the ratio of time scales. The typical time scale for a local velocity perturbation to be transported convectively by the flow along the body is  $t_{{\rm adv}}\sim L/U$, whereas  the typical time scale for this perturbation to diffuse away from the body due to viscosity is $t_{{\rm diff}}\sim \rho L^2/\eta$. We see therefore that $\Re={t_{{\rm diff}}}/{t_{{\rm adv}}}$, and a low Reynolds number flow is one for which fluid transport is dominated by viscous diffusion.

We can also interpret $\Re$ as a ratio of forces on the body. A typical viscous stress on a bluff body is given by
$\sigma_{\rm viscous} \sim \eta U/L$, leading to a typical viscous force on the body of the form $f_{\rm viscous}\sim \eta U L $. A typical inertial stress is given by a Bernoulli-like dynamic pressure, $\sigma_{\rm inertial} \sim \rho U^2$, and therefore an inertial force  $f_{\rm inertial} \sim \rho U^2 L^2 $.  We see that  the Reynolds number is given by $\Re=  f_{\rm inertial}/f_{\rm viscous}$, and therefore in a low-Reynolds number flow the forces come primarily from viscous drag.

A fourth interpretation, more subtle, was offered by Purcell~\cite{purcell77}. He noted that, for a given fluid,  ${\cal F}=\eta^2/\rho$ has units of force, and that any body  acted upon by the force ${\cal F}$ will experience a Reynolds number of unity, independent of its size. Indeed, it is easy to see that $\Re=f_{\rm viscous}/{\cal F}$ and $\Re=(f_{\rm inertial}/{\cal F})^{1/2}$, and therefore a body with a Reynolds number of one will have $f_{\rm inertial}=f_{\rm viscous}={\cal F}$. A body moving at low Reynolds number experiences therefore forces smaller  than ${\cal F}$ (${\cal F} \approx 1$ nN for water).

What are the Reynolds numbers for swimming microorganisms~\cite{childress81}?
In water ($\rho \approx 10^3$ kg/m$^3$, $\eta \approx 10^{-3}$ Pa$\cdot$s), a  swimming bacterium such as {\it E. coli} with $U \approx 10$ $\mu$m/s and $L \approx 1$--$10$ $\mu$m has a Reynolds number $\Re\approx 10^{-5}$--$10^{-4}$. A human spermatozoon with $U \approx 200$ $\mu$m/s and $L \approx 50$ $\mu$m moves with  $\Re\approx 10^{-2}$. The larger ciliates, such as {\it Paramecium}, have $U \approx 1$ mm/s and $L \approx 100$ $\mu$m, and therefore $\Re\approx 0.1$~\cite{brennen77}. At these low Reynolds numbers, it is appropriate to study the limit $\Re=0$, for which  the Navier-Stokes equations~(\ref{NS}) simplify to the Stokes equations
\begin{equation}\label{Stokes}
-\nabla p + \eta \nabla^2 \u = 0 , \quad 
\nabla \cdot \u = 0.
\end{equation}
Since swimming flows are typically unsteady, we implicitly assume the typical frequency $\omega$ is small enough so that the  ``frequency Reynolds number" $\rho L\omega^2/\eta$ is also small.
Note that  Eq.~(\ref{Stokes}) is 
linear and independent of time, a fact  with important consequences for locomotion, as we discuss below.

Before closing this subsection, we point out an important property of Stokes flows called the reciprocal theorem. It is a principle of virtual work which takes a particularly nice form thanks to the linearity of Eq.~(\ref{Stokes}). Consider a volume of fluid $V$, bounded by a surface $S$ with outward normal $\n$, in which you have two solutions to Eq.~(\ref{Stokes}),
 $\u_1$ and  $\u_2$. If the stress fields of the two flows are  $\bsi_1$ and $\bsi_2$, then the reciprocal theorem states that the mixed virtual works are equal:
\begin{equation}\label{reciprocal}
\int\!\!\!\int_S \u_1\cdot \bsi_2 \cdot \n \,\d S=\int\!\!\!\int_S \u_2\cdot \bsi_1 \cdot \n \,\d S.
\end{equation}

%%%%%%%%%%%%%%%%%%%%%%%%%%%%%%%%
\subsection{Motion of solid bodies}
\label{motion}
When a solid body submerged in a viscous fluid is subject to a external force smaller than ${\cal F}$, it will move with a low Reynolds number.  What is its trajectory? Since Eq.~(\ref{Stokes}) is linear, 
the relation between kinetics and kinematics is linear.
Specifically, if the solid body is subject to an external force $\F$, and an external torque $\L$, it will move with velocity $\U$ and rotation rate $\Om$ satisfying \begin{equation}\label{resistance}
\left(
\begin{array}{c}
\F   \\
 \L   \\
\end{array}
\right)
=
\left(\begin{array}{ll}
\aA  &  \B  \\
 \B^T&  \C  \\
\end{array}\right)
\cdot 
\left(\begin{array}{c}
\U  \\
 \Om  \\
\end{array}\right),
\end{equation}
or the inverse relation
\begin{equation}\label{mobilities}
\left(
\begin{array}{c}
\U   \\
 \Om   \\
\end{array}
\right)
=
\left(\begin{array}{ll}
\M  &  \N  \\
 \N^T &  \OO  \\
\end{array}\right)
\cdot 
\left(\begin{array}{c}
\F  \\
 \L  \\
\end{array}\right).
\end{equation}
The matrix in Eq.~(\ref{resistance}) is the ``resistance" matrix of the body, and the matrix of Eq.~(\ref{mobilities}) is the ``mobility'' matrix. The reciprocal theorem~(\ref{reciprocal}) forces these matrices  
to be symmetric~\cite{happel}. 
Dimensionally, since low-Re stresses scale as $\sim \eta U/L$, the sub-matrices scale as
$[\aA]\sim \eta L$, 
$[\B]\sim \eta L^2$, 
$[\C]\sim \eta L^3$, 
and similarly 
$[\M]\sim (\eta L)^{-1}$, 
$[\N]\sim (\eta L^2)^{-1}$, 
$[\OO]\sim (\eta L^3)^{-1}$.  
For most problems, the details of the geometry of the body make these matrices impossible to calculate analytically.
The simplest example is that of a solid sphere of radius $a$, for which we have isotropic translational and rotational drag, $\M=\aA^{-1}=(6\pi\eta a)^{-1}{\bf 1}$, and $\OO=\C^{-1}=(8\pi\eta a^3)^{-1}{\bf 1}$; the cross-couplings $\M$ and $\B$ vanish by symmetry.

Three important features of Eqs.~(\ref{resistance}-\ref{mobilities}) needed to be emphasized for their implications for locomotion.  The first important property is drag anisotropy: The 
matrices $\aA$, $\M$, $\C$, and $\OO$ 
need not be isotropic (proportional to ${\bf 1}$).  As we discuss in \S\ref{life}, drag anisotropy 
is a crucial ingredient without which  biological locomotion could not occur at low Reynolds number. For  a simple illustration, consider
a slender prolate spheroid of major axis $a$ and minor axis $b$ with $a\gg b$. If $\e$ denotes the direction along the major axis of the spheroid, we have $\aA=A_\parallel \e \e +A_\perp ({\bf 1}-\e\e) $, with  $A_\parallel\approx 4\pi\eta a /[\ln(2a/b)-1/2]$ and $A_\perp\approx 8\pi\eta a[\ln(2a/b)+1/2]$.

Secondly, there exist geometries for which the matrices $\B$ and $\N$ are non-zero: chiral bodies, which lack a mirror symmetry plane. In that case, there is the possibility of driving translational motion through angular forcing---this strategy is employed by bacteria with rotating helical flagella 
(see \S\ref{actuation}). 

Thirdly, these matrices are important as they allow calculate the diffusion constants of solid bodies.  The fluctuation-dissipation theorem states that, in thermal equilibrium at temperature $T$, the translational diffusion constant of a solid body is given by the Stokes-Einstein relationship  $\D = k_B T\, \M$, where $k_B$ is Boltzmann's constant, while the rotational diffusion constant is given by  $\D_R = k_BT\, \OO$. The typical time scale  for a body to move by diffusion along its own size is $\tau_D\sim L^2/ [\D]$, while $\tau_R\sim[\D_R]^{-1}$ is the typical time scale for the reorientation of the cell by thermal forces alone. For a non-motile {\it E. coli} bacterium 
at room temperature, we have $[\D]\sim$ 0.1 $\mu$m$^2$/s in water;  while the time scale for thermal reorientation of the cell axis, $[\D_R]^{-1}$, is a few minutes.

\subsection{Flow singularities}
\label{singularities}

Since the Stokes' equations, Eq.~(\ref{Stokes}), are linear,  traditional mathematical methods to solve for flow and pressure  fields can rely on linear superposition. The Green's function to  Stokes flow with a   Dirac-delta forcing of the form ${\delta}(\x- 
\x') \F$ is given by
\begin{eqnarray}\label{Oseen}
\label{u} \u(\x)&=&\G(\x-\x')\cdot \F,\,\,{\rm with}\,\,\,\, \G (\r) 
= \frac{1}{8\pi\eta}\left(\frac{{\bf 1}}{r}+\frac{\r \r}{r^3}\right), 
\quad r=|\r|,\\
p(\x)& = & {\bf H}(\x-\x')\cdot \F,\,\,{\rm with}\,\,\,\, {\bf H} (\r) 
= \frac{\r}{4 \pi r^3}\cdot
\end{eqnarray}
The tensor $\G$ is known as the Oseen tensor, and the fundamental  
solution, Eq.~(\ref{Oseen}),  is termed a stokeslet \cite{hancock53}.  
Physically, it represents the flow field due to a point force, $\F$,  
acting on the fluid at the position $\x'$ as a singularity. The  
velocity field is seen to decay in space as $1/r$, a result which can  
also be obtained by dimensional analysis. Indeed, for a three-dimensional force ${\bf F}$ acting on the fluid, and by linearity of  
Stokes' flow, the flow velocity has to take the form $u \sim F f 
(\theta,r,\eta)$, where $\theta$ is the angle between the direction of  
$\F$ and $\r$, and where $F$ is the magnitude of the force.  
Dimensional analysis leads to $u\sim g(\theta) F /\eta r $ with a $1/r 
$ decay.

An important property of the  stokeslet solution for locomotion is  
directional anisotropy. Indeed, we see from Eq.~(\ref{u}) that if we  
evaluate the velocity in the direction parallel to the applied force,  
we obtain that $u_\parallel ={F}/{4\pi\eta r}$, whereas  the   
velocity in the direction perpendicular to the force is given by $u_ 
\perp={F}/{8\pi\eta r}$. For the same applied force, the flow field  
in the parallel direction is therefore twice that in the  
perpendicular direction ($u_\parallel = 2 u_\perp$). Alternatively,  
to obtain the same velocity, one would need to apply a force in the  
perpendicular direction twice as large as in the parallel direction  
($F_\perp= 2 F_\parallel  $). Such anisotropy, which is reminiscent  
of the anisotropy in the mobility matrix for long slender bodies (\S 
\ref{motion}; see also \S\ref{RFT})  is at the origin of the drag-based propulsion method employed by swimming microorganisms  (see \S 
\ref{thrust}).

 From the fundamental solution above, Eq.~(\ref{Oseen}), the complete  
set of singularities for viscous flow can be obtained by  
differentiation  \cite{chwang75}. One derivative leads to force-dipoles, with flow fields decaying as $1/r^2$. Two derivatives leads  
to source-dipole (potential flow also known as a doublet), and force-quadrupoles, with velocity decaying in space as $1/r^3$.  Higher-order singularities are easily obtained by subsequent differentiation.

A well-chosen distribution of such singularities can then be used to  
solve exactly Stokes' equation in a variety of geometry. For example,  
the Stokes flow past a sphere is a combination of a stokeslet and  a
source-dipole at the center of the sphere \cite{batchelor1967}.  For  
spheroids, the  method was pioneered by Chwang and Wu \cite 
{chwang75}, and we refer to Refs.~\cite{kimbook,pozrikidis_BI} for   a 
textbook treatment. A linear superposition of singularities is also  
at the basis of the boundary integral method to computationally solve  
for Stokes flows using solely velocity and stress information  at the  
boundary  (see Refs.~\cite{poz,pozrikidis_BI}).

When a flow field is described by a number of different flow  
singularities, the singularity with the slowest spatial decay
  is the one  that dominates  in the far field. Since a cell swimming  
in a viscous fluid at low-Reynolds numbers is force- and torque-free  
(Eq.~\ref{noforce} below), the flow singularities that describe point-forces (stokeslets) and point-torques (antisymmetric force-dipole, or  
rotlets) cannot be included in the far-field description. As a  
result, the flow field far from a swimming cell is in general well  
represented by a symmetric force-dipole, or stresslet \cite 
{batchelor70_2}.  Such far-field behavior has important consequences  
on  cell-cell hydrodynamic interactions as detailed in \S\ref{cellcell}.

%%%%%%%%%%%%%%%%%%%%%%%%%%%%%%%%
%%%%%%%%%%%%%%%%%%%%%%%%%%%%%%%%

\section{Life at low Reynolds number}
\label{life}

We now consider the general problem of self-propelled motion at  low Reynolds number. 
We call a body a ``swimmer" if by deforming its surface it is able to sustain movement through fluid in the absence of external (non-hydrodynamic) forces and torques. Note that the ``body" includes appendages such as the cilia covering a \textit{Paramecium} or the helical flagella of \textit{E. coli}.

%%%%%%%%%%%%%%%%%%%%%%%%%%%%%%%%
\subsection{Reinterpreting the Reynolds number}

We first offer an alternative interpretation of the Reynolds number in the context of swimming motion. Let us consider a swimmer of mass $m$ and size $L$ swimming with 
velocity $U$ through a viscous fluid of density $\rho$ and viscosity $\eta$. Suppose the 
swimmer suddenly stops 
deforming its body; it will then decelerate
according to Newton's law 
$ma=f_{{\rm drag}}$. What is the typical length scale $d$ over  
which the swimmer will 
coast  
due to the inertia of its movement? For motion at high Reynolds number, as in the case of a human doing the breaststroke, the typical drag is $f_{{\rm drag}}=f_{{\rm inertial}}\sim\rho U^2L^2$, leading to a deceleration $a\sim\rho U^2L^2/m$. The swimmer coasts a length $d\sim U^2/a\sim m/(\rho L^2)$.  If the swimmer has a  
density $\rho_\mathrm{s}\sim m/L^3$, we see that the dimensionless coasting distance is given by the ratio of densities, $d/L\sim \rho_\mathrm{s}/\rho$. A human swimmer in water can cruise for a couple of meters. In contrast, for motion at low Reynolds number, the drag force has the viscous scaling, $f_{{\rm drag}}=f_{{\rm viscous}}~\eta UL$, and the swimmer can coast 
a distance ${d}\approx L\, \Re \,{\rho_\mathrm{s}}/{\rho}$, where $\rho_\mathrm{s}\sim m/L^3$ is the density of the swimmer. For a swimming  bacterium such as {\it E. coli}, this argument leads to $d\approx 0.1$\,nm~\cite{purcell77}. 
For $\Re<1$, The Reynolds number can therefore be interpreted as a nondimensional cruising distance.

A consequence of this analysis is that in a world of low-Reynolds number, the response of the fluid to the motion of boundaries is instantaneous. This conclusion was anticipated by our second interpretation of the Reynolds number (\S\ref{fluids}), where we saw that in the limit of very low $\Re$, velocity perturbations diffuse rapidly relative to the rate at which fluid particles are carried along by the flow.
To summarize, the rate at which the  momentum of a low-$\Re$ swimmer is changing is completely negligible when compared to the typical magnitude of the  forces from the surrounding viscous fluid. As a result, Newton's law becomes a simple statement of instantaneous balance between external and fluid forces and torques
\begin{equation}\label{noforce}
\F_{\rm ext}(t) + \F(t)={\bf 0},\quad \L_{\rm ext}(t)+\L(t)={\bf 0}.
\end{equation}
In most cases, there is no external forces, and  $\F_{\rm ext}(t)={\bf 0}$. Situations 
where $\L_{\rm ext}(t)$ is non-zero include the locomotion of nose-heavy or bottom-heavy cells \cite{pedley92};  
in all other cases we will assume  $\L_{\rm ext}(t)={\bf 0}$.

%%%%%%%%%%%%%%%%%%%%%%%%%%%%%%%%
\subsection{The swimming problem}

Mathematically, the swimming problem is stated as follows. Consider a body submerged in a viscous fluid. In a reference frame fixed with respect to some arbitrary reference point in its body, the swimmer 
deforms its surface in a prescribed time-varying fashion given by a velocity field on its surface, $\u_S(t)$. The velocity field $\u_S(t)$  is  the ``swimming gait''. A swimmer is a deformable body by definition, but it may be viewed at every instant as a solid body with
with 
unknown velocity $\U(t)$  and rotation  rate $\Om(t)$. The instantaneous velocity on the swimmer's surface is therefore given by $\u=\U + \Om\times \x + \u_S$, which provides the boundary conditions needed to solve  Eq.~(\ref{Stokes}).
The unknown values of  $\U(t)$ and $\Om(t)$ are determined by satisfying Eq.~(\ref{noforce})

The mathematical difficulty of  
solving the swimming problem arises  
from having to solve for the Stokes flow with unknown boundary condition; in that regard, low-$\Re$ swimming is reminiscent of an eigenvalue problem. A great simplification  was derived by Stone and Samuel~\cite{stone96}, who applied  the reciprocal theorem,  Eq.~(\ref{reciprocal}), to the swimming problem. Recall that the reciprocal theorem involves two different flow problems for the same body. Let  $\u$ and $\bsi$ the velocity and stress fields we seek in the swimming problem. For the second flow problem, suppose 
$\hat\u$  and $\hat{\bsi}$  are the velocity and stress fields for the dual problem of instantaneous solid body motion of the swimmer with velocity $\hat\U$ and rotation rate $\hat{\Om}$.  This problem correspond to subjecting the shape, $S(t)$, to an external force, $\hat \F$, and torque, $\hat\L$.  
Applying the reciprocal theorem~(\ref{reciprocal}), we obtain \cite{stone96}
\begin{equation}\label{SS}
\hat\F \cdot \U + \hat\L \cdot \Om = -\int\!\!\!\int_{S(t)}\u_S\cdot \hat{\bsi}\cdot \n\,\d S.
\end{equation}
Equation~(\ref{SS}) shows explicitly how the swimming velocity $\U$ and rotation rate $\Om$ may be found in terms of the gait $\u_S$, given the solution to the dual problem of the flow induced by the motion of the rigid body with instantaneous shape $S(t)$, subject to force $\hat\F$ and torque $\hat\L$. Note that since $\hat\F$ and $\hat\L$ are arbitrary, Eq.~(\ref{SS}) provides enough equations to solve for all components of the swimming kinematics. Note also that for squirming motion, where the shape of the swimmer surface remains constant ($\u_S\cdot\n=0$),
Eq.~(\ref{SS}) simplifies further. For a spherical squirmer of radius $a$~ \cite{stone96}, we have the explicit formulas~\cite{stone96}
\begin{equation}\label{SSSS}
 \U  = -\frac{1}{4\pi a^2}\int\!\!\!\int_{S}\u_S\,\d S,
 \quad
 \Om=-\frac{3}{8\pi a^3}\int\!\!\!\int_{S}\n\times \u_S  \,\d S.
\end{equation}

%%%%%%%%%%%%%%%%%%%%%%%%%%%%%%%%
\subsection{Drag-based thrust}
\label{thrust}
\begin{figure}[t]
\begin{center}
 \includegraphics[width=0.7\textwidth]{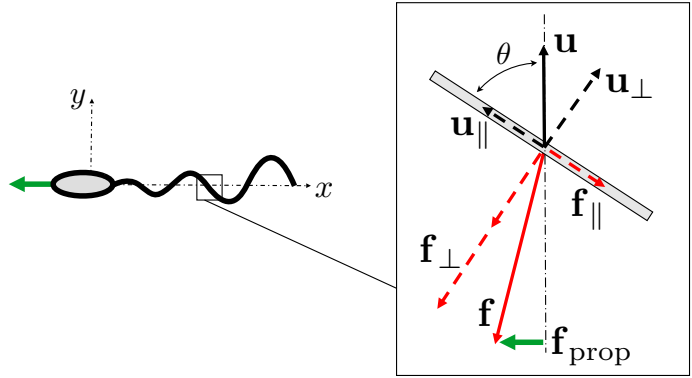}
\end{center}
 \caption{Physics of drag-based thrust: The drag anisotropy for slender filaments provides a means to generate forces perpendicular to the direction of the local actuation (see text for notation).}
\label{drag}
\end{figure}

Most biological swimmers exploit the motion of slender appendages (``flagella'') for locomotion.  This limit of slender bodies allows us to provide a physical, intuitive way to understand the origin of locomotion through drag; the specifics of biological and artificial flagellar actuation will be discussed in \S\ref{actuation}.

The fundamental property allowing for drag-based thrust of slender filaments is their drag anisotropy, as introduced in \S\ref{motion}. Indeed, consider a thin filament immersed in a viscous fluid which is motionless but for flows induced by the deformations of the filament. 
The shape of the filament is described by its tangent vector $\t(s)$ at distance $s$ along the filament, and its instantaneous deformation is described by the velocity field $\u(s,t)$, where $t$ is time. 
For asymptotically slender filaments, (see \S\ref{classics}), as in the case of prolate spheroids, the local viscous drag force per 
unit length opposing the motion of the filament is 
\begin{equation}\label{local_drag}
\f  = -\xi_\parallel \u_\parallel - \xi_\perp \u_\perp,
\end{equation}
where  $\u_\parallel$ and $\u_\perp$ are the projections of the local velocity on the directions parallel and perpendicular to the filament, {\it i.e.} $\u_\parallel=(\u\cdot \t)\t$ and $\u_\perp =\u-\u_\parallel$;  $\xi_\parallel$ and $\xi_\perp$ are the 
corresponding drag coefficients (typically $\xi_\perp/ \xi_\parallel\approx 2$). 

The origin of drag-based thrust relies on  the following two physical ideas: (a) The existence of drag anisotropy means that propulsive  forces can be created at a right angle with respect to the local direction of motion of the filament, and (b) a filament can deform in time-periodic way and yet create non-zero time-averaged propulsion.  
To illustrate these ideas, consider the beating filament depicted in Fig.~\ref{drag}. Any short segment of the filament may be regarded as straight and moving with velocity $\u$ at an angle $\theta$ with the local tangent (Fig.~\ref{drag}, inset). This velocity resolves into components  $u_\parallel=u\cos\theta$ and $u_\perp=u\sin\theta$, leading to drag per unit length components $f_\parallel=-\xi_\parallel u_\parallel=-\xi_\parallel u\cos\theta$ and
$f_\perp= - \xi_\perp u_\perp=- \xi_\perp u\sin\theta $.
For isotropic drag, $\xi_\parallel=\xi_\perp$, and the force on the filament has the same direction as the velocity of the filament; however, if $\xi_\parallel\neq\xi_\perp$, the drag per unit length on the filament includes a component $\f_{\rm prop}$ which is  
perpendicular to the direction of the velocity, 
\begin{equation}\label{prop}
\f_{\rm prop}=(\xi_\parallel-\xi_\perp)u\sin\theta\cos\theta\,\e_x.
\end{equation}

In addition, in order to generate a net propulsion from a time-periodic movement, we 
see from Eq.~(\ref{prop}) that both the filament velocity $u$ and its orientation angle $\theta$ need to vary periodically in time. 
For example, actuation with a given $\u$ and $\theta$, followed by the change  $\u\to-\u$ and $\theta\to\pi-\theta$, leads to a a propulsive force with a constant sign; 
in contrast, 
actuation in which only $\u\to-\u$ changes periodically leads to zero average force.

It is important to realize  that  this result relies on two ideas 
associated with two different length scales. One is purely {\it local}, and states that with the appropriate geometry and actuation, a force can be created  in the direction perpendicular the motion of the filament. This conclusion  relies explicitly on the properties of Stokes flows, and in a world with isotropic viscous friction ($\xi_\perp=\xi_\parallel$), locomotion would not be possible \cite{gray55,becker03}. The second idea is a {\it global} constraint that says that the periodic actuation of the filament needs to be sufficiently subtle 
to generate non-zero forces on average; this property is generally  known as Purcell's scallop theorem~\cite{purcell77}.

%%%%%%%%%%%%%%%%%%%%%%%%%%%%%%%%
\subsection{The scallop theorem}

As pointed out above, the Stokes equation---Eq.~(\ref{Stokes})---is
linear and independent of time.  
These properties lead to {\it kinematic reversibility}, an important and well-known symmetry property associated with the motion of any body  at zero Reynolds
number~\cite{happel,leal}.
Consider the motion of a solid body with an instantaneous prescribed
velocity, $\U$, and rotation rate, $\Om$, together with the  flow
field surrounding it. If we apply the scaling
$\U\to\alpha\U$ and $\Om\to\alpha\Om$, then by linearity, the entire
flow and pressure field transform as 
 $\u\to \alpha\u$ and $p
\to \alpha p$. Consequently,  the instantaneous flow streamlines   
remain identical,
and the fluid stresses undergo a simple linear scaling, resulting in
the symmetry $\F\to\alpha\F$ and $\L\to \alpha\L$ for the force and
torque acting on the body.
In particular, when $\alpha=-1$, this means that an instantaneous
reversing of the forcing does not modify the flow patterns, but only
the direction in which they are occurring.

When applied to low-Reynolds number locomotion, the linearity and
time-independence of Stokes equation of motion lead to two important
properties~\cite{purcell77}. The first one is that of {\it rate
independence}:  If  a body undergoes surface deformation, the
distance travelled by the swimmer between two different surface
configurations does not depend on the rate at which the surface
deformation occurs but only on its geometry (i.e. the sequence of
shapes the swimmer is going through between these two configurations).

A mathematical proof of this statement can be outlined as follows. We
consider for simplicity swimmers with no rotational motion (an
extension to the case $\Om\neq{\bf 0}$ is straightforward). Consider
a body deforming its surface between two different configurations
identified by time $t_0$ and  $t_1$. We denote by $\r_S$ the
positions of points on the surface of the swimmer.
  From Eq.~(\ref{SS}), we know that the instantaneous speed of
locomotion is given by a general integral of the form
\begin{equation}\label{U}
\U(t)= \int\!\!\!\int_{f(\r_S)}\dot \r_S \cdot \g(\r_S)\,\d S,
\end{equation}
where we have used %that 
$\u_S=\dot \r_S = \d \r_S /\d t$. The net
motion of the swimmer between
$t_0$ and $t_1$ is therefore given by
\begin{equation}\label{dX}
\Delta \X=\int_{t_0}^{t_1} \U(t)\,\d t.
\end{equation}
Now consider the same succession of swimmer shapes, but occurring  at
a different rate. We describe it by a mapping $t'=f(t)$, with $t_i=f
(t_i)$ for $i=0,1$, such that the shape $\r_S'(t')$ is the same as
the shape $\r_S(t)$ for all times.  We now have
\begin{equation}
\Delta \X'=\int_{t_0}^{t_1} \U'(t')\,\d t'
=\int_{t_0}^{t_1} \U'(f(t))f'(t)\,\d t,
\end{equation}
where
\begin{equation}
\U'(t')f'(t)
= \int\!\!\!\int_{f(\r_S')}\frac{\d  \r_S'}{\d t'} f'(t) \cdot \g
(\r_S)\,\d S
= \int\!\!\!\int_{f(\r_S)}\frac{\d  \r_S}{\d t}  \cdot \g(\r_S)\,\d S
\end{equation}
using the chain rule. We see therefore that $\U'(t')f'(t)=\U(t)$, and
therefore $\Delta \X=\Delta \X'$. The net distance traveled by the
swimmer does not depend on the rate at which it is being deformed,
but only on the geometrical sequence of shape.
One consequence of this property is that many aspects of low-Reynolds
number locomotion  can be addressed %with 
using a purely geometrical point
of view~\cite{shapere87,shapere89_1,shapere89_2,yariv06,avron08}.

The second important property of swimming without inertia is the
so-called {\it scallop theorem}: If the sequence of shapes displayed by
a  swimmer deforming in a time-periodic fashion is identical when
viewed after a time-reversal transformation (a class of surface deformation
termed ``reciprocal deformation''), then the swimmer cannot move on
average. This second property puts a strong geometrical constraint on
the type of swimming motion which will be effective at low Reynolds
numbers.

An outline of the proof can be offered as follows (again, we consider
purely translational motion for simplicity). Let us consider a
swimmer deforming its body  between times $t_0$ and $t_1$,  and a
sequence of shape described by $\r_S$. We assume that $\r_S(t_0)=\r_S
(t_1)$ so that we are looking at the swimming motion over one period
of surface deformation.
The net distance traveled by the swimmer is given by Eqs.~(\ref{U}--\ref{dX}). We now consider the motion between $t_2$ and $t_3$
obtained by time-reversal  symmetry of the first motion; we describe
it by a temporal mapping $t'=f(t)$, with $t_2=f(t_1)$ and $t_3=f(t_0)
$, defined such that the shape $\r_S'(t')$ is the same as the shape $
\r_S$(t). In that case, using similar arguments at those used to
demonstrate the first property above, we see that
\begin{equation}
\Delta \X'=\int_{t_2}^{t_3} \U'(t')\,\d t'
=\int_{t_1}^{t_0} \U(t)\,\d t
=-\int_{t_0}^{t_1} \U(t)\,\d t
=-\Delta \X,
\end{equation}
and reversing the sequence of shape leads therefore to the opposite
distance traveled. However, since the body deformation is reciprocal,
the sequence of shape between $t_2$ and $t_3$ is the same as between
$t_0$ and $t_1$, and therefore the distance traveled should be the
same independently of the direction of time: $\Delta \X'=\Delta \X$.
By combining the two results, we see therefore  that $\Delta \X={\bf
0}$: Reciprocal motion cannot be used for locomotion at low Reynolds
numbers. Note here that in order to demonstrate this result,
we do not need to assume anything about the geometry of the fluid
surrounding the swimmer, so the scallop theorem remains valid near
solid walls, and more generally in confined environments.

In his original article, Purcell illustrated this result by using the
example of a scallop, a mollusk that opens and closes its shell in a
time period fashion. A low-Reynolds number scallop undergoes a  
reciprocal deformation, and therefore cannot swim in the absence of inertia
(independent of the rate  of opening and closing)\footnote{A  
real scallop actually swims at high Reynolds number, a regime for which the constraints of the theorem of  
course do not apply.}.  Another  example  of a reciprocal deformation is a
dumbbell, made of two solid spheres separated by time-periodic distance.
More generally, bodies with a single degree of freedom deform in a
reciprocal fashion, and cannot move on average.

Successful swimmers must display therefore non-reciprocal body kinematics. In his original paper, Purcell proposed a simple example of non-reciprocal body deformation, a two-hinged body composed of three rigid links rotating out-of-phase with each other, now refereed to as Purcell's swimmer  \cite{becker03}. Another elementary example  is a trimer, made of three rigid spheres whose separation distances vary in a time-periodic fashion with phase differences \cite{najafi04}. More examples are discussed in  \S\ref{artificial}. Note that, mathematically, the presence of non-reciprocal kinematics is a necessary but not sufficient  condition to obtain propulsion. A simple counterexample is a two swimmers which are mirror-images of each other and  arranged head-to-head; although the kinematics of the two bodies taken together is non-reciprocal, the mirror symmetry forbids net motion of their center of mass.

For biological bodies deforming in a continuous fashion, the
prototypical non-reciprocal deformation is a wave.
Consider a continuous filament of length $L$ deforming with small
amplitude $y(x,t)$ (i.e. for which $  |{\p y}/{\p x}|\ll 1$); in that
case, the propulsive force generated along the filament, Eq.~(\ref
{prop}), is given by
\begin{equation}
\F_{\rm prop}\approx (\xi_\perp-\xi_\parallel)\int_0^L \left(\frac{\p
y}{\p t}\frac{\p y}{\p x}\right)\d x\,\e_x.\label{Fpropeqn}
\end{equation}
If the filament deforms as a planar wave  traveling in the $x$-
direction, $y(x,t)=f(x-ct)$, the force is given by
$\F_{\rm prop}= c(\xi_\parallel-\xi_\perp)\int f^2(\eta)\d \eta\,\e_x
$ and propulsion is seen to occur in the direction opposite to that
of the wave ($cF_{\rm prop}<0$). Mathematically, a wave-like
deformation allows the product $({\p y}/{\p t}\cdot{\p y}/{\p x})$ to keep
a constant sign between $x=0$ and $x=L$, and therefore all portions
of the filament  contribute to generating propulsion. In general, all
kinds of three-dimensional wave-like deformations lead to
propulsion, in particular helical waves of flexible filaments~\cite
{chwang71}.

Finally, it is worth emphasizing that the scallop theorem is
strictly valid in the limit where all the relevant Reynolds numbers
in the swimming problem  are set to zero. Much recent work has been
devoted to the breakdown of the theorem with inertia, and the
transition  from the Stokesian realm to the Eulerian realm is found
to be either continuous or discontinuous depending on the spatial
symmetries in the problem considered \cite
{childress04,childress_conf,vandenberghe04,alben05,vandenberghe06,lu06,lauga_purcell}.

%%%%%%%%%%%%%%%%%%%%%%%%%%%%%%%%
%%%%%%%%%%%%%%%%%%%%%%%%%%%%%%%%
\section{Historical studies, and further developments}
\label{classics}

In this section we turn to the first calculations of the swimming velocities of model microorganisms. We consider two simple limits: 
(1) propulsion by small amplitude deformations of the surface of the swimmer, and (2) propulsion by the motion of a slender filament. Although these limits are highly idealized, our calculations capture essential physical aspects of swimming that are present in more realistic situations.

\subsection{Taylor's swimming sheet}
\label{TaylorSheetSection}

In 1951, G. I.  Taylor asked how a microorganism could propel itself using viscous forces alone, rather than imparting momentum to the surrounding fluid as fish do~\cite{taylor51}. To answer this question, he calculated the flow induced by propagating transverse waves of small amplitude on a sheet immersed in a viscous fluid. In this subsection, we review Taylor's calculation~\cite{taylor51}. The sheet is analogous to the beating flagellum of a spermatazoon, but since the flow is two-dimensional, the problem of calculating the induced flow is greatly simplified. The height of the sheet over the plane $y=0$ is  
\begin{equation}
h=b\sin(kx-\omega t),\label{sheet}
\end{equation}
where the $x$-direction is parallel to the direction of propagation of the wave, $b$ is the amplitude, $k$ is the wavenumber, and $\omega$ is the frequency of the oscillation. Note that we work in the reference frame in which the material points of the sheet move up and down, with no $x$-component of motion. The problem is further simplified by the assumption that the amplitude is small compared to the wavelength $2\pi/k$. Note that the motion of Eq.~(\ref{sheet}) implies that the sheet is extensible. If the sheet is intensible, then the material points of the sheet make narrow figure eights instead of moving up and down; nevertheless the extensible and inextensible sheets have the same swimming velocity to leading order in $bk$. 

To find the flow induced by the traveling-wave deformation, solve the Stokes equations with no-slip boundary conditions at the sheet,
\begin{equation}
\u(x,h(x,t))=-b\omega\cos(kx-\omega t)\e_y,\label{TaylorSheetBC}
\end{equation}
with an unknown but uniform and steady flow far from the sheet,
\begin{equation}
\lim_{y\rightarrow\infty} \u(x,y)=-\U.\label{TaylorSheetBCFar}
\end{equation}
Since we work in the rest frame of the sheet, $\U$ is the swimming velocity of the sheet in the laboratory frame, in which the fluid is at rest at $y\rightarrow\infty$. Although it turns out in this problem that the leading order swimming speed is steady in time, other situations lead to unsteady swimming speeds. In all cases we are free to use non-inertial frames---even rotating frames---without introducing fictitious forces, since inertia may be disregarded at zero Reynolds number.

Although $\U$ is unknown, Taylor found that no additional conditions are required to determine $\U$; instead, there is a unique value of $\U$ consistent with the solution to the Stokes equations and the no-slip boundary condition~(\ref{TaylorSheetBC}). It is also important to note that although the Stokes equations are linear, the swimming speed $\U$ is not a linear function of the amplitude $b$, since $b$ enters the no-slip boundary condition both on the right-hand side of Eq.~(\ref{TaylorSheetBC}) and implicitly on the left-hand side through Eq.~(\ref{sheet}). In fact, symmetry implies that the swimming speed $\U$ must be an even function of $b$. Replacing $b$ by $-b$ amounts to translating the wave~(\ref{sheet}) by half a wavelength. But any translation of the wave cannot change the swimming speed; therefore, $\U$ is even in $b$. 

Taylor solved the swimming problem by expanding the boundary condition~(\ref{TaylorSheetBC}) in $bk$, and solving the Stokes equations order by order. We will consider the leading term only, which as just argued is quadratic in $b$. Since the swimming velocity $\U$ is a vector, it must be proportional to the only other vector in the problem, the wavevector $\k=k\e_x$. For example, if we were to consider the superposition of two traveling waves on the sheet, propagating in different directions, we would expect the swimming direction to be along the vector sum of the corresponding wavevectors. Dimensional analysis determines the remaining dependence of $\U$ on the parameters of the problem: $\U\propto\omega\k b^2$. Taylor's calculation yields the proportionality constant, with sign:
 \begin{equation}
 \U=-\frac{1}{2}\omega\k b^2.\label{TaylorSheetSpeed}
 \end{equation}
Note that dimensional considerations require the swimming speed to be independent of viscosity. This result holds due to our somewhat unrealistic assumption that the waveform~(\ref{sheet}) is prescribed, independent of the load. However, the rate $W$ that the sheet does work on the fluid does depend on viscosity. The net force per wavelength exerted by the sheet on the fluid vanishes, but by integrating the local force per area ($\sim\mu\omega bk$) against the local velocity ($\sim\omega b$), Taylor found 
\begin{equation}
W=\eta\omega^2 k b^2.\label{Taylor51work}
\end{equation}
Note that only the first-order solution for the flow is required to calculate $W$.
In \S\ref{actuation} we consider more realistic models that account for the internal mechanisms that generate the deformation of the swimmer. Such models can predict a viscosity-dependence in the swimming speed, since the shape of the beating filament may depend on viscosity~\cite{camalet99,camalet00:njp,FuWolgemuthPowers2008}. And in \S\ref{complex} we show how the speed of a swimmer in a complex fluid can depend on material parameters, even for the swimming problem with prescribed waveform. 

According to Eq.~(\ref{TaylorSheetSpeed}), the swimmer moves in the direction opposite to the traveling wave. It is instructive to also consider the case of a longitudinal wave, in which the material points in the frame of the sheet undergo displacement $\delta(x,t)=b\sin(kx-\omega t)$, yielding the no-slip boundary condition
\begin{equation}
\u(x+\delta(x,t),y=0)=-b\omega\cos(kx-\omega t)\e_x.\label{TaylorSheetLongBC}
\end{equation}
For a longitudinal wave, the swimming velocity is in the \textit{same} direction as the traveling wave. 

\begin{figure}[t]
\begin{center}
  \includegraphics[width=0.9\textwidth]{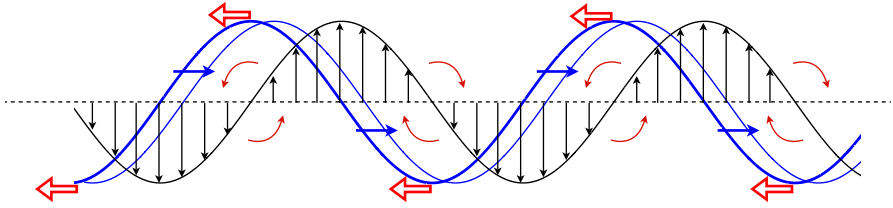}
\end{center}
  \caption{Physical interpretation of the swimming direction for  
Taylor's swimming sheet. The vertical displacements (black vertical arrows and  
black line) along the sheet as the wave of deformation is traveling  
to the right (blue lines and rightward arrows) leads to creation of flow  
vorticity of alternating signs (curved red arrows), which induces  
leftward motion of the sheet (straight red leftward arrows).}
\label{fig:sheet}
\end{figure}

The direction of swimming of Taylor's sheet can be understood on the  
basis of the following simple arguments. Let us consider a sheet  
deforming as a sine-wave propagating to the right (Fig.~\ref 
{fig:sheet}). Consider the vertical displacements occurring along the  
sheet as a result of the propagating wave. During a small interval of  
time, the original wave (Fig.~\ref{fig:sheet}, thick blue line) has  
moved to the right (Fig.~\ref{fig:sheet}, thin blue line), resulting  
in vertical displacements given by a profile which is  $\pi/2$ out of  
phase with the shape of the sheet. Indeed,  wherever the sheet has a  
negative slope, the material points go up as the wave progresses to  
the right, whereas everywhere the sheet has a positive slope, the  
material points go down as a result of the wave. The resulting  
distribution of vertical velocity along the sheet is illustrated in
Fig.~\ref{fig:sheet} by the black line and vertical arrows. This velocity  
profile forces the surrounding fluid, and we see  
that the fluid acquires vorticity of alternating sign along the sheet  
(illustrated by the curved red arrows in Fig.~\ref{fig:sheet}). The  
vorticity is seen to be positive  near the sheet crests, whereas it  
is negative near the sheet valleys.   The longitudinal flow  
velocities associated with this vorticity distribution allow us to  
understand the swimming direction. In the case of positive vorticity,  
the   velocity of the induced vortical flow is to the left at the  
wave crest, which is the current position of the sheet. In the case  
of negative vorticity, the induced flow velocity is to the left in  
the valleys, which is also where the sheet is currently located. As a  
consequence, the longitudinal flow field induced by the transverse  
motion of the sheet leads to flow contribution which are to the left  
in all cases, and the sheet is seen to swim to the left (straight red  
leftward arrows). Note that if the sheet were not free to move, then it would  
create an external flow field that cancels the sheet-induced flow,  
and the net flow direction would therefore be  to the right---the  
sheet acts as a pump.

There are many generalizations to Taylor's 1951 calculation. Taylor himself considered the more realistic geometry of an infinite cylinder with a propagating transverse wave~\cite{taylor52}. In this case, there is a new length scale $a$, the radius of the cylinder, and the calculation is organized as a power series in $b/a$ rather than $bk$. In the limit $ak\rightarrow0$, the swimming velocity has the same form as the planar sheet~\cite{taylor52}. With a cylinder, we can study truly three-dimensional deformations of a filament, such as helical waves.   A helical wave can be represented by the superposition of two linearly polarized transverse waves, with perpendicular polarizations and a phase difference of $\pi/2$. If these waves have the same amplitude, speed, and wavelength, then the swimming velocity is twice the velocity for a single wave. Although the hydrodynamic force per unit wavelength acting on the waving filament vanishes, there is a nonvanishing net hydrodynamic torque per unit wavelength, which is ultimately balanced by the counter-rotation of the head of the organism~\cite{taylor52}.

The Taylor sheet calculation may also be extended to finite objects. For example, to model the locomotion of ciliates such as \textit{Opalina} and \textit{Paramecium},  Lighthill introduced the ``envelope model," in which the tips of the beating cilia that cover the cell body are represented by propagating surface waves~\cite{Lighthill1952,Blake1971a,Blake1971c}. Perhaps the simplest version of the envelop model is the two-dimensional problem of an undulating circle in the plane, which may equivalently be viewed as an infinite cylinder with undulations traveling along the circumferential direction~\cite{Blake1971c}.  Unlike the problem of a rigid cylinder towed through liquid at zero Reynolds number, the undulating cylinder does not suffer from the Stokes paradox~\cite{batchelor1967,landauFM}, since the total force on the cylinder is zero. 
And unlike the Taylor sheet problem, where the swimming speed emerges self-consistently, the condition of vanishing force is required to determine the swimming speed of the undulating cylinder. The problematic solutions that lead to the Stokes paradox are the same ones that lead to a net force, as well as a diverging kinetic energy, and are therefore eliminated in the swimming problem~\cite{Blake1971c}. 

In three dimensions, the swimming speed is also determined by the condition of vanishing total force~\cite{Lighthill1952,Blake1971a}, but since the solution to the problem of towing a sphere with an external force is well-behaved, we may also consider solutions with nonzero force. These solutions must be well-behaved if we are to apply  reciprocal theorem of \S\ref{fluids}, which gives perhaps the shortest route to calculating the swimming speed~\cite{stone96,Ehlers_etal1996}. We can also gain additional insight into why the swimming speed for a prescribed deformation of the surface is independent of viscosity. Using the linearity of Stokes flow, at any instant we may decompose the flow field generated by the swimmer into a ``drag flow" and a ``thrust flow," $\mathbf{v}=\mathbf{v}_\mathrm{d}+\mathbf{v}_\mathrm{t}$~\cite{childress81}. The drag flow is the flow induced by freezing the shape of the swimmer and towing it at velocity $\U$ with a force $\F$, to be determined. The thrust flow is the flow induced by the swimmer's motion at that instant when it is prevented from moving by an anchoring force $\F_\mathrm{anchor}$, which is determined by the shape and and rate of change of shape of the swimmer at that instant. Superposing the two flows, and adjusting $\F$ to cancel $\F_\mathrm{anchor}$ yields the swimming speed $\U$. Since the linearity of Stokes flow implies that both $\F$ and $\F_\mathrm{anchor}$ depend linearly on viscosity, the swimming velocity $\U$ does not depend on viscosity. Note that the same conclusion follows from an examination of the reciprocal theorem formula, Eq.~(\ref{SS}).

Finally, in the sheet calculation, it is straightforward to include the effects of inertia and show that if flow separation is disregarded, the swimming speed decreases with Reynolds number, with an asymptotic value at high Reynolds number of half the value of Taylor's result (\ref{TaylorSheetSpeed})~\cite{Reynolds1965,Tuck1968}. At zero Reynolds number, the effect of a nearby rigid wall is to increase the swimming speed as the gap between the swimmer and the wall decreases, for prescribed waveform~\cite{Reynolds1965}. However, if the swimmer operates at constant power, the swimming speed decreases as the gap size decreases~\cite{Reynolds1965}.

\subsection{Local drag theory for slender rods}
\label{RFT}
 
All the calculations of the previous subsection are valid when the amplitude of the deflection of the swimmer is small. These calculations are valuable since they allow us to identify qualitative trends in the dependence of the swimming velocity on geometric and, as we shall see in \S\ref{complex}, material parameters. However, since real flagella undergo large-amplitude deformations, we cannot expect models based on small-amplitude deformations to give accurate results. Fortunately, we may develop an alternative approximation that is valid for large deformations by exploiting the fact that real flagella are long and thin. The idea is to model the flow induced by a deforming flagellum by replacing the flagellum with a line of singular solutions to Stokes flow of appropriate strength. In this subsection and the following subsection we develop these ideas, first in the simplest context of local drag theory, also known as resistive force theory, and then using the more accurate slender-body theory.

\begin{figure}[t]
\begin{center}
 \includegraphics[width=0.6\textwidth]{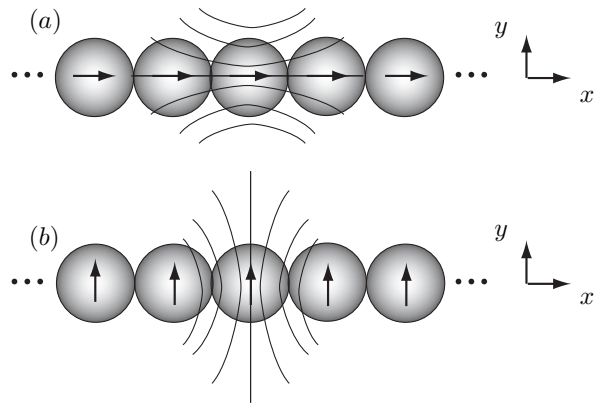}
\end{center}
 \caption{Model for a rod subject to an external force along the rod (a) or perpendicular to the rod (b). The arrows at the centers of the spheres represent external forces. The flow field of the central sphere only is shown. Each sphere induces a similar flow field that helps push the other spheres along. }
\label{RFTfig}
\end{figure}

To introduce local drag theory, we develop an intuitive model for calculating the resistance matrix of a \textit{straight} rigid rod of length $L$ and radius $a$.  Our model is not rigorous, but it 
captures the physical intuition behind the more rigorous theories described below.  Suppose the rod is subject to an external force $\F_\mathrm{ext}$. Suppose further that this force is uniformly distributed over the length of the rod with a constant force per unit length. Our aim is to find an approximate form for the resistance matrix, or equivalently, the mobility matrix, with errors controlled by the small parameter $a/L$. To this end, we replace the rod with $N=L/(2a)$ spheres equally spaced along the $x$-axis, with positions $\x_j=(2aj, 0,0)$. According to our assumption of uniformly distributed force, each sphere is subject to an external force $\F_\mathrm{ext}/N$. If there were no hydrodynamic interactions among the spheres, then each sphere would move with velocity, $\u=\F_\mathrm{ext}/(6\pi\eta aN)$, and the mobility matrix would be isotropic. In fact, the motion of each sphere induces a flow that helps move the other spheres along (Fig.~\ref{RFTfig}). 

To calculate the flow $\u_j$ induced by the $j$th sphere, recall from \S\ref{fluids} that the flow induced by a moving sphere is the superposition of a stokeslet and a source-dipole. Since we seek the mobility to leading order, we only keep the far-field terms; thus
\begin{equation}
\u_j(\x)=\frac{1}{8\pi\eta |\x-\x_j|}\left(\mathbf{1}+\e_x\e_x\right)\cdot(\F_\mathrm{ext}/N).\label{usubi}
\end{equation}
Since each sphere moves with the local flow, we identify the velocity of the $i$th sphere with the superposition of the flow induced by the force $\F/N$ at $\x_i$ and the flows induced by the forces at all the other spheres:
\begin{equation}
\u(\x_i)=\F_\mathrm{ext}/(6\pi\eta a N)+\sum_{j\neq i}\u_j(\x_i).\label{uofxj}
\end{equation}
Since $a\ll L$, we replace the sum over $i$ by an integral, noting that the density of spheres is $1/(2a)$, and taking care to exclude a small segment around $x_i$ from the region of integration:
\begin{equation}
\u(\x_i)=\frac{\F_\mathrm{ext}}{6\pi\eta a N}+
\frac{1}{8\pi\eta}\int^{'}
\frac{1}{|x_i-x|}(\mathbf{1}+\e_x\e_x)\cdot\frac{\F_\mathrm{ext}}{N}\,\frac{\d x}{2a},
\end{equation}
where the prime on the integral indicates that the region of integration is from $-L/2$ to $L/2$ except for the interval $(x_i-2a,x_i+2a)$. Evaluation of the integral yields
\begin{equation}
\u(\x_i)=\frac{\F_\mathrm{ext}}{6\pi\eta a N}+
\frac{1}{4\pi\eta}\ln\left(\frac{L}{4a}\right)\left(\mathbf{1}+\e_x\e_x\right)\cdot\frac{\F_\mathrm{ext}}{2aN},
\end{equation}
where we have disregarded end effects by assuming $|x_i|\ll L$. Keeping only the terms which are leading order in $\ln(L/a)$, and using the fact that $\u(x_i)$ is constant for a rigid rod, we find 
\begin{equation}
\u=\frac{\ln(L/a)}{4\pi\eta}(\mathbf{1}+\e_x\e_x)\cdot\f_\mathrm{ext},
\end{equation}
where $\f_\mathrm{ext}=\F_\mathrm{ext}/(2Na)$ is the externally imposed force per unit length. In our model, the only forces acting between any pair of spheres is the hydrodynamic force: There are no internal cohesive forces. Therefore, drag per unit length $\f=-\f_\mathrm{ext}$, and 
\begin{equation}
f_\perp=-\xi_\perp u_\perp,\quad f_{||}=-\xi_{||}u_{||},\label{spheresRFT}
\end{equation}
where $\perp$ and $||$ denote the components perpendicular and parallel to the $x$-axis, respectively, and $\xi_\perp=2\xi_{||}=4\pi\eta/\ln(L/a)$. Once again, we encounter the anisotropy already mentioned for slender bodies (\S\ref{motion}) and stokeslets (\S\ref{singularities}) that is necessary for drag-based thrust (\S\ref{thrust}).

In our derivation of Eq.~(\ref{spheresRFT}), we assumed zero deformation since the filament was straight. Turning now to deformed filaments, suppose that the filament is gently curved, $\kappa a\ll1$, where $\kappa^2=|\partial^2\r/\partial s^2|^2$, and $\r(s)$ gives the position of the of the centerline of the filament with arclength coordinate $s$. In the limit of very small curvature, it is reasonable to assume that the viscous force per unit length acting on  the curved filament is the same as the viscous force per unit length acting on a straight rod of the same length. Since local drag theory is an expansion in powers of $1/\ln(L/a)$, it is valid for filaments that are ``exponentially thin." That is, to make $1/\ln(L/a)$ of order $\epsilon$ with $\epsilon\ll1$, we need $a/L\sim\exp(-1/\epsilon)$. Below in \S\ref{sbt_section} we introduce slender-body theory, which has the advantage of being accurate for thin ($a/L\sim\epsilon^p$) rather than  exponentially thin filaments.

Slender-body theory also more accurately captures the hydrodynamic interactions between distant parts of a curved filaments. To see the limitations of the resistive force theory coefficients of Eq.~(\ref{spheresRFT}), consider a rigid ring of radius $R$ and rod diameter $2a$, falling under the influence of gravity in a very viscous fluid. Suppose the plane of the ring is horizontal. Compare the sedimentation rate of the ring with that of a horizontal straight rod with length $L=2\pi R$. In both cases, each segment of the object generates a flow which helps push the other segments of the object down. But the segments of the ring are closer to each other, on average, and therefore the ring falls faster. Using the coefficients of local drag theory from Eq.~(\ref{spheresRFT}) would lead to the same sedimentation rate for both objects. This argument shows the limitations of our local drag theory. One way to improve our theory is to use a smooth distribution of stokeslets and source-dipoles to make a better approximation for the flow induced by the motion of the rod. Applying this approach to a sine wave with wavelength $\lambda$ leads to~\cite{hancock53,lighthill76} 
\begin{eqnarray}
\xi_\perp&=&\frac{2\pi\eta}{\ln(2\lambda/a)-\frac{1}{2}}\\
\xi_{||}&=&\frac{4\pi\eta}{\ln(2\lambda/a)+\frac{1}{2}}.\label{Hancockeqns}
\end{eqnarray}
In Ref.~\cite{lighthill76}, Lighthill refined the arguments of~\cite{hancock53} and gave more accurate values for $\xi_\perp$ and $\xi_{||}$. Despite the limitations of local drag theory, we will see that it is useful for calculating the shapes of beating flagella and the speeds of swimmers. 

In the remainder of this section, we describe some of the applications of local drag theory to the problem of swimmers with prescribed stroke. To keep the formulas compact, we work in the limit of small deflections, although local drag theory is equally applicable to thin filaments with large deflections. Consider the problem of a spherical body of radius $a$ propelled  by a beating filament with a planar sine wave~\cite{chwang71}
\begin{equation}
h(x,t)=b\sin(k x-\omega t).\label{hsine}
\end{equation}
As in our discussion of the Taylor sheet, \S\ref{TaylorSheetSection}, we work in the frame of the swimmer. Thus, the problem is to find the flow velocity $\U$ that yields zero net force and moment on the swimmer. To simplify the discussion, we suppose that external forces and moments are applied to the head to keep it from rotating or moving in the $y$-direction.  In real swimmers, there is a transverse component of the velocity and a rotation, which both play an important role in determining the swimmer's trajectory and the shape of the flagellum~\cite{yundt75,lauga07_pre}.

 Equation~(\ref{spheresRFT}) gives the viscous forces per unit length acting on the filament. The total force per unit length has a propulsive component, Eq.~(\ref{Fpropeqn}), arising from the deformation of the filament, and also a drag component, arising from the resistance to translating the swimmer along the $x$ direction.
Integrating this force per length to find the total $x$-component of force,  writing the drag force on the sphere as $\xi_0aU=6\pi\eta aU$, and balancing the force on the sphere with the force on the filament yields
\begin{equation}
U=\frac{(\xi_\perp-\xi_{||})\int_0^L\dot h h'\, \mathrm{d}x}{\xi_{||}L+\xi_0a}.\label{swimspeed1}
\end{equation}
Note that only the perpendicular component of the rod velocity leads to propulsive thrust; the motion of the rod tangential to itself  hinders swimming.  Inserting the sinusoidal waveform (\ref{hsine}) into Eq.~(\ref{swimspeed1}) and averaging over a period of the oscillation yields 
\begin{equation}
\langle U\rangle=-\frac{\xi_\perp-\xi_{||}}{2\xi_{||}}\frac{\omega k b^2}{1+(\xi_0a)/(\xi_{||}L)}.\label{Ufilament}
\end{equation}
The form of Eq.~(\ref{Ufilament}) is similar to the result for a swimming sheet, Eq.~(\ref{TaylorSheetSpeed}); when $L\gg a$ and $\xi_\perp=2\xi_{||}$, the two expressions are identical. For $L\gg a$, the swimming speed is independent of $L$ for fixed $k$ and $b$, since lengthening the filament increases the drag and propulsive forces by the same amount.

Since the swimmer has finite length, we can define the efficiency as the ratio of the power required to drag the swimmer with a frozen shape though the liquid at speed $U$ to the average rate of work done by the swimmer. In other words, the efficiency is the ratio of the rate of useful working to the rate of total working. To leading order in deflection, the efficiency $e$ is given  for arbitrary small deflection $h$ by
\begin{equation}
e=\frac{(\xi_{||}L+\xi_0a)U^2}{\xi_\perp\int_0^L\langle\dot h^2\rangle\, \mathrm{d}x}\label{efficiencyEqn}
\cdot
\end{equation}
For the sinusoidal traveling wave~(\ref{hsine}) with $\xi_\perp=2\xi_{||}$, 
\begin{equation}
e=\frac{k^2b^2/2}{1+(\xi_0a)/(\xi_{||}L)}.\label{efficiencyEqn2}
\end{equation}
For small deflections, $kb\ll1$, and the hydrodynamic efficiency is small. Note that the total efficiency is given by the product of the hydrodynamic efficiency and the efficiency of the means of energy transduction.

\begin{figure}[t]
\begin{center}
\includegraphics[width=0.6\textwidth]{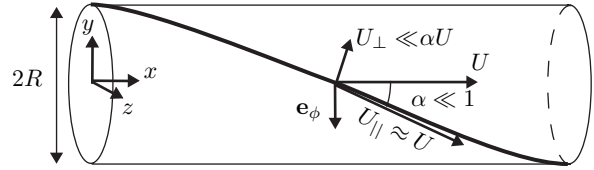}
\end{center}
 \caption{Forces on a helical segment pulled through a viscous fluid with speed $U$. Half of a helical pitch is shown. The helix is prevented from rotating by an external moment along $x$. }
\label{helixresistance}
\end{figure}

We now turn to the case of rotating helix, such as the flagellar filament of \textit{E. coli}. The body of the cell is taken to be a sphere of radius $a$. For simplicity, suppose the radius $R$ of the helix is much smaller than the pitch of the helix, or equivalently, that the pitch angle $\alpha$ is very small (Fig.~\ref{helixresistance}). Expressions relevant for large amplitudes may be found in Refs.~\cite{Magariyama_etal1995,chattopadhyay06}. We also assume that the axis of the helix always lies along the $x$-axis, held by external moments along $y$ and $z$  if necessary. The helix is driven by a rotary motor embedded in the wall of the body, turning with angular speed $\Omega_\mathrm{m}$ relative to the body. The helix rotates with angular speed $\Omega$ in the laboratory frame, and the body must counter-rotate with speed $\Omega_\mathrm{b}$ to ensure the total component of torque along $x$ of the swimmer vanishes. The angular speeds are related by $\Omega_\mathrm{b}+\Omega=\Omega_\mathrm{m}$.

To calculate the force and moment acting on the helical filament, use Eq.~(\ref{resistance}), simplified to include only components along $x$-axis:
\begin{equation}\label{xresistance}
\left(
\begin{array}{c}
F   \\
M  \\
\end{array}
\right)
=
\left(\begin{array}{ll}
A  &  B  \\
 B&  C  \\
\end{array}\right)
\cdot 
\left(\begin{array}{c}
U  \\
 \Omega  \\
\end{array}\right),
\end{equation}
where $F$ and $M$ are the external force and moment, respectively, required to pull the helix with speed $U$ and angular rotation rate $\Omega$. To leading order in $\alpha$, and with the assumption that $L\gg a, R$, the resistance coefficients are approximately
\begin{eqnarray}
A&\approx&\xi_{||}L\\
B&\approx&-(\xi_\perp-\xi_{||})\alpha R L\\
C&\approx&\xi_\perp R^2 L.\label{ABC}
\end{eqnarray}
These values are deduced by applying Eqs.~(\ref{spheresRFT}) to special cases of motion.  The coefficient $A$ may be found by considering the case of a helix pulled at speed $U$ but prevented from rotating. Likewise, $C$ may be found by examining a helix that rotates about $x$ but does not translate along $x$. To find $B$, we may examine the moment required to keep a helix pulled with speed $U$ from rotating (Fig.~\ref{helixresistance}). Or we may also examine the force required to keep a helix rotating at angular speed $\Omega$ from translating. The equivalence of these two calculations is reflected in the symmetry of the resistance matrix, which ultimately stems from the reciprocal theorem, Eq.~(\ref{reciprocal}). Note that the sign of the coupling $B$ between rotational and translational motion is determined by the handedness of the helix. 
 
To find the swimming speed $U$ and rate of filament rotation, we equate the external forces and moments on the filament to the forces and moments acting on the cell body, 
\begin{eqnarray}
F&=&-\xi_0aU\label{Fbody}\\
M&=&-\xi_\mathrm{r}a^3\Omega_\mathrm{b},\label{Mbody}
\end{eqnarray}
where we have introduced a resistance coefficient $\xi_\mathrm{r}=8\pi$ for the rotation of the sphere. Solving for the three unknowns $U$, $\Omega$, and $\Omega_\mathrm{b}$, we find
\begin{equation}
U\approx\alpha\frac{\xi_\perp-\xi_{||}}{\xi_{||}}\left(\frac{\xi_\mathrm{r}}{\xi_\perp}\right)\left(\frac{a^3}{RL}\right)\Omega_\mathrm{m},\label{Uhelixeqn}
\end{equation}
$\Omega_\mathrm{b}=\Omega_\mathrm{m}+\mathcal{O}(a/L)$, and $\Omega\approx\Omega_\mathrm{m}(\xi_\mathrm{r}/\xi_\perp)a^3/(R^2L)$. The resistance $\xi_0$ of the body  does not enter since we assume $L\gg a$. The speed is linear in $\alpha$ since the sign of the speed is given by the handedness of the helix, which is given by the sign of $\alpha$. Note the contrast with the planar wave. If $a=0$, then the velocity (\ref{Uhelixeqn})  vanishes: Propulsion by means of a rotating helix requires a  body. However,  a planar wave does not require a body for propulsion [see Eq.~(\ref{Ufilament})]. 
For the helical filament, the swimming speed decreases with increasing $L$ for fixed body size $a$ and fixed motor speed $\Omega_\mathrm{m}$, since  torque balance forces the filament to rotate more slowly with increasing $L$. Also, since the resistance to rotation of a helix scales as $R^2$, the helix rotates sufficiently faster as the radius decreases to overcome the reduced thrust force implied by the linear dependence of $B$ on $R$, leading to a speed that increases like $1/R$ for decreasing $R$.
It is important to note that the motor speed $\Omega_\mathrm{m}$ of a swimming bacterium is not directly observable with current techniques; typically the approach just described is used to make a prediction for the relation between observables such as $U$ and $\Omega$, which is then compared with measurements~\cite{chattopadhyay06,LiTang2006}

\subsection{Slender-body theory}
\label{sbt_section}

The local drag theory illustrated in the previous section allows  an  
intuitive presentation of the scaling laws for flagella-based  
locomotion.  It turns out however that this theory is quantitatively  
correct only for exponentially slender filaments.  Let  $\lambda$ be  
the typical length scale along the flagellum on which its variations  
in shape occur, such as the wavelength, and $a$ the flagellum radius.  
The local drag theory assumes that $1/\log (\lambda/a) \ll 1$.
Since real biological flagella have aspect  ratios on the order $a/ 
\lambda \sim 10^{-2}$, an improved modeling approach is necessary  
\cite{lighthill76,johnson79}.
The idea, termed slender-body theory and  pioneered by Hancock \cite 
{hancock53}, is to take advantage of the slenderness of the filaments  
and replace the solution for the  dynamics of the three-dimensional of  
the filament surface by that of its centerline using an appropriate  
distribution of flow singularities. Two different approaches to the  
method have been proposed.

The first  approach consists of solving for the flow as a natural  
extension of the local theory, and approximating the full solution  
as  a series of logarithmically small terms \cite 
{cox70,batchelor70,tillett70}. Physically, the flow field close to  
the filament is locally two-dimensional, and therefore diverges  
logarithmically away from the filament because of Stokes' paradox of  
two-dimensional flows \cite{leal}. The flow far from the filament is  
represented by a line distribution of stokeslets of unknown  
strengths, and diverges logarithmically near the filament  due to the  
singular nature of the stokeslets (the logarithmic divergence is due  
to the line integration of $1/r$ stokeslet terms). Matching these two  
diverging asymptotic behaviors allows the determination of the stokeslet  
strengths order by order as a series of terms of order $1/(\log  
\lambda/a)^n$. The leading-order term in this series, of order $1/\log 
( \lambda/a)$, is the local drag theory, and gives a stokeslet  
distribution proportional to the local velocity.  The next order term  
is in  general non-local and provides the stokeslet strength
at order $1/(\log \lambda/a )^2$ as an integral equation on the  
filament shape and velocity. Terms at higher order can be generated  
in a systematic fashion  \cite{cox70,batchelor70,tillett70}. This  
approach to slender-body hydrodynamics is the logical extension to  
the local drag theory, and all the terms in the expansion can be   
obtained analytically which makes it appealing. However, the major  
drawback to this approach is that each term in the expansion is only  
smaller than the previous term  by a factor $1/(\log\lambda / a)$, so  
a large number of terms is necessary in order to provide an accurate  
model for the flow.

A second approach, asymptotically more accurate but technically more  
involved, consists in bypassing the logarithmically-converging series  
by deriving directly the integral equation satisfied by the (unknown)  
distribution of singularities along the filament. Such approach has  
been successfully implemented using matched asymptotics \cite 
{keller76-jfm} or uniform expansions  \cite{geer76}, and leads to   
results accurate at order  $a/\lambda$. This approach is usually  
preferable since instead of being only logarithmically correct it is  
algebraically correct. This improved accuracy comes at a price  
however, and at each instant an integral equation must be solved  
to compute the force distribution along the filament.  An improvement  
of the method was later proposed by accurately taking into account  
end effects and a prolate spheroidal cross-section, with an accuracy  
of  order  $(a/\lambda)^2\log(a/\lambda)$ \cite{johnson80}.

\begin{figure}[t!]
\begin{center}
  \includegraphics[width=0.9\textwidth]{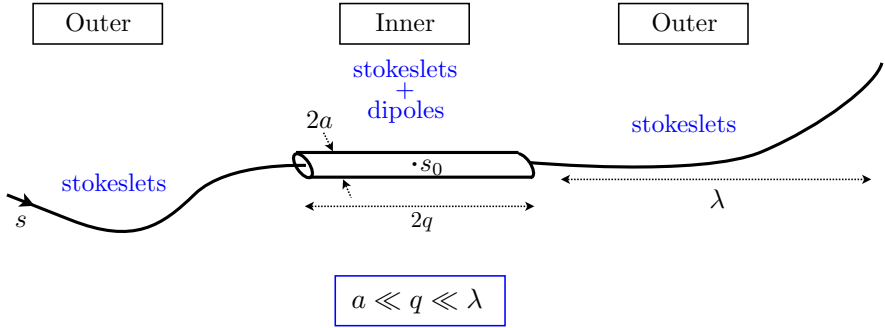}
\end{center}
  \caption{Lighthill's singularity construction for the slender-body  
theory of flagellar hydrodynamics. To compute the flow field  at a  
point $s_0$ along the flagellum, Lighthill picks an intermediate  
length scale  $q$ such that   $a\ll q \ll \lambda$ and represents the  
flow by a distribution of stokeslests at a distance further away than  
$q$ from $s_0$ (outer problem) and a distribution of stokeslets and  
source-dipoles within $q$ of $s_0$ (inner problem).  The dipole  
strengths in the inner problem are obtained by requiring that the  
result be independent of  $q$.}
\label{lighthill_SBT}
\end{figure}

In his John von Neumann lecture, Lighthill proposed an alternate  
method for the derivation of such integral equations. Instead of using  
asymptotic expansions, he used physical arguments to derive the type  
and strength of the singularities located along the filament \cite 
{lighthill76,lighthill96_theorem}. By analogy with the flow past a  
sphere, Lighthill first proposed that a line distribution of  
stokeslets and source-dipoles should be appropriate to represent the  
flow field induced by the motion of the filament. He then  
demonstrated that the strength of the dipole distribution should be  
proportional to the stokeslets strengths using the following argument  
(see Fig.~\ref{lighthill_SBT}). Consider a location $s_0$ along the  
filament. By assumption of slenderness, it is possible to find an  
intermediate length scale $q$ along the filament such that $a\ll q\ll  
\lambda$. The flow field on the surface of the filament at the  
position $s_0$ is then the sum of the flow due to the singularities  
within a distance $q$  from $s_0$ (``inner'' problem) and those  
further away than $q$ from $s_0$ (``outer'' problem). Since $q\gg a 
$, the contribution at $s_0$ from the outer problem is primarily  
given by a line distribution of stokeslets (the source dipoles decay  
much faster in space). In the inner problem, Lighthill then showed  
that it was possible to analytically determine the strength of the  
dipoles to ensure that the complete solution (sum of the inner and  
outer problems) was independent of the value of $q$. The dipole  
strength is found to be proportional to the stokeslet strength, and   
the resulting value of the velocity of the filament at $s=s_0$ is  
given by the integral equation
\begin{equation}\label{lighthill_integral}
\u(s_0)=\frac{\f_\perp(s_0)}{4\pi \eta} + \int_{|\r_0-\r|>\delta} \G 
(\r_0-\r)\cdot \f (s)\,  \d s.
\end{equation}
In Eq.~(\ref{lighthill_integral}), $\f$ is the local strength of the  
(unknown) stokeslet distribution (dimensions of force per unit  
length),  $\G$ is the Oseen tensor from  Eq.~(\ref{u}), $\delta$ is a  
length scale that appears in the analysis ($\delta =\frac{1}{2}a\sqrt 
{\mathrm{e}}$), and $\f_\perp$ represents the normal component of the  
stokeslet distribution, {\it i.e.} $\f_\perp=({\bf 1}-{\bf t}{\bf t})  
\cdot\f$ if ${\bf t}$ is local the tangent to the filament. Note that  
Lighthill's slender-body analysis is less mathematically rigorous  
than those presented in Refs.~\cite{keller76-jfm,geer76}, and  
consequently gives results which are only valid at order $(a/\lambda)^ 
{1/2}$ \cite{childress81}. His derivation provides however important  
physical insight into a subject, the topic  of flow singularities,  
that is usually very  mathematical. The resulting integral  
formulation, Eq.~(\ref{lighthill_integral}), is relatively simple to  
implement numerically, and can also be used to derive ``optimal''  
resistance coefficients for the local drag theory (see \S\ref{RFT}).
His modeling approach has been extended for filament motion near a  
solid boundary \cite{gueron92,gueron93,gueron97,gueron99}, and an  
alternative approach based on the method  of regularized flow  
singularities \cite{cortez01} has also been devised \cite{cortezPC}.

%%%%%%%%%%%%%%%%%%%%%%%%%%%%%%%%
%%%%%%%%%%%%%%%%%%%%%%%%%%%%%%%%
\section{Physical actuation}
\label{actuation}
In this section we describe common mechanisms for the generation of non-reciprocal swimming strokes. In addition to biological mechanisms such as rotating helices or beating flagella, we also describe simple mechanisms that do not seem to be used by any organism. The non-biological setups are useful to study since they deepen our understanding of the biological mechanisms. For example, the modes of an elastic rod driven by transverse oscillations at one end are useful for understanding the shape of a beating flagellum driven by motors distributed along its entire length. An important theme of this section is the fluid-structure interaction for thin filaments in viscous liquid.

\subsection{Boundary actuation }

\begin{figure}[t]
\begin{center}
 \includegraphics[width=0.7\textwidth]{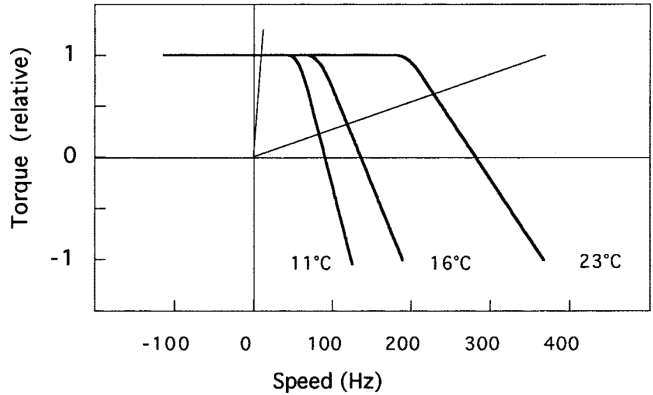}
\end{center}
 \caption{The torque-speed characteristic for the flagellar motor of \textit{E. coli} for three different temperatures (bold lines). Also shown (thin lines) are the load lines for a cell body with radius of 1\,$\mu$m  (left), and a minicell with effective radius of about $0.3$\,$\mu$m (right). Figure reprinted from Ref.~\cite{berg03}, courtesy of  Howard Berg 
 and by permission from Annual Reviews, copyright 2003.}
\label{EcoliMotorfig}
\end{figure}

We begin with the case of boundary actuation, in which an elastic filament is driven by a motor at its base. The rotary motor of the bacterial flagellum is a prime example of such a biological actuating device~\cite{berg00,berg03,macnab84,schuster94}. The steady-state relation between motor torque and motor speed is shown in Fig.~\ref{EcoliMotorfig}~\cite{berg03}. At low speeds, the motor torque is roughly constant; at higher speeds it decreases linearly with speed, reaching zero torque at about 300\,Hz at 23$^\circ$\,C. To determine the speed of the motor from the motor torque-speed relation, use torque balance and equate the  motor torque with the load torque. By the linearity of Stokes flow, the load torque is linear in rotation speed. In the experiments used to make the graph of Fig.~\ref{EcoliMotorfig}, the flagellar filament of \textit{E. coli} was tethered to a slide, and the rotation of the body was observed. A typical body of 1\,$\mu$m radius has a substantial resistance, leading to the steep load curve on the left of Fig.~\ref{EcoliMotorfig} and a correspondingly low rotation speed. A smaller load, such as that of a minicell, leads to a load curve with smaller slope, and higher rotation speed. The torque-speed characteristic $M(\Omega_\mathrm{m})$ allows us to go beyond the artifice of the previous section where we calculated the swimming speed $U$ in terms of the motor speed $\Omega_\mathrm{m}$. Solving Eqs.~(\ref{xresistance}) and (\ref{Fbody}--\ref{Mbody}) along with $M=M(\Omega_\mathrm{m})$ yields the swimming speed in terms of the geometrical parameters of the flagellum, the cell body, the drag coefficients, and the properties of the motor. 

In the previous section we described how a rotating helix generates propulsion. Since the flagellar filaments of \textit{E. coli} and \textit{S. typhimurium} are relatively stiff, a helical shape is necessary to escape the constraints of the scallop theorem as a  straight rod rotating about its axis generates no propulsion. Indeed, mutant \textit{E. coli} with straight flagella do not swim~\cite{MatsuuraKamiyaAsakura1978}. If the rate of rotation of a straight but flexible rod is high enough for the hydrodynamic torque to twist the rod through about one turn, then the rod will buckle into a gently helical shape that can generate thrust~\cite{wolgemuth00,lim04,wada06}. However, the high twist modulus of the filament and the low rotation rate of the motor make this kind of instability unlikely in the mutant strains with straight flagella. On the other hand, a rotating helix with the dimensions of a flagellar filament experiences much greater hydrodynamic torque since the helical radius (microns) is much greater than the filament radius ($\approx10$\,nm). The handedness of the helix also breaks the symmetry of response to the sense of rotation of the motor: Counterclockwise rotation of a left-handed helix in a viscous fluid tends to decrease the pitch of a helix, whereas clockwise rotation tends to increase the pitch~\cite{kim_powers2005}. There is no noticeable difference between the axial length of rotating and de-energized flagella for counterclockwise rotation~\cite{Bergetal2004private}; calculations of the axial extension~\cite{kim_powers2005} based on estimates of the bending stiffness of the flagellar filament~\cite{takano_et_al2003} are consistent with this observation. However, the hydrodynamic torque for clockwise rotation is sufficient to trigger polymorphic transformations, in which a right-handed helical state invades the left-handed state by the propagation of a front~\cite{macnab77_curly,hotani1982}.

Now consider the case of an elastic filament driven by a mechanism that oscillates the base of the filament in the direction normal to the tangent vector of the filament. Although we know of no organism that uses this mechanism to swim, study of this example has proven instructive.  In early work,  Machin~\cite{machin58} pointed out that the overdamped nature of low Re flow leads to propagating waves of bending with exponential decay of the amplitude along the length of the filament. Since the observed beating patterns of sperm flagella typically have an amplitude that increases with distance from the head, Machin concluded that there must be internal motors distributed along the length of the flagellum that give rise to the observed shape. This problem has served as the basis for many subsequent investigations of the fluid-structure interaction in swimming~\cite{machin58,machin63,WigginsGoldstein,lowe03,lagomarsino03_filament,yu06,lauga07_pre,kosa07,FuWolgemuthPowers2008}, and has even been applied to the determination of the persistence length of actin filaments~\cite{riveline97,Wiggins:Biophys}. Therefore, we give a brief overview of the most important elements of the problem.

\begin{figure}[t]
\begin{center}
 \includegraphics[width=0.8\textwidth]{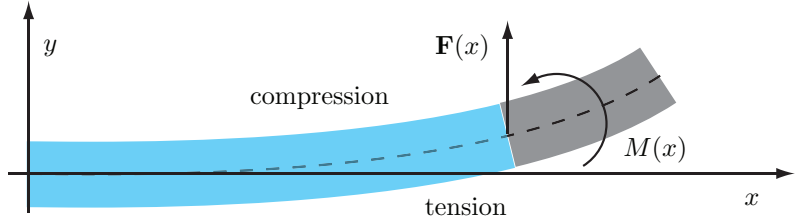}
\caption{A bent elastic rod. The darkly shaded part of the rod exerts a force $\mathbf{F}(x)$ and moment $M(x)$ on the lightly shaded part through the cross-section at $x$.}
\label{bentrodfig}
\end{center}
\end{figure}
Consider a thin rod of length $L$ constrained to lie in the $xy$-plane, aligned along the $x$-axis in the absence of external loads. We will consider small deflections $h(x)$ from the straight state. When the rod is bent into a curved shape, the part of the rod on the outside of the curve is under tension, while the part of the rod on the inside of the curve is under compression (Fig.~\ref{bentrodfig}). Therefore, the section of the rod to to right of $x$ exerts a moment $M=A\kappa(x)\approx A\partial^2 h/\partial x^2$ on the section to the left of $x$, where $A$ is the bending modulus and $\kappa(x)$ is the curvature of the rod at $x$~\cite{landau_lifshitz_elas}. Working to first order in deflection, balance of moments on an element of length $\mathrm{d}x$ of the rod implies
\begin{equation}
\frac{\partial M}{\partial x}+F_y=0, \label{mombaleqn}
\end{equation}
where $F_y=-A\partial^3h/\partial x^3$ is the $y$-component of the force exerted through the cross-section at $x$. Thus, if the rod has a deflection $h(x)$, then an elastic force $f_y\mathrm{d}x=(\partial F_y/\partial x)\mathrm{d}x=-(A\partial^4h/\partial x^4)\mathrm{d}x$ acts on the element of length $\mathrm{d}x$ at $x$. Balancing this elastic force with the transverse viscous force from resistive force theory yields a hyper-diffusion equation
\begin{equation}
\xi_\perp\frac{\partial h}{\partial t}=-A\frac{\partial^4 h}{\partial x^4}.
\label{hyperdiffusion-eqn}
\end{equation}
The shape of the rod is determined by solving Eq.~(\ref{hyperdiffusion-eqn}) subject to the appropriate boundary conditions, which are typically zero force and moment at  the far end, $x=L$. At the near end, common choices are oscillatory displacement $h(0,t)=b\cos(\omega t)$ with clamping $\partial h/\partial x|_{x=0}=0$, or oscillatory angle $\partial h/\partial x|_{x=0}=\theta\cos(\omega t)$ with $h(0,t=0)$, where $b/L\ll 1$ and $\theta\ll1$~\cite{machin58,WigginsGoldstein}.

The appearance of $\partial h/\partial t$ in Eq.~(\ref{hyperdiffusion-eqn}) causes the breakdown of kinematic reversibility: Even for a reciprocal actuation such as $h(0,t)=b\cos(\omega t)$, Eq.~(\ref{hyperdiffusion-eqn}) implies that the rod shape is given by propagating waves. Physically, the breakdown of kinematic reversibility occurs because flexibility causes distant parts of the rod to lag the motion of the rod at the base. We saw in \S\ref{fluids} that zero-Re flow is effectively quasistatic since the diffusion of velocity perturbations is instantaneous. When the filament is flexible, the time it takes for perturbations in shape to spread along the rod scales as $\xi_\perp L^4/A$. Since the shape of the rod does not satisfy kinematic reversibility, the flow it induces does not either, and a swimmer could therefore use a waving elastic rod to make net progress.

The wavelength and the decay length of the propagating waves is governed by a
penetration length, $\ell=[A/(\xi_\perp\omega)]^{1/4}$. Sometimes this length is given in the dimensionless form of the ``Sperm number," $\mathrm{Sp}=L/\ell=L(\omega\xi_\perp/A)^{1/4}$. If the rod is waved rapidly, the penetration length is small $\ell\ll L$ and propulsion is inefficient since most of the filament has small defection and contributes drag but no thrust. For small frequencies, the rod is effectively rigid $\ell\gg L$, and there is no motion since kinematic reversibility is restored. Thus, we expect the optimum length for propulsion is $\ell\approx L$, since at that length much of the rod can generate thrust to compensate the drag of pulling the filament along $x$~\cite{WigginsGoldstein,lauga07_pre}. Note that our discussion of flexibility may be generalized to other situations; for example, the deformation of a flexible wall near a swimmer is not reversible, leading to a breakdown of the scallop theorem even for a swimmer that has a reciprocal stroke~\cite{trouilloud08}. In this case the average swimming velocity decays with a power of the distance from the wall, and therefore this effect is relevant in confined geometries.

An interesting variation on the Machin problem is to rotate a rod which is tilted relative to its rotation axis. If the rod is rigid, then it traces out the surface of a cone. But if it is flexible, then the far end will lag the base, and the rod has a helical shape. As long as the tilt angle is not too small, this shape may be determined without considering effects of twist~{ \cite{klapper96,goldstein98}. If the driving torque rather than speed is prescribed, there is a transition at a critical torque at which the shape of the rod abruptly changes from gently helical to  a shape which is much more tightly wound around axis of rotation, with a corresponding increase in thrust force~\cite{manghi06,qian08,coq08}.

\begin{figure}[t]
\begin{center}
\includegraphics[width=0.5\textwidth]{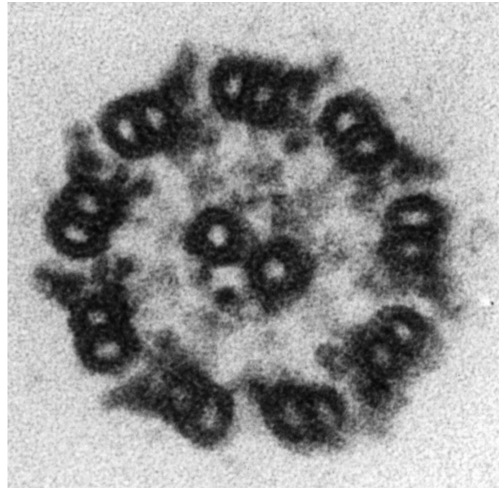}
\caption{Cross-section section of an axoneme from wild type \textit{Chlamydomonas}, courtesy of Steve King. The diameter is approximately 200\,nm. 
Note that the plane containing the active motors is perpendicular to the plane defined by the beating filament.}
\label{axonemeXsection}
\end{center}
\end{figure}

\subsection{Distributed actuation}

We now consider distributed actuation, in which molecular motors are distributed along the length of the filament. Eukaryotic flagella and cilia use distributed actuation. Figure~\ref{axonemeXsection} shows a cross-section of the axoneme, the core of a eukaryotic flagellum. As mentioned in the introduction, the axoneme consists of nine microtubule doublets spaced along the circumference of the flagellum, with two microtubules running along the center. In this review we restrict our attention to the case of planar beating, although many sperm flagella exhibit helical beat patterns, and nodal cilia have a twirling, rotational beat pattern~\cite{Nonaka_etal1998,HilfingerJulicher2008}. The bending of the eukaryotic flagellum arises from the relative sliding of neighboring microtubule doublets~\cite{satir68,summers71,brokaw72,brokaw89}. The sliding is caused by the action of ATP-driven dynein motors, which are spaced every 24\,nm along the microtubles~\cite{camalet99}. Since the relative sliding of the microtubules at the end near the head is restricted~\cite{Riedel-Kruse_etal2007}, and since each microtubule doublet maintains its approximate radial position due to proteins in the core of the flagellum, the filament must bend when the motors slide microtubule doublets. For example, in Fig.~\ref{sliding}, motors have slid the lower doublet to the right relative to the upper doublet.

The simplest approach to understanding how the sliding of the microtubules generates propulsion is to prescribe a density of sliding force and deduce the shape of the flagellum and therefore the swimming velocity from force and moment balance. This approach is taken in Refs.~\cite{FuPowersWolgemuth2007,FuWolgemuthPowers2008}, where the effects of viscosity and viscoelasticity are studied. A more complete model would account for how the coordination of the dynein motors arises. Over the years, several different models for this coordination have been suggested. Since sea urchin sperm flagella continue to beat when they have been stripped of their membranes with detergent
~\cite{GibbonsGibbons1972}, it is thought that the motor activity is not coordinated by a chemical signal but instead arises spontaneously via the mechanics of the motors and their interaction~\cite{Brokaw1975,JulicherProst1997}. A detailed discussion of molecular motors and the different mechanisms that have been put forth for controlling the beat pattern would take us too far afield~\cite{MuraseHinesBlum1989,Lindemann1994a,Lindemann1994b,Lindemann2002,Brokaw2002}. Instead we review regulation by load-dependent motor detachment rate as presented by Riedel-Kruse and collaborators, who showed that this mechanism is consistent with observations of the flagellar shape~\cite{Riedel-Kruse_etal2007}. 

\begin{figure}[t]
\begin{center}
 \includegraphics[width=0.6\textwidth]{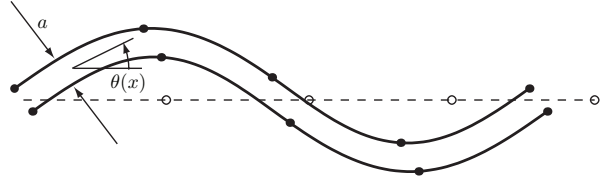}
\caption{Simple model for a planar flagellum actuated by sliding filaments. The dotted line is the center of the flagellum when it is straight. The solid lines represent the two microtubule doublets that the active motors slide relative to each other. The two solid curves and the dotted line all have the same length; the dots divide each curve into quarters as measured along the contour.}
\label{sliding}
\end{center}
\end{figure}

The first step is to simplify the problem by considering planar beating and projecting the three-dimensional flagellum of Fig.~\ref{axonemeXsection} onto a planar flagellum such as that shown in Fig.~\ref{sliding}. Since the lateral spacing between the two filaments is a constant, $a$, the amount that the top filament slides by the bottom is given by
\begin{equation}
\Delta(x)=\Delta_0+a\left[\theta(x)-\theta(0)\right],\label{slidingeqn}
\end{equation}
where $\Delta_0$ is the sliding displacement at the base of the flagellum where $x=0$, and  $\theta(x)$ is the angle the tangent of the flagellum centerline makes with the $x$-axis. In Fig.~\ref{sliding}, $\Delta_0=0$, and  $\Delta<0$ except at either end of the flagellum. We will forbid sliding at the base, but it is straightforward to allow such sliding in our model, and measurements of the shape of bull sperm flagella suggest that $\Delta(0)\approx 55$\,nm~\cite{Riedel-Kruse_etal2007}.
Equation~(\ref{slidingeqn}) in the case $\Delta_0=0$ is most easily deduced when the two filaments form the arcs of concentric circles, but it holds for more general curved shapes as well. Also, we will work in the limit of small deflection, but Eq.~(\ref{slidingeqn}) is valid even for large deflections.

Now consider the forces between the two filaments. Let $f(x,t)$ denote the force per unit length along $x$ that the bottom filament exerts on the top filament. This force per unit length could arise from passive resistance to sliding as well as motors. The internal moment acting a cross section has a contribution from bending and an additional piece from $f$: 
\begin{equation}
M=A\frac{\partial^2 h}{\partial x^2}-a\int^L_x f\,\mathrm{d}x.\label{Mneweqn}
\end{equation}
Moment balance on an element of the flagellum yields the transverse force acting on a cross section,
$F_y=-A\partial^3h/\partial x-a f$, which with force balance yields the equation of motion for the filament:
\begin{equation}
\xi_\perp\frac{\partial h}{\partial t }=-A\frac{\partial^4 h}{\partial x}-a\frac{\partial f}{\partial x}.\label{activehyperdiff}
\end{equation}
For example, if the motors are not active and there is only elastic resistance to sliding, $f=-K\Delta$, then the shape equation is
\begin{equation}
\xi_\perp\frac{\partial h}{\partial t }=-A\frac{\partial^4 h}{\partial x}+Ka^2\frac{\partial^2 h}{\partial x^2}.\label{activepassivehyperdiff}
\end{equation}
Even if the bending stiffness $A$ vanishes, the elastic resistance to sliding will damp out any perturbation of the straight shape. More generally, the passive resistance will have a viscous as well as an elastic component: $f=-K\Delta-\lambda\dot\Delta$.

Now consider an active component to $f$. As in the case of the bacterial rotary motor, the speed of a dynein motor is thought to decrease with load. One end of the dynein is strongly attached to a microtubule doublet, while the other end attaches to and detaches from the neighboring doublet. The motor only does work when attached. The proposal of Refs.~\cite{JulicherProst1997,camalet99,camalet00:njp} is that oscillations arise spontaneously since the detachment rate increases with load. To see that a load-dependent detachment rate leads to positive feedback, consider a collection of motors sliding one filament past another. If there is a perturbation in which the sliding rate increases, then the load on each individual motor must decrease, according to the force-speed motor characteristic. On the other hand, the decrease in force per motor leads to an reduction in the detachment rate, which in turn leads to a greater total sliding force, and ultimately, an increase in sliding velocity. Working in the frequency domain, we may define a susceptibility $\chi$ via $\tilde f=-\chi(\omega)\tilde \Delta$. For the passive elements mentioned above, 
 $\chi=K+\mathrm{i}\omega\lambda$.  A simple quantitative model with load-dependent detachment rate leads to a susceptibility $\chi$ that can have a negative effective elastic constant or a negative effective drag term~\cite{Riedel-Kruse_etal2007}. By using Eq.~(\ref{activehyperdiff}) to relate the shape of the beating flagellum to the susceptibility $\chi$, Riedel-Kruse and collaborators showed that the passive components of $\chi$ were small compared to the active components; in other words, the forces and moments induced by the motors are balanced by viscous drag and bending forces, rather than internal resistance to sliding. Since the oscillations arise spontaneously in this model, the calculation of the beating filament shape amounts to solving an eigenvalue problem~\cite{
camalet99,camalet00:njp}. 

We close this section by noting that there are several other distinct means of actuation. 
The bacterium \textit{Spiroplasma} has an internal helical ribbon that is though to undergo contractions that in turn cause the body to change shape. The shape change amounts to the propagation of kinks down the body, where the kinks are generated by a change in the handedness of the cell body~\cite{ShaevitzLeeFletcher2005,wada07}. 
Spirochetes such as \textit{Leptospiracaeae} have internal flagella that wrap around the periplasmic space between the cell body and an outer sheath. The flagellar filaments emerge from motors at either pole of the cell, but the rotating flagella distort the body in a nonreciprocal way that leads to locomotion~\cite{Charon_etal1984,GoldsteinCharon1990}.  These swimmers have inspired recent table-top experiments in which rigid superhelices sediment in a very viscous fluid~\cite{JungMareckFauciShelley2007}. The helical shape causes the superhelices to rotate as they fall. An important finding of Ref.~\cite{JungMareckFauciShelley2007} is that resistive force theory give poor quantitative results for the rotation speed, even getting the sense of rotation wrong for tightly coiled superhelices. 
\section{Hydrodynamic interactions}
\label{interactions}

\subsection{Interactions between cells} 
\label{cellcell}

Microorganisms swimming in viscous fluids typically do so within semi-dilute or dense cell populations. As an organism is swimming, it sets up a flow which will be felt by the cells nearby, possibly  
affecting the dynamics at the level of the entire population. For example, spermatozoa involved in human reproduction  
may swim in population sizes as high as millions of cells~\cite{suarez06}. Bacterial suspensions are known to display so-called ``bacterial turbulence,'' where large-scale intermittent motion in the forms of swirls and jets is set up when the cells become sufficiently concentrated~\cite{mendelson99,wu00,dombrowski04,tuval05,cisneros07,sokolov07,Wolgemuth2008}. Even for small numbers of cells, hydrodynamic interactions are suspected to play an important role, in particular in reproduction. Such is the case for spermatozoa of the wood mouse which aggregate, and  thereby swim faster~\cite{moore02}. The pairing of the  opossum spermatozoa enable them to swim more  efficiently in very viscous fluids~\cite{moore95}. Fishfly spermatozoa cluster in dense bundles for similar reasons~\cite{hayashi98}. Recently, sea urchin sperm cells were observed to arrange into  periodic vortices~\cite{riedel05}.

\begin{figure}[t]
\begin{center}
 \includegraphics[width=0.7\textwidth]{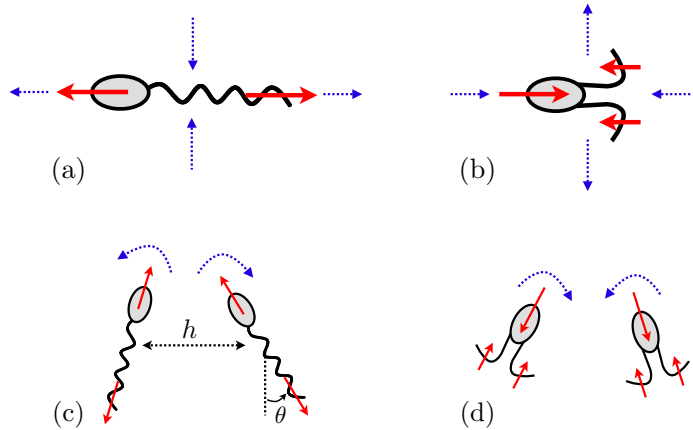}
\caption{The flow field created by a swimmer at low Reynolds number. 
(a): Cells which are pushers have a positive force-dipole ($p>0$, see text) and induce a flow field directed away from the cells along their swimming direction (repulsion) and a flow field directed toward the cells along their side (attraction) (red solid arrows represent local forcing from the cell on the surrounding fluid); 
(b): Pullers have a negative force dipole ($p<0$), inducing an attractive flow field along their swimming direction and a repulsive flow field along their side; (c) Two pushers on a converging course reorient each other, tending toward a configuration with cells parallel and swimming side-by-side ($h$ is the separation distance between the cells, and $\theta$ is the angle between the cell direction and the direction normal to their separation);  (d)  Two pullers on a diverging course reorient each other, tending toward a configuration in which the cells are antiparallel, swimming away from each other.}
\label{fig:cellcell}
\end{center}
\end{figure}

Let us first consider the interactions between two swimming cells. The flow field created by the first organism (cell A) will in general lead to two types of passive hydrodynamic interactions with its neighbor (cell B). First, cell B will feel the velocity field created by cell A, and will 
be carried along by this flow 
as a result. In addition, cell B will also feel the gradients in the velocity field created by cell A, which tend to change the orientation of B and and 
thereby affect its future swimming direction. 

The first type of hydrodynamic interaction can be intuitively understood by considering the far-field flow created by a swimming cell (Fig.~\ref{fig:cellcell}). As discussed in \S\ref{singularities}, since a cell is force-free, the velocity in the far field is a force dipole decaying typically as $1/r^2$.  Two different types of force-dipoles  can in general arise, leading to  significantly different physical pictures. 
Consider a microorganism with an elongated body and 
let $\e$ be a unit vector attached to the cell and pointing along the swimming direction, which is also along the elongated direction of the cell. Usually, the force dipole $\bf p$ will also be in the swimming direction, ${\bf p }= p \e$, but it can have either a positive or negative sign. 
Cells for which $p>0$ are called ``pushers,'' and include typical swimming spermatozoa, or bacteria such as {\it E. coli}. Pushers repel fluid from the body along the long axis, and draw fluid in to the sides
(Fig.~\ref{fig:cellcell}a; the red solid arrows represent local forcing from the cell on the surrounding fluid).  Cells with $p<0$ are called ``pullers,'' the prototypical example of which is the alga {\it Chlamydomonas} (see also the  artificial swimmer in Ref.~\cite{dreyfus05}). Pullers draw 
fluid in along the elongated 
direction, and push fluid out from the sides (Fig.~\ref{fig:cellcell}b).
Mathematically, the flow induced at a distance $\r $ from the  dipole ${\bf p }= p \e$  is given by
\begin{equation}\label{dipoleU}
{\bf u}(\r)=\frac{p}{8\pi\eta r^3}\left[3 \cos^2\theta-1 \right]{\bf r},
\end{equation}
where $\theta$ is the angle between the dipole direction, $\e$, and the flow position, $\r$. Physically, the dipole strength scales as $|p|\sim \eta U L ^2 $ where $U$ is the swimming speed of the cell, and $L$ its typical length. 
From Eq.~(\ref{dipoleU}), we see that two similar cells swimming side by side ($\theta=\pi/2$) experience a relative velocity scaling as $\Delta u_{{\rm side}}\sim -p /\eta r^2$. Therefore two pushers attract each other, while two pullers repel each other   \cite{guell88}. Conversely, the relative velocity of the two swimmers are aligned along their swimming direction ($\theta=0$) scales as
$\Delta u_{{\rm aligned}}\sim p /\eta r^2$, and the sign of the resulting attraction/repulsion is reversed.

The second effect of hydrodynamic interactions between two cells is the reorientation  due to velocity gradients. 
The vorticity field induced by the flow set up by  the cell, Eq.~(\ref{dipoleU}),  is given by 
\begin{equation}\label{dipoleO}
\bom =\nabla\times \u =\frac{3p}{4\pi\eta}\frac{(\e\cdot\r)(\e \times  \r)}{r^5}\cdot
\end{equation}
A sphere subject to this flow will rotate at a rate of half the vorticity, $\Om=\frac{1}{2}\bom$, to leading order in the ratio of the sphere size and the separation of the sphere from the microorganisms inducing the flow~\cite{kimbook}.
For interacting cells, which are typically not spherical, 
an additional component for $\Om$ arises from the  symmetric part of the rate of strain, ${\bf E}=\frac{1}{2}(\nabla \u + \nabla \u ^T)$,  
since elongated cells 
tend to align with the principal axis of strain, $\Om\sim \e\times ({\bf E}\cdot \e)$, with a sign that depends on the geometry of cells (typically positive for prolate cells, and negative for oblate cells)  \cite{kimbook,pedley92}. As was the case for the attraction/repulsion between cells, both $\bom$ and $\bf E$ change sign with $p$, and  qualitatively different rotational behaviors are expected to occur for pushers vs. pullers. Nearby pushers on a converging course induce flow fields on each other that reorient them in the side-by-side configuration (Fig.~\ref{fig:cellcell}c). If the cells are separated by a distance $h$, and are oriented with an angle $\theta$ with respect to the distance perpendicular to their separation, the reorientation takes place 
with a rotation rate $\Omega \sim -{p \theta}/{\eta h^3}$. In contrast, pullers induce the opposite flow field, which leads to a reorientation of the cells 
in the elongated direction (Fig.~\ref{fig:cellcell}d). Interestingly, for both pushers and pullers, the final configuration is one for which cells induce attracting flow fields on each other. As discussed below, similar results govern the orientation of cells near boundaries.

The results above describe appropriately the  leading-order hydrodynamic interactions between cells. Higher-order effects can be considered with various levels of modeling accuracy, two of which we address here.  First, there is an active component to hydrodynamic interactions. The physical reason is the following: In the flow field created by cell A, cell B sets up its own disturbance flow (just as a solid body sets up a disturbance flow when located in a shear flow), which then influences the velocity and orientation of cell A. This is a weaker effect (decaying faster in space), but important for artificial swimmers with no permanent dipoles \cite{pooley07_2}. Higher-order ``reflections'' can be considered in a similar manner. In addition,  the flow induced by a cell is only dipolar at leading order, and includes also in general higher singularities (decaying faster), such as source-dipoles and force quadrupoles. The quadrupolar contribution is important because it dominates the pair velocity correlations, as recently demonstrated experimentally for swimming {\it E. coli}  \cite{liao07}.

Beyond the simple physical picture presented above, a few studies have looked in detail at hydrodynamic interactions between more realistic models of  swimming cells. Interactions between two cells were studied analytically and numerically for two different cases, spherical squirming (swimming motion where all surface deformation occurs tangentially to the swimmer surface) \cite{ishikawa06,ishikawa_paramecia06} and swimming with a single helical flagellum \cite{ishikawa07_bacteria}. Detailed experiments were also carried out to study hydrodynamic interactions between the protozoan {\it Paramecium} \cite{ishikawa_paramecia06}. In all cases,  the far-field physical picture described above is correct, but the details of hydrodynamic interactions at short range are also important, and in many cases lead to an instability of the side-by-side configuration for pushers, and unsteady  three-dimensional cell trajectories.

Beyond the dilute limit, dense cell suspensions display remarkable complex dynamics. Oriented  suspensions of swimmers are found to be long-wavelength unstable \cite{simha02,saintillan07}, with persistence of short-range order \cite{saintillan07}, leading to nonlinear states of jets and swirls  \cite{saintillan08} similar to those observed experimentally 
 \cite{mendelson99,wu00,dombrowski04,tuval05,cisneros07,sokolov07} (see Ref.~\cite{ishikawa08} for a two-dimensional study). Isotropic suspensions have also been found to be unstable \cite{saintillan08}.  As a result of the large-scale motion in dense suspensions, swimming cells undergo a  remarkable decrease in  effective diffusivity \cite{hernandez-ortiz05,ishikawa_pedley_diffusion07,mehandia08,underhill08}.

One final important feature of cell suspensions is their rheological characteristics  \cite{hatwalne04,ishikawa_pedley_rheology07}. In the limit where cells do not interact with each other hydrodynamically, the response of a cell population to an external shear can be quantified using Batchelor's theory for suspensions of force-free bodies \cite{batchelor70_2}. For simplicity, let us consider spherical cells with radius $a$, described by force dipoles ${\bf p}=p\e$, and distributed with  volume fraction $c\ll 1$ in a Newtonian fluid of viscosity $\eta$.
If an external shear flow is applied to the suspension, with shear rate $\dot\gamma$,  the  effective viscosity of the population, $\eta_{{\rm eff}}$, is different from the background viscosity of the Newtonian fluid at leading order in $c$ by the amount
\begin{equation}\label{etaeff}
\frac{\eta_{{\rm eff}}}{\eta} = 1 + \left[\frac{5}{2}  + \frac{3\tau_s \langle  e_1e_2 \rangle}{4\pi \eta \dot \gamma}\right]c + O(c^2).
\end{equation}
In Eq.~(\ref{etaeff}), the coefficient $5/2$ is the Einstein contribution to the viscosity \cite{batchelor1967}, $\langle...\rangle$ denotes  averages over the cell population,  the directions $1$ and $2$ refer to the external flow and shear directions respectively, and $\tau_s= p/a^3$ is the typical active stress created by the swimming cell. For swimming {\it E. coli}, $p\approx 0.1$--$1\times10^{-18}$\,Nm \cite{berke08} and $a\approx 1$--$10$\,$\mu$m, so $\tau_s\approx10^{-4}$--$1$\,Pa. Since the direction $\e$ of each  cell  will rotate as a result of the external shear, the effective viscosity of the population,  Eq.~(\ref{etaeff}), is unsteady and shear-dependent. When $\tau_s\ll \eta \dot \gamma$, the viscosity is dominated by the passive  (Einstein) response of the cells, whereas when $\tau_s\gg \eta \dot \gamma$ the rheology is expected to be governed by active stresses. In addition,  anisotropy in the distribution of swimming cells (e.g. orientationally ordered state  \cite{hatwalne04}) leads in general to normal stress differences with normal stress coefficients, $\Psi_1$ and $\Psi_2$, scaling as \cite{oldroyd1950,bird76,birdvol1,birdvol2,tanner88,larson88,doi88,bird95,larson99} 
\begin{equation}
\Psi_1  =  \frac{3\tau_s}{4\pi \dot\gamma^2} \langle e_1e_1-e_2e_2 \rangle c,\quad 
\Psi_2  =  \frac{3\tau_s}{4\pi  \dot\gamma^2}\langle e_2e_2-e_3e_3\rangle c.
\end{equation}
A detailed study of the rheological characteristics of non-dilute suspensions of cells was carried out for spherical squirmers \cite{ishikawa_pedley_rheology07}. Hydrodynamic interactions give rise to time-varying cell-cell configurations, and affect therefore the rheological properties at order $O(c^2)$  \cite{batchelor70_2}.

\subsection{Interactions between cells and boundaries}
\label{cellwall}

Just as other nearby cells influence the dynamics of a swimming microorganism, the presence of boundaries, and more generally confinement, significantly  impacts cell locomotion.  In addition to affecting the  concentration of chemical species that influence the motility of microorganisms~\cite{fletcher82}, boundaries modify the hydrodynamic stresses acting on the cells, and near-wall motility is therefore both  biologically and physically different from bulk motility. Biological locomotion near boundaries includes surface-associated bacterial infections~\cite{cooper88,harkes92b}, biofilm formation~\cite{costerton95,vanloosdrecht90}, spermatozoa locomotion at the uterotubual junction~\cite{suarez06} and surface-associated behavioral change~\cite{harshey03}.  In this section, we focus on the fluid mechanics of locomotion near walls. 

Four distinct aspects of cell locomotion are modified by the presence of nearby boundaries.
The first is the change in the swimming speed near a wall, which was addressed theoretically by a number of studies \cite{Reynolds1965,katz74,katz75,katzblake,brennen77}. Since viscous 
drag increases as 
a body comes closer to a boundary, it 
might be expected that a cell would slow down. However, since the propulsion method is also drag-based, a closer look is necessary. Indeed, for the Taylor sheet geometry, we mentioned in \S\ref{TaylorSheetSection} that the presence of a nearby wall speeds up the swimmer with a prescribed waveform. 
Here consider this matter further for a swimmer that has no head, and swims using planar waves on a flagellum. 
As was shown in  Eq.~(\ref{swimspeed1}), for a given waveform of the flagellum, the swimming speed of the cell is an increasing function of the ratio between the perpendicular ($\xi_\perp$) and parallel ($\xi_\parallel$) drag coefficients. Both coefficients are found to increase near a wall, but $\xi_\perp$ increases faster than $\xi_\parallel$, so that the ratio $\xi_\perp/\xi_\parallel$ increases, and so does the swimming speed. Physically, for  fixed waveform, the drag-based propulsive force generated by the flagellum increases near the wall, and so does the resistive drag on the swimmer, but the propulsion increase is stronger, and therefore the swimming speed goes up. Associated with the increase in the speed, there is an increase of the rate of  working that the swimmer has to provide in order to maintain the same waveform near the wall \cite{Reynolds1965,katz74,katz75,katzblake,brennen77,fauci95}. If alternatively the swimmer is assumed to swim with constant power, then the presence of a boundary leads in general  to a decrease of the swimming speed, leading to  a decrease of the swimming efficiency  except for some special flagellar waveforms  \cite{Reynolds1965,katz74}.

\begin{figure}[t]
\begin{center}
 \includegraphics[width=0.9\textwidth]{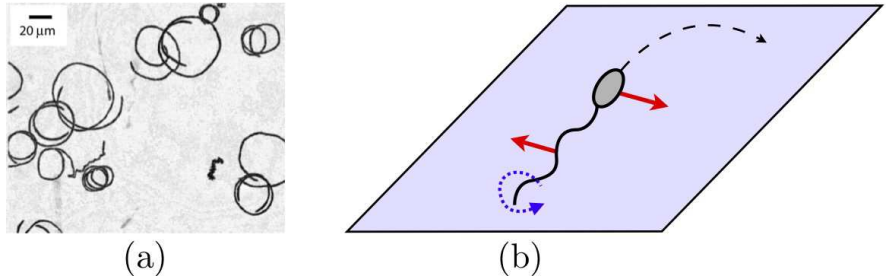}
\caption{Circle swimming of bacteria near boundaries. 
(a): Circular trajectories for smooth-swimming  {\it E. coli} bacteria near a glass surface; superposition of 8 seconds of bacteria position (picture courtesy of  Willow DiLuzio); 
(b): Physical mechanism for circle swimming; the rotation of the bacterial flagella near the surface
(blue dotted arrow) induces a net force on the flagella parallel to the surface but perpendicular to the flagella axis
(red arrows, solid); an equal and opposite force is acting on the cell body due to its counter-rotation, resulting in a wall-induced torque acting on the cell, and a circular swimming trajectory (black arrow, dashed).
}
\label{fig:circle}
\end{center}
\end{figure}

The second type of wall-influence on the swimming kinematics of some microorganisms is a change in their trajectories. This is most famously the case for swimming bacteria which have helical flagella,   such as {\it E. coli}, and change their swimming trajectory from straight to circular near a surface (Fig.~\ref{fig:circle}a). In an infinite fluid {\it E. coli} swims, on average, in a straight line, with a swimming speed found by balancing flagellar propulsion by  drag  (see \S\ref{RFT}). 
The chiral shape  of the flagella  is important for propulsion generation, but since the propulsive force is axisymmetric  when  averaged over on period of flagella rotation, the motion occurs on  a straight line. Near a wall,  the chiral propulsion mechanism  leads a breaking of the (time-averaged) axisymmetry of the propulsive force because  new non-zero components arise in the motility matrix of the helix.  For a helix parallel to a surface, the presence of a boundary leads to a non-zero coupling between the  rotation of the helix around its axis (Fig.~\ref{fig:circle}b, blue dotted arrow), and the force on the helix in the direction  perpendicular to the helix axis and parallel to the surface (Fig.~\ref{fig:circle}b, red solid arrow). In other words, when the helical flagella rotate, they create a net force on the cell at a right angle with respect to the motion and parallel to the surface. There is an exact and opposite force acting on the cell body, which rotates in the opposite direction as the flagella, and the net effect is a wall-induced torque (Fig.~\ref{fig:circle}b, red solid arrows). If the cell were to continue swimming in a straight line, it would have to apply a net torque on the surrounding fluid. 
Since a swimming  bacterium is in fact torque-free, the cell cannot swim straight but instead rotates at a rate such that the viscous torque from that rotation exactly balances the wall-induced torque, and therefore swims along circles on the surface (Fig.~\ref{fig:circle}b, black dashed arrow).  For a cell using a left-handed helix for propulsion, such as {\it E. coli}, this effect leads to swimming-to-the-right  \cite{ramia93,frymier95,frymier97,vigeant97,diluzio05,lauga06}.
\begin{figure}[t]
\begin{center}
 \includegraphics[width=0.7\textwidth]{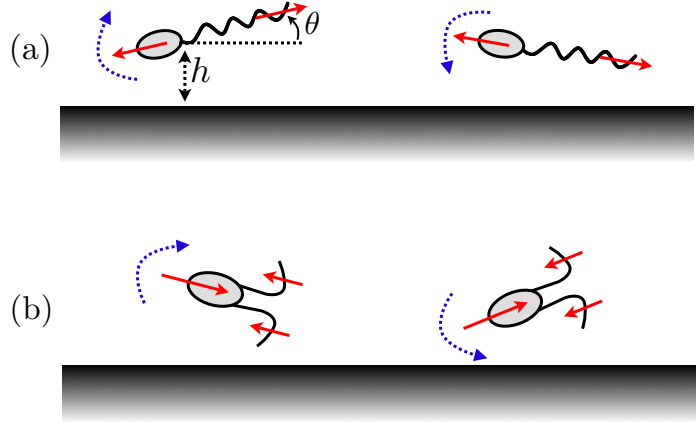}
\caption{Wall-induced rotation of swimming cells. A swimming cell is located at a distance $h$ from a solid surface, and at an angle $\theta$ with respect to the direction of the surface. (a): Pushers are reoriented hydrodynamically in the direction parallel to the surface (equilibrium, $\theta=0$); (b): Pullers are reoriented in the direction perpendicular to the surface (equilibrium, $\theta=\pm\pi/2$).}
\label{fig:wall}
\end{center}
\end{figure}

The third influence of boundaries on swimming cells is analogous to the attraction and re-orientation induced by cell-cell interactions and discussed in \S\ref{cellcell}. Consider a single cell moving near a solid wall. As the cell is swimming, it sets up a dipolar flow field, but this flow field in general does not satisfy the no-slip boundary condition on the wall. The flow in the presence of the wall is then the linear superposition of the infinite-fluid flow, plus any image flow field created on the other side of the surface,  necessary  to enforce the boundary condition on the wall exactly. Image flow fields for the fundamental singularities of Stokes flows were pioneered by Blake  \cite{blake71_image,blake74_image}, and a textbook treatment is offered in Refs.~\cite{kimbook,poz}. Physically, the approach is similar to the method of images in electrostatics, but with the complication that in fluid mechanics the no-slip boundary conditions is vectorial, whereas the constant-potential boundary conditions in electrostatics are scalar. Because of the image flow field, a cell described by a dipole strength $p$, located at a distance $h$ from the surface,  and pointing at an angle $\theta$ from the surface direction   (see notation on Fig.~\ref{fig:wall}a) is subject to the gradients of the image flow field,  and as a result  rotates with speed $\Omega \sim -{p \theta}/{\eta h^3}$ in the direction parallel to the surface and perpendicular to the cell body  \cite{berke08}.  The $1/h^3$ scaling originates from the leading order vorticity of the image flow field, which is also dipolar. The rotation occurs as if the cell is interacting hydrodynamically with a mirror-image cell located on the other side of the surface, and the rotation rate is therefore analogous to that quantified by Eq.~(\ref{dipoleO})\footnote{The effect of a flat boundary is mathematically equivalent to the presence of a mirror-image cell instantaneously located on the other side of the surface if the surface is a no-shear interface (e.g. a free surface). If instead it is a no-slip surface, the analogy is not quite exact mathematically, but it remains qualitatively correct.}. If the cell is a pusher,  the wall-induced rotation rate, $\Omega=\d \theta / \d t$,  tends to align the swimming cell in the direction parallel to the surface ($\theta=0$, Fig.~\ref{fig:wall}a). As a result of this parallel configuration, the  cells will  swim  in a side-by-side configuration with their image cell, and are therefore  attracted to the surface with an attractive speed scaling as $u_\perp \sim{p}/{\eta h^2} $ (see Eq.~\ref{dipoleU}). This physical picture explains the  accumulation of swimming cells near surfaces observed in many biological experiments \cite{rothschild63,winet84,winet84a,fauci95,cosson03,woolley03,hernandez-ortiz05,berke08}. In contrast, cells which are pushers are rotated in the opposite direction, and their  stable configuration is instead  at a right angle with respect to the surface ($\theta=\pm \pi/2$, Fig.~\ref{fig:wall}b). In a confined environment, these cells are therefore always swimming toward one surface, a result which also leads to their accumulation. The physical origin is however different, and instead of being attracted by an image cell as pushers are, pullers simply  swim into the wall.

The fourth  hydrodynamic effect of boundaries, less studied, is a potential reduction of cell-cell hydrodynamic interactions near solid surfaces. Indeed, in many cases, a flow singularity at a distance $h$ from a  solid surface is canceled out in the far field by its image on the other side of the surface, and the overall flow decays faster in space than the original singularity (at distances further away than $h$ from it). For example, a stokeslet near  a wall has a dipolar behavior in the far field if the stokeslet is parallel to the surface, and a quadrupolar decay if it is perpendicular to it \cite{blake71_image,blake74_image}. Swimming cells typically behave as force-dipoles in an infinite fluid, but when near a solid surface, configurations exist where the total flow field (dipole + images) ends up decaying   as $1/r^3$ or $1/r^4$ , and hydrodynamic interactions with other cells are  weaker as a result. 
In that case, the distance $h$ between the swimming cell and the wall acts therefore as an effective cut-off for cell-cell interactions.

%%%%%%%%%%%%%%%%%%%%%%%%%%%%%%%%
\subsection{Interactions between flagella}
\label{flagellumflagellum}

\subsubsection{Eukaryotic flagella and cilia.}

We now decrease our length scales, and discuss hydrodynamic interactions between cellular appendages. The first historical evidence of important hydrodynamic effects between nearby eukaryotic flagella were observations  by James Gray of in-phase beating of spermatozoa flagella, later reproduced by Taylor  \cite{taylor51}. Similar observations have been made since then, such as the  synchronous beating of Fishfly spermatozoa  \cite{hayashi98}. Why would organisms phase-lock in such a manner? Exploiting the swimming sheet model, Taylor showed that two swimmers with the same prescribed waveform (Fig.~\ref{fig:cilia}a) dissipate the least amount of mechanical energy for swimming when the waves are exactly in phase (Fig.~\ref{fig:cilia}b), and the dissipation goes up monotonically with phase difference. Using high-amplitude numerical simulations, these results were later revisited \cite{fauci90}. Two sheets with same waveform but different phases  are seen to swim  with different velocity, and their phase difference $\phi$ evolves until they are either perfectly in phase ($\phi=0$) or perfectly out of phase ($\phi=\pi$), with both configurations being stable \cite{fauci90}. 

Two nearby swimmers therefore  display phase-locking, but the locked state can be that of maximum dissipation, in contrast with what is observed experimentally. For real eukaryotes, the flagellar waveform is however not fixed  but is found as a solution to a mechanical problem, as explained in \S\ref{actuation}: The shape arises from a balance between internal forcing inside the flagella (molecular motors), passive elastic (possibly viscoelastic) resistance from the axoneme, and viscous resistance from the outside fluid. When there is a second swimmer located nearby, the fluid forces are modified, and so is the mechanical force balance determining the flagella shape. Through this neighbor-induced change in the waveform, cells are expected to be able to perfectly phase lock, as is observed in the synchronization of nonlinear oscillators \cite{strogatz00}.
\begin{figure}[t]
\begin{center}
\begin{center}
 \includegraphics[width=0.8\textwidth]{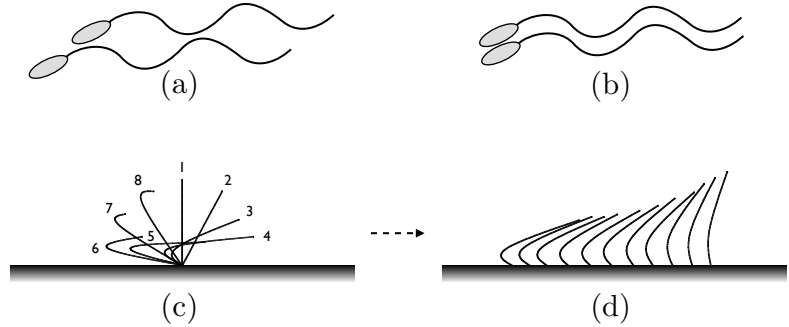}
\end{center}
\caption{Hydrodynamic interactions between eukaryotic flagella and cilia. (a): Nearby swimming cells with arbitrary phase difference; (b) Phase-locked nearby swimming cells; (c) The asymmetric beat pattern of a cilium  (sequence 1 to 10); (d) Schematic  representation of the metachronal waves of collective beating displayed by arrays of cilia.}
\label{fig:cilia}
\end{center}
\end{figure}

Most of the  work related to hydrodynamics-induced waveform change and phase locking was concerned with    cilia, which are short eukaryotic flagella (a few microns long), found in densely packed arrays and involved in many  important   biological functions, including sensing, fluid transport, locomotion, and development  \cite{blake74,brennen77,childress81,sleigh88,braybook,salathe07,smith08}. An individual cilium deforms in a non-reciprocal fashion,  with a typical high-friction power strokes  (sequence 1 to 3 in Fig.~\ref{fig:cilia}c) and a low friction recovery stroke (sequence 4 to 8 in Fig.~\ref{fig:cilia}c).   The internal actuation of each cilium is independent from that of its neighbors, and they do not communicate with each other except through the fluid. However, when they are  closely-packed  on surfaces, cilia arrays display collective behavior, termed {\it metachronal waves}.  The deformation of each cilium is locked in phase with that of its neighbor,  with a constant (small) phase difference, leading to propagating waves of deformation  (Fig.~\ref{fig:cilia}d) \cite{sleigh88}. 

The physical origin of 
coordinated beating is the central question of ciliary dynamics, which 
a number of theoretical studies have attempted to answer~\cite{gueron92,gueron93,gueron97,gueron99,lenz06,vilfan06,guirao07,niedermayer08}.
Two different approaches have been proposed. In the  first one,  the mechanics of each cilium is modeled in the most accurate way, and numerical simulations are used  to compute the collective beating  \cite{gueron92,gueron93,gueron97,gueron99}. The crucial ingredient in that approach is to correctly model the internal load-dependent  force generation in the axoneme (without load dependence, or feedback, there is no phase-locking as can be expected  \cite{niedermayer08}). With that approach, it is found that two cilia starting randomly end up beating in perfect synchrony within two beating cycles \cite{gueron97}. If instead there is a large number of cilia, waves arise naturally as a result of hydrodynamic interactions  \cite{gueron97}. Subsequent work showed that as the waves develop from arbitrary initial conditions the rate of work done by the cilia as they are beating is decreasing \cite{gueron99}. Physically,  because of viscous drag, it is energetically advantageous for  one cilium to beat in the presence of a neighboring cilium with similar phase.

The second approach  considers  simplified models for the dynamics of the cilia,  providing analytical insight into the necessary conditions for phase-locking~\cite{lenz06,vilfan06,guirao07,niedermayer08}. A first study considered a regular lattice of cilia, where their direction (the direction of the beating plane) is assumed to obey a balance between rotational Brownian motion and rotation induced by  the flow created by all other cilia. For small enough temperature, a transition is observed between a state where no net flow occurs on average, to a state where all cilia point in the same direction and a net flow is created~\cite{guirao07}. Further insight is provided by considering a simplified load-dependent internal molecular engine. In that case,  metachronal waves  arise only if a constant phase shift is assumed to exist between each cilium and its neighbor \cite{guirao07}. Motivated by nodal flows in development~\cite{smith08}, a second study considered cilia whose tips perform three-dimensional trajectories over a surface. Each cilium is modeled by a sphere subject to an active load-dependent force, and interacting hydrodynamically with a second cilium. Depending on the  relative position and orientation of the two cilia models, in-phase ($\phi=0$) or out-of-phase locking ($\phi=\pi$)   arise from random initial conditions~\cite{vilfan06}. A similar model with two sphere-like cilia rotating due to an applied torque  near a wall was recently proposed. In that case, in-phase locking is obtained provided that the circular trajectory of each cilium is allowed to vary in response to hydrodynamic interactions~\cite{niedermayer08}. 

%%%%%%%%%%%%%%%%%%%%%%%%%%%%%%%%
\subsubsection{Bacterial flagella.}
\label{bundling}

Hydrodynamic interactions between flagella also play a pivotal role for bacterial locomotion. In that case, the phenomenon of interest is flagellar bundling \cite{anderson75,macnab77}. Wild-type swimming bacteria, such as {\it E. coli}, typically display ``run-and-tumble'' behavior during their locomotion. During runs, the bacterium swims along a roughly straight path, and its flagellar filaments are bundled together tightly behind the cell (Fig.~\ref{fig:bundling}a and d). Near the end of a run, one or or more motors reverses, and the corresponding filaments unwind from the bundle. Viscous stresses lead to polymorphic transitions which turn the cell, ultimately leading to a random reorientation once the motors reverse again and the full bundle forms~\cite{turner00}  (Fig.~\ref{fig:bundling}b). The process by which all the flagella, which are randomly distributed along the cell's surface, come together (Fig.~\ref{fig:bundling}c) to form a perfect aggregate at the end of a tumbling  event (Fig.~\ref{fig:bundling}d) is termed bundling. Since the flagella do not communicate except through the fluid, it has long be postulated that bundling occurs only through hydrodynamic interactions.

\begin{figure}[t]
\begin{center}
\begin{center}
 \includegraphics[width=0.9\textwidth]{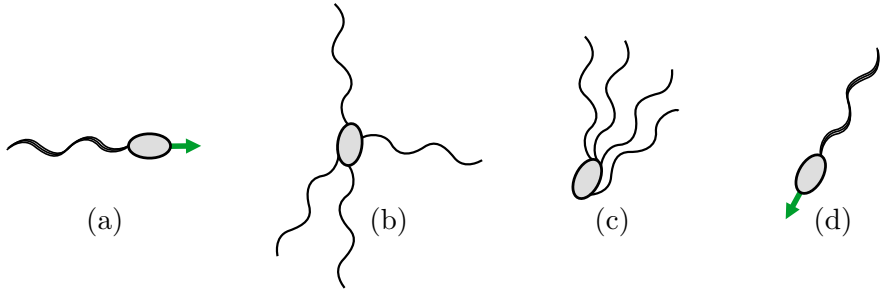}
\end{center}
\caption{Bundling of bacterial flagella. During swimming, the bacterial flagella are gathered in a tight bundle behind the cell as it moves through the fluid (a and d). During a tumbling event, the flagella come out the bundle (b), resulting in a random reorientation of the cell before the next swimming event. At the conclusion of the tumbling event, hydrodynamic interactions lead to  the relative attraction of the flagella (c), and their synchronization to form a perfect bundle (d).}
\label{fig:bundling}
\end{center}
\end{figure}

During bundling, two different physical mechanisms are involved: (1) the  attraction between the rotating flagella, and (2) their subsequent phase-locking inside the bundle.  Concerning the attraction between the rotating flagella, two different scenarios have been proposed. Both mechanisms play a role. One is purely kinematic, and relies on the simple observation that as the cell starts swimming, the drag on all the flagella naturally sweeps them behind the cell~\cite{anderson75}. Under this scenario, the flagella are not so much attracted by each other, but are simply passively dragged behind the cell body. The wrapping of the filaments in a helical shape is then achieved in a passive manner by the body rotation, which is necessary in order to achieve overall torque balance (see \S\ref{RFT})~\cite{anderson75,powers02}. The second attraction scenario relies on hydrodynamic interactions between the flagella, in which each helix induces a flow which causes the other to bend and twist about its neighbors. The geometry of the helices is critical for this scenario. The role of this geometry has been studied with macro-scale experiments using flexible helices in air~\cite{macnab1977}. In these experiments, the bundling was forced by means of guides at the distal ends of the helices. It was found that left-handed helices rotating counter-clockwise as viewed from the distal end (the same handedness and rotation sense as a bacterium on a run) can maintain a steadily rotating bundle, without jams, when the helices are wrapped around each other in a right-handed sense. Later, these experiments were extended with polymeric helices in highly viscous silicon oil~\cite{kim03,kim04_PIV}. There it was shown that induced flow naturally causes tight bundling of pairs of left-handed helices driven counter-clockwise.  The mirror image of this arrangement but no other combination also led to bundling~\cite{kim03}. The time for bundling in the experiment was observed to scale with the rotation speed of the helix, rather than the relaxation time of flexible helices in a viscous fluid.
Bundling has been also been studied computationally using a regularized version of slender-body theory  \cite{flores05}. For three left-handed helical flagella  driven at constant torque and rotating  counter-clockwise, hydrodynamic interactions lead to a bundling of the filaments, although not as tight a bundle as observed experimentally \cite{flores05}.

 The issue of synchronization between  helices driven with constant torques was addressed numerically \cite{kim04,reiehert05}. Perhaps surprisingly, two nearby rotating helices are seen to not synchronize if the helices are perfectly rigid  \cite{kim04}, and some mechanical flexibility is necessary for flow-based synchronization to occur \cite{reiehert05}. In bacteria, this flexibility is likely to be provided by the flagellar hook which connects the rotary motor to the helical flagellum \cite{block89,block91}. The contact forces that arise between different helices as the bundle forms likely also play a role in synchronization.

%%%%%%%%%%%%%%%%%%%%%%%%%%%%%%%%
\section{Swimming in complex fluids}
\label{complex}

Biological fluids are often laden with polymers and therefore have non-Newtonian rheology. For example, mucus is found at the cervix and throughout the rest of the mammalian female reproductive tract~\cite{suarez06}. The nature of the cervical mucus influences the likelihood of fertilization in humans: There is a good correlation between the hydration of the cervical mucus and the incidence of pregnancy~\cite{Bigelow_etal2004}. Although cervical mucus is a complex, heterogeneous gel, some simple trends have been observed. There is an inverse relation between the degree of viscoelasticity of the mucus and the ability of sperm to penetrate the mucus~\cite{wolf77_1,wolf77_2,wolf77_3,wolf78}.  Human sperm flagella in cervical mucus have a higher beat frequency, smaller amplitude, and shorter wavelength than when in semen~\cite{katz78}. The swimming speed is the same in both media, but the sperm swim along straighter paths in cervical mucus~\cite{katz78}. In addition to the mucus at the cervix and in the uterine cavity, sperm must also penetrate the matix coating the ovum, known as the cumulus oophorus, which is a viscoelastic actin-based gel~\cite{dunn76}. Spermatozoa can undergo an internal change known as hyperactivation in which the beat pattern changes from a symmetric to asymmetric form; hyperactivated cells have better motility in viscoelastic media~\cite{suarez91,suarez92,suarez96,ho01,suarez03}.

Rheological measurements show that cervical mucus is highly viscous, with a viscosity of 100\,Pa\,s~\cite{hwang69}. The viscoelasticity is characterized by a relaxation time of $\tau=1$--$10$\,s~\cite{tam80,eliezer74,litt76,wolf77_3} and an elastic modulus of $0.1$--$10$\,Pa\,s~\cite{litt76,wolf77_3}. These properties depend strongly on the phase of the menstrual cycle as well as hydration. Since the typical beat frequency of a flagellum is $\omega=20$--50\,Hz~\cite{brennen77}, the Deborah number $\omega\tau$ is easily larger than one, indicating that viscoelastic effects are important. 

There are several other important examples of swimming or transport in complex fluids. The cilia that line the human upper airway lie in a thin layer of of Newtonian liquid, which in turn is coated with a high-viscosity layer of mucus~\cite{sleigh88}. Again, since the Deborah number for this system is large~\cite{lauga07}, elastic effects are important. Bacteria also encounter viscoelastic fluids: The ulcer-causing bacterium \textit{Helicobacter pylori} swims  through mucus lining the stomach~\cite{MontecuccoRappuoli2001}, and spirochetes move through the connective tissue of the host during infection~\cite{WolgemuthCharonGoldstein2006}.

In our discussion of swimming in a complex fluid, we use continuum mechanics to model the fluid. This simplification is necessary for making progress, but it is important to realize that the size scale of the microstructure of the fluid can be comparable to the size of the swimmer, and therefore a different approach may be necessary to accurately capture the interactions between the swimmer and the polymers. Complex fluids display a vast array of non-Newtonian effects, such as stress relaxation, normal stress differences, and shear-rate dependent viscosity~\cite{bird76,birdvol1,birdvol2,tanner88,larson88,doi88,bird95,larson99}. Our approach is to illustrate some of the distinctive properties of swimmers in complex fluids by focusing on one class of models, fluids with fading memory. These models apply to polymer solutions. 

When a polymer solution is subject to shear, the polymers stretch out. The resulting loss of entropy of the polymers leads to an effective elastic force tending to recover the initial configuration of the polymers. The balance of the entropic force and the viscous drag on the polymer sets the time scale $\lambda$ over which the fluid has memory. The appearance of this new time scale gives polymeric liquids a completely different character relative to viscous Newtonian fluids. First, the property of kinematic reversibility is lost, even if the Reynolds number is vanishingly small, since growth or decay of stress can lag the change in shape of a swimmer by the time scale $\lambda$. Second, the new time scale implies that the constitutive relation for a polymer solution depends on the rate of change of stress with time, which automatically leads to nonlinear terms. This fact may be seen by invoking material frame indifference, a fundamental assumption of continuum mechanics~\cite{oldroyd1950,birdvol1}, or by deriving the continuum theory directly from a microscopic theory of the polymers in solution~\cite{doi88,larson99}.
Many of the properties of Stokes flow that we have invoked in our study of swimming have relied on the linearity of Stokes flow. The nonlinearity of the equations of motion for a polymeric liquid generally implies that the scallop theorem does not hold. The loss of linearity spoils superposition, and therefore we may not use the singular solutions of \S\ref{singularities} to construct a slender body theory. Hydrodynamic interactions between distant cells will have a different character than in Stokes flow, since the far field form of the velocity field due is changed. These features of polymeric liquids make the study of swimming in complex fluids a daunting challenge, but also an area of opportunity. Except for a few early works~\cite{chaudhury79,fulford98,ross74}, the area is largely unexplored.

To examine these issues, we review the calculation~\cite{lauga07} of the swimming velocity of the Taylor sheet of \S\ref{TaylorSheetSection} for a fluid described by the Oldroyd-B model~\cite{birdvol1}. In this model, the deviatoric stress $\btau= p\mathbf{1}+\bsi$ is the sum of a contribution from the polymer solute $\btau^\mathrm{p}$ and a contribution from the Newtonian solvent $\btau^\mathrm{s}=2\eta_\mathrm{s}\mathbf{E}$, where  ${\bf E}=\frac{1}{2}(\nabla \u + \nabla \u ^T)$. The polymer contribution relaxes to the viscous stress over the time scale $\lambda_1$,
\begin{equation}
\btau^\mathrm{p}+\lambda_1\stackrel{\triangledown}{\btau^\mathrm{p}}=2\eta_\mathrm{p}\mathbf{E},\label{taudecaypolymer}
\end{equation}
where 
\begin{equation}
\stackrel{\triangledown}{\btau}
=\frac{\partial \btau}{\partial t}
+\mathbf{u}\cdot\nabla{\btau}
-{\btau}\cdot\nabla\mathbf{u}
-(\nabla\mathbf{u})^{\mathrm{T}}\cdot{\btau} \label{u-convected}
\end{equation}
is the upper-convected derivative of $\btau$. The upper-convected derivative of a tensor $\btau$ is the expression in general coordinates of the rate of change of the tensor calculated in a frame that translates and deforms with the local fluid velocity~\cite{oldroyd1950,birdvol1}. Using $\btau=\btau^\mathrm{p}+\btau^\mathrm{s}$ and eliminating $\btau^\mathrm{p}$ from Eq.~(\ref{taudecaypolymer}) yields the Oldroyd-B model,
\begin{equation}
\btau+\lambda_1\stackrel{\triangledown}{\btau}=2\eta(\mathbf{E}+\lambda_2\stackrel{\triangledown}{\mathbf{E}})\label{OldroydB}
\end{equation}
where $\eta=\eta_\mathrm{p}+\eta_\mathrm{s}$ and $\lambda_2=\eta_\mathrm{s}\lambda_1/\eta<\lambda_1$. The equation of motion is $-\nabla p+\nabla\cdot\btau=0$ with $\nabla\cdot\mathbf{u}=0$.
Note that the explicit presence of the time derivatives as well as the nonlinear terms in Eqn.~(\ref{OldroydB}) spoils kinematic reversibility and violates the assumptions of the scallop theorem.

The solution to the swimming problem for a sheet with prescribed traveling wave (\ref{sheet}) proceeds just as in the Newtonian case. The velocity field, stress, and boundary conditions are expanded to second order in amplitude $b$, and the governing equations are solved subject to the boundary conditions (\ref{TaylorSheetBC}) and (\ref{TaylorSheetBCFar}). To first order, the flow field is identical to the flow field of the Stokes problem. However, due to the relaxation time $\lambda$, the first-order component of the stress field $\btau$ has a lag relative to the Stokes stress field. Through the nonlinear terms of Eq.~(\ref{OldroydB}), this stress field drives a second-order flow that leads to the swimming speed
\begin{equation}
|\mathbf{U}|=\frac{1}{2}\omega k b^2\frac{1+\omega^2\lambda_1^2\eta_\mathrm{s}/\eta}{1+\omega^2\lambda_1^2}\cdot\label{Oldroydspeed}
\end{equation}
Since $\eta_\mathrm{s}<\eta$, the swimmer in an Oldroyd-B fluid is slower than in a Stokes fluid.
Note that unlike the Stokes, case, the swimming speed (\ref{Oldroydspeed}) depends on material parameters such as $\lambda_1$, $\eta_\mathrm{s}$, and $\eta$. The Oldroyd-B constitutive equation is inapplicable to flows with large extension rates. Nevertheless, the result (\ref{Oldroydspeed}) continues to hold for more accurate fading memory models such as FENE-P, the Johson-Segalman-Oldroyd model, and the Giesekus model~\cite{lauga07}. The expression for the speed, Eq.~(\ref{Oldroydspeed}), also continues to hold for a cylindrical filament with a traveling wave, in the limit in which the radius of the cylinder is much smaller than the lateral displacement of the cylinder~\cite{FuPowersWolgemuth2007,FuWolgemuthPowers2009}.

The nonlinear dependence of the swimming speed on the frequency signals the breakdown of the scallop theorem. To see why, first consider the Newtonian limit $\lambda_1\rightarrow0$, in which all elastic effects vanish and the speed is proportional to $\omega$. Now consider a reciprocal motion situation in which a traveling wave of wave number $k$ and frequency $\omega$ travels rightward on the sheet for a period $(2\pi/\omega)/3$, and then leftward for a period $2(2\pi/\omega)/3$~\cite{FuWolgemuthPowers2009}. The sheet will have zero net displacement after this process. Now consider this waveform for a sheet in an Oldroyd-B fluid. Suppose the sheet has periodically been executing these motions long enough that transients from startup from rest have died away. Since the speed now depends nonlinearly on $\omega$, the translation in each segment of the motion is different: The sheet moves faster during the leftward motion with the smaller frequency. Thus, there is a net displacement. This argument disregards the memory of the fluid, since the stress during each stroke is effected by the previous stroke. Accounting for these memory effects leads to a slightly smaller net displacement, but it still has the same qualitative dependence on $\omega\lambda_1$.

The properties of the medium can affect not only the speed of a swimmer with a prescibed stroke, but also the  shape of a beating flagella. For example, as the viscosity of the solution increases, the wave form of a human sperm flagellum flattens along most of its length, with most of the deflection taking place at the distal end~\cite{Ishijima1986}.
Thus it natural to model the dependence of the shape on material parameters such as viscosity and relaxation time. As we learned in \S\ref{actuation}, the small-amplitude shape is determined by an equation which is first-order in amplitude. Thus, we only need the 
fluid force to first order. For the Oldroyd-B fluid, the calculations for the filament geometry shows that this fluid force has the form of resistive force theory, with complex frequency-dependent effective viscosity~\cite{FuWolgemuthPowers2008}. Denoting Fourier components with a tilde, the shape equation for an Oldroyd-B fluid is  
\begin{equation}
\frac{\mathrm{i}\xi_\perp}{1+\mathrm{i}\lambda_1\omega} \tilde h=-A\frac{\partial^4 \tilde h}{\partial x}-a\frac{\partial \tilde f}{\partial x},\label{complexshapeeqn}
\end{equation}
where $f$ is the sliding force density. By prescribing the force density, imposing the boundary conditions, and using linear superposition (valid to first order), we may calculate the shape of the filament as a function of material parameters. The result is that this model gives patterns that look qualitatively similar to the experimental observations~\cite{FuWolgemuthPowers2008}. Since the shape of the filament determines the swimming velocity, the dependence of shape on relaxation time gives an addition correction to the swimming velocity of the prescribed sheet.

A natural next step is to extend the study of swimming to large amplitude deflections. Recent numerical work on large-amplitude peristaltic pumping of an Oldroyd-B fluid two dimensions  has shown similar qualitative differences with the Newtonian case, such as reduced pump rate, dependence on material parameters, and loss of kinematic reversibility~\cite{TeranFauciShelley2008}. The application of numerical methods such as these to swimmers will be crucial for understanding finite-size effects such as the role of the shape of the cell body or the end of the flagellum, since large normal stresses can develop at such regions of high curvature.

%%%%%%%%%%%%%%%%%%%%%%%%%%%%%%%%
%%%%%%%%%%%%%%%%%%%%%%%%%%%%%%%%

\section{Artificial swimmers and optimization}
\label{artificial}

\begin{figure}[t]
\begin{center}
\begin{center}
 \includegraphics[width=0.95\textwidth]{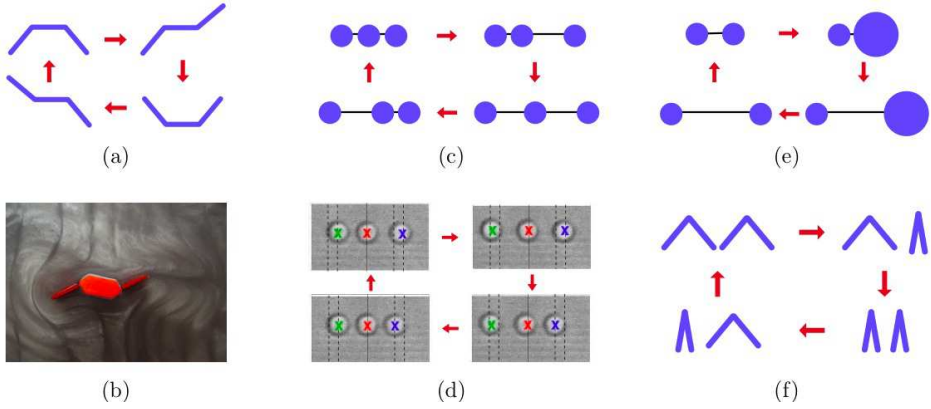}
\end{center}
\caption{Artificial swimmers with discrete degrees of freedom: 
(a) Purcell's three-link swimmer (red arrows indicate change in conformation of the swimmer shape); 
(b) Macro-scale experimental realization of Purcell's swimmer (Photo courtesy of Brian Chan and Peko Hosoi); 
(c) Three-sphere swimmer;
(d) Micro-scale realization of a three-sphere pump \cite{leoni08};
(e) Two-sphere swimmer with shape change;
(f) Hydrodynamic interactions between two reciprocal swimmers.
Picture in Fig.~\ref{fig:discrete}d adapted from Ref.~ \cite{leoni08} courtesy of Marco Cosentino Lagomarsino.
}
\label{fig:discrete}
\end{center}
\end{figure}

In this final section, we look beyond the biological realm and survey the bioengineering advances in design and optimization of artificial swimmers and bio-inspired systems.

\subsection{Designing artificial swimmers}

A number of designs have been proposed and tested for artificial swimming devices at low Reynolds number. They can  be sorted in three different categories.  The first category includes all swimmers which deform their shape with only a  
finite number of degrees of freedom,  actuated in a non-reciprocal fashion (and therefore,  at least two degrees of freedom are needed). The original example  is Purcell's three link swimmer, which posseses two hinges varying in time with phase differences (Fig.~\ref{fig:discrete}a and b)  \cite{purcell77,becker03,tam07,avron08}. A second related design is that of a three-sphere swimmer (Fig.~\ref{fig:discrete}c): The distances separating the
spheres vary in time and with phase differences, leading to locomotion \cite{najafi04,najafi05,dreyfus05,golestanian08,golestanian08_2,golestanian08_3}.  The extension to N spheres was also proposed theoretically \cite{felderhof06}.  The three-sphere design was implemented experimentally using optical tweezers, and the out-of-phase motion of the three colloidal  beads leads to fluid  pumping  (Fig.~\ref{fig:discrete}d)  \cite{leoni08}.  A third design only needs two spheres, but is coupled with a change in shape of one of the spheres, leading to non-reciprocal kinematics (Fig.~\ref{fig:discrete}e). As the sphere increases in size, it serves as an anchor against which the swimmer can push, and rectify a time-periodic change in size  \cite{avron05:pushme}. A related idea using rotating two-sphere swimmers, was also proposed \cite{ogrin08}. Finally, because there is no ``many-scallop" theorem \cite{koiller96}, two reciprocal non-swimmers can exploit hydrodynamic interactions to swim collectively  (Fig.~\ref{fig:discrete}f). The resulting collective swimming speeds depend on the separation distance between the swimmers as well as  their relative position and orientation   \cite {laugabartolo08,alexander08}.

\begin{figure}[t]
\begin{center}
\begin{center}
 \includegraphics[width=0.8\textwidth]{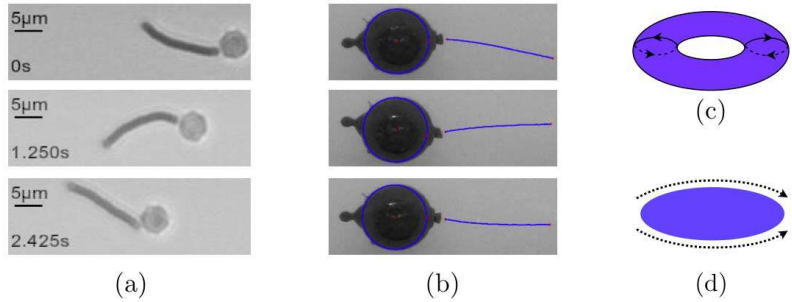}
\end{center}
\caption{Artificial swimmers with continuous deformation:   
(a) Synthetic flagella made of actuated micron-scale magnetic filaments \cite{dreyfus05_nature};
(b) Macro-scale swimming by boundary actuation of an elastic filament  \cite{yu_unpublished};
(c) Toroidal swimming;
(d) Swimming by surface treadmilling.
Picture in Fig.~\ref{fig:continuous}a adapted from  Ref.~\cite{howse07} courtesy of Remy Dreyfus and by permission from
Macmillan Publishers Ltd, copyright 2005.}
\label{fig:continuous}
\end{center}
\end{figure}

The second category of artificial swimmers includes bodies deforming in a continuous fashion. In an experimental breakthrough, a team has been able to exploit the response of paramagnetic filaments to external magnetic fields to  construct the first artificial micron-long swimmer (Fig.~\ref{fig:continuous}a)  \cite{dreyfus05_nature}, thereby motivating a number of theoretical studies  \cite{cebers05,gauger06,roper06,keaveny08,roper08}. In this work, a AC magnetic field is applied in the direction perpendicular to the  filaments, and the presence of a body (here, a red blood cell), breaks the right-left symmetry in the continuous response of the filament. The result is a wave of deformation traveling from the tip of the filament to the body it is attached to, and locomotion flagella-first \cite{dreyfus05_nature}. A second type of swimmer with continuous deformation exploits the boundary actuation of elastic filaments to generate propulsive forces (see \S\ref{actuation}), and locomotion (Fig.~\ref{fig:continuous}b). In a macro-scale experiment, the forces arising from such elastic swimming were measured \cite{yu06}, and the actuation method was implemented successfully to obtain swimming   \cite{kosa07,yu_unpublished}. A third type of swimmers with continuous deformation  are those with a special mode of surface deformation called ``surface treadmilling'', for which the shape of the swimmer is fixed and where its body undergoes a continuous tangential displacement along its surface (Fig.~\ref{fig:continuous}c,d). In his original article, Purcell proposed such design in the shape of a torus undergoing continuous surface rotation (Fig.~\ref{fig:continuous}c)  \cite{purcell77}, an idea recently analyzed in detail \cite{thaokar07,thaokara08,leshansky08}. If the body is slender and displays directed tangential displacement all along its surface, so that  a material source is present on one side of the body,  and a sink  on the other side (Fig.~\ref{fig:continuous}d), the resulting  locomotion can occur with arbitrarily high efficiency \cite{leshansky07}.

The final category of artificial swimmers use chemical reactions to power locomotion,  a case investigated both experimentally  \cite{paxton04,fournier-bidoz05,kline05,mano05,paxton06, ruckner07,howse07,leoni08,tao08} and theoretically  \cite{golestanian05,golestanian07,brady08}. The prototypical setup is  a body composed of two materials, one that is inert, and one that is a catalyst  or  reactant  for a chemical reaction. The presence of a chemical reaction leads to an imbalance in concentration for some of the reactants and/or products to the reaction, which leads to an imbalance of osmotic pressures on the body, and results in directed propulsion. An example of the trajectory of such chemical swimmer is illustrated in Fig.~\ref{fig:chemical}.

\begin{figure}[t]
\begin{center}
\begin{center}
 \includegraphics[width=0.7\textwidth]{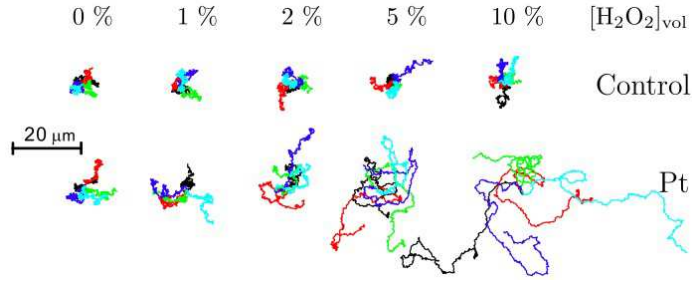}
\end{center}
\caption{Locomotion powered by chemical reactions  \cite{howse07}. Polystyrene colloidal spheres in a solution of water and hydrogen peroxide  ([H$_2$O$_2$]) with increasing concentration. 
Top: Control experiment (inert spheres);
Bottom: Experiment where the spheres are half-coated by platinum, a catalyst for the reduction of [H$_2$O$_2$] into oxygen and water, which leads to directed swimming of the spheres (coupled to Brownian rotational diffusion). Picture reproduced from  Ref.~\cite{howse07} courtesy of Ramin Golestanian by permission from the American Physical Society,  copyright 2007.}
\label{fig:chemical}
\end{center}
\end{figure}

%%%%%%%%%%%%%%%%%%%%%
\subsection{Exploiting low-$\Re$ locomotion}

The physical mechanisms of propulsion at low Reynolds number  can be exploited in a number of  ways beyond biological or synthetic locomotion.  In a pioneering experiment,  suspensions of swimming 
bacteria were seen to provide an effective high-temperature thermal bath for suspended inert particles  \cite{wu00}. Similar results were obtained  for diffusive mixing of two fluids containing swimming cells  \cite{kim04_diffusion}, and for the motion of colloidal particles above surfaces covered with attached bacteria  (Fig.~\ref{fig:bath}) \cite{darnton04}. A second  utilization of swimming is  cargo-towing, with potential applications in biomedical devices. Experiments were performed with solid bodies covered with attached bacteria, with successful translational and rotational towing \cite{darnton04}. A related method was also devised to transport micron-size objects at the single-cell level \cite{weibel05}, and a theoretical framework for towing by swimming now exists \cite{raz08}. A final application of flagella motion is pumping. Indeed, swimming and pumping are dual problems, and in general a tethered swimmer acts as a pump  \cite{raz07}.  One approach, studied numerically, uses elastic filaments attached to a solid surface and actuated by external time-varying torques  \cite{kim06:pumping}. Experiments were also conducted on surfaces covered with filaments made of self-oscillating gels and displaying wave-like motion \cite{tabata02}.

%%%%%%%%%%%%%%%%%%%%%

\subsection{Optimization}

Motivated by the optimal tuning of synthetic micro-swimmers, as well as possible insight into evolutionary processes, significant work as been devoted to the optimization of locomotion at low Reynolds number. Since time can be 
scaled out of all low Reynolds number swimming problems in Newtonian fluids, optimizing swimming speeds is not, in general, a well-posed mathematical problem, and an additional form of normalization is required. The traditional approach is to define a swimming efficiency as the ratio of the useful to the total rate of working of the swimmer against the viscous fluid (see \S\ref{RFT}). The optimization problem becomes a maximization problem for the efficiency \cite{lighthill75}, and is equivalent to finding the swimmer with the largest swimming speed for a given amount of mechanical energy available.

\begin{figure}[t]
\begin{center}
\begin{center}
 \includegraphics[width=0.75\textwidth]{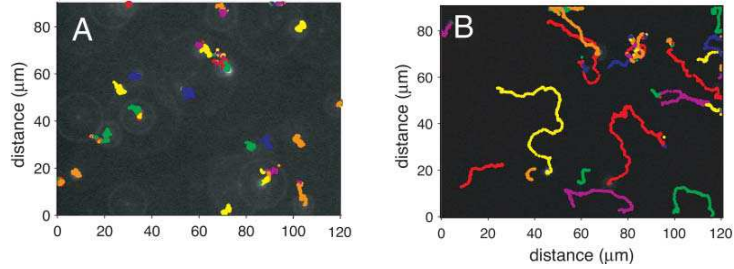}
\end{center}
\caption{Enhanced diffusion near a carpet of  swimming cells \cite{darnton04}. Trajectory of  fluorescent particles over a 10\,s interval when located above a flat surface coated with attached bacteria (two-dimensional view from above the surface).
(A): The particles are located $80$\,$\mu$m above the surface, and move by  Brownian motion.
(B): When the particles are located  3\,$\mu$m above the surface, their effective diffusion is significantly larger than Brownian motion~\cite{darnton04}. Picture reproduced from  Ref.~\cite{darnton04} courtesy of Kenny Breuer and by permission from 
the Biophysical Society,  copyright 2004.}
\label{fig:bath}
\end{center}
\end{figure}

Two types of optimization questions can be asked, related to either the waveform or the overall geometry of the swimmer. The first problem was formulated by Lighthill as the following: If a flagellum is being distorted as a planar traveling wave, what would be its optimal waveform \cite{lighthill75}?  Assuming an infinitely long flagellum, and within the framework of the local drag theory, the optimized flagellum is a sawtooth wave, where each branch of the sawtooth makes an angle $\pm \theta$ with the horizontal, with $\theta=\arccos \left[({1+({\xi_\parallel/\xi_\perp})^{1/2}})^{-1/2}\right]$  \cite{lighthill75}.  Since $\xi_\parallel/\xi_\perp\approx 1/2$, the optimal angle for the sawtooth is  $\theta\approx 40^\circ$, leading to an optimal value for the product of the wave amplitude, $b$, and the wavenumber, $k$, as $bk\approx 1.3$. The optimal value for $\theta$ is insensitive to the exact value of the ratio $\xi_\parallel/\xi_\perp$, or on the possible presence of a cell body  \cite{lighthill75}. Numerical simulations  for  sawtooth patterns revealed the optimal criterion of Lighthill to remain valid for finite-size swimmers \cite{pironneau74}. In addition, flagella with the optimal sinusoidal deformation have very similar geometrical characteristics: $bk\approx 1.26$ using the local drag theory   \cite{pironneau_inbook},  $bk\approx 1$ using slender body theory  \cite{higdon79a}. Similar results are obtained by numerical  optimization of the flagellar waveform  \cite{tamPhD}.

The existence of an optimal angle between the direction of swimming and the direction of the flagellum can be understood on the basis of the physical picture we introduced for drag-based propulsion (Fig.~\ref{drag}). Indeed, we see from 
Eq.~(\ref{prop}) that the propulsive force arising from the local motion of the flagellum scales with its orientation as 
$f_{\rm prop}\sim\sin\theta\cos\theta$. If the work done by the flagellum was not a function of its orientation, the optimal angle would therefore be the one for which $\sin 2\theta\sim 1$, so $\theta\sim 45^\circ$. Because $\xi_\perp > \xi_\parallel$, some energy can be saved by decreasing the angle, thereby promoting tangential  over perpendicular motion, which explains why the optimal angle is close to, but below,  45$^\circ$. In the case of planar waves, the optimal flagellum waveform is therefore kinked (sawtooth), and the slope angle alternates between $+\theta$ and $-\theta$. In the case of a three-dimensional helical filament, a constant angle between the local tangent along the flagellum and the swimming direction can be accommodated while keeping a smooth waveform if (and only if) the flagellum takes the shape of  a helix.
In that case, the optimal angle was determined numerically using slender body theory to be  $\theta= 45^\circ$ \cite{higdon79b}  (see also Refs.~\cite{chwang75,lighthill96_helical}).

A second design feature which can be optimized is the overall geometry of the swimming cell, in particular the ratio of the length $L$ of its flagellum and the radius $a$ of its  body, which is assumed to be spherical for simplicity. In the case of swimmers exploiting planar waves, this problem can only be studied if hydrodynamic interactions between the body and the flagellum are properly taken into account, as otherwise the presence of a body always decreases the swimming efficiency (see Eq.~\ref{efficiencyEqn2}). This problem was analyzed in detail using slender body theory for sinusoidal waveforms, and the optimal flagellum length to body size ratio  was found to be $L/a\approx 25 $ \cite{higdon79a}. Subsequent work determining the optimal flagellar  waveforms showed that the optimal ratio is close to $L/a\approx 24$  \cite{tamPhD}. In the case of helical waves, an optimal body size can be determined even in the absence of hydrodynamic interactions. Indeed, we saw in \S\ref{RFT} that cells with helical flagella need a body in order to be able to move. A small body  prevents  the swimmer from moving and is not efficient, whereas a large body presents too large of a drag to the overall cell, and is not efficient either: An optimal size must therefore exist.  When hydrodynamic interactions are also taken into account, the optimal ratio has been determined numerically to be $L/a\approx 10$  using slender-body theory \cite{higdon79b} and $\approx 12 $ using three-dimensional boundary elements  \cite{fujita01} (see also Ref.~\cite{chwang75}).  Other biologically-relevant optimization problems that have been addressed include the optimization of feeding current for tethered cells  \cite{higdon79}, and the influence of the body shape on the optimal flagellar waveform  \cite{tamPhD} and overall cell geometry \cite{fujita01}.

Finally, work has been devoted to the optimization of simple (artificial) swimmers. For bodies with discrete degrees of freedom, the actuation of Purcell's swimmer (Fig.~\ref{fig:discrete}a)  was optimized numerically \cite{tam07,avron08}. More formal mathematical work was devoted to the optimization of the  three-sphere swimmer (Fig.~\ref{fig:discrete}c) using a control-theory framework \cite{alouges08}. For swimmers with continuous degrees of freedom, the geometrical approach to  swimming at low Reynolds number  was used to derive optimal low-amplitude swimming by surface deformation of spheres and cylinders \cite{shapere87,shapere89_1,shapere89_2}, and was later extended to large-amplitude deformation of  two-dimensional bodies \cite{avron04:opt}.

\section{Outlook}

In this paper, our aim has been to illustrate the theoretical  
framework necessary to understand biological and synthetic locomotion  
at low Reynolds number. Since new experimental methods and setups  
promise to reveal the mechanisms for biological locomotion with ever  
more quantitative detail, future work in the field is likely to be  
significant. We believe in particular that there is great opportunity  
for theorists, since---as we have emphasized throughout the review---simple calculations are usually sufficient to gain fundamental insight  
into the mechanisms of locomotion. Detailed numerical computations will also  
play a crucial role, since many of the issues we discussed, such as  
hydrodynamic interactions in dense suspensions of swimmers and  
locomotion in complex fluids, involve nonlinear processes. A further  
challenge will be to integrate the understanding of basic mechanisms  
across multiple scales, from the levels of molecular motors to  
individual cells to large populations of cells. Finally, many of the  
implications of the ideas sketched in this review have yet to be  
realized in important areas such as the ecology of marine bacteria,  
the formation of bacterial biofilms, and the mechanics of reproduction.

\ack{It is a pleasure to acknowledge the many useful discussions on locomotion at low Reynolds number that we have had over the years with our colleagues: Denis Bartolo, Howard Berg, Alison Berke, Michael Berns, Jacy Bird, Kenny Breuer, Willow DiLuzio, Henry Fu, Ray Goldstein, David Gonzalez-Rodriguez, Alexander Grosberg, Peko Hosoi, Greg Huber, John Kessler, MinJun Kim, MunJu Kim, Stephan Koehler, Marcos, Thibaud Normand, Vivek Shenoy, Mike Shelley, Linda Shi, Todd Squires, Srikanth Srigiriraju, Roman Stocker, Howard Stone, John Toner, Renaud Trouilloud, Linda Turner, Annemarie Van Parys, George Whitesides, Chris Wiggins, Charles Wolgemuth and Tony Yu.

This work was supported in part by National Science Foundation Grants Nos.
CTS-0624830 (EL), CBET-0746285 (EL), NIRT-0404031 (TRP) and  DMS-0615919 (TRP). The authors thank the Aspen Center for Physics, where this work was conceived.
}
%\newpage

\bibliographystyle{unsrt}
\bibliography{reviewTP}

\begin{thebibliography}{100}

\bibitem{bergbook}
H.~C. Berg.
\newblock {\em {\it E. coli} in {M}otion}.
\newblock Springer-Verlag, New York, NY, 2004.

\bibitem{suarez06}
S.~S. Suarez and A.~A. Pacey.
\newblock Sperm transport in the female reproductive tract.
\newblock {\em Human Reprod. Update}, 12:23--37, 2006.

\bibitem{childress81}
S.~Childress.
\newblock {\em Mechanics of {S}wimming and {F}lying}.
\newblock Cambridge University Press, Cambridge U.K., 1981.

\bibitem{ellington84}
C.~P Ellington.
\newblock {\em The Aerodynamics of Hovering Insect Flight}.
\newblock The Royal Society, London, U.K., 1984.

\bibitem{vogel96}
S.~Vogel.
\newblock {\em Life in Moving Fluids}.
\newblock Princeton University Press, Princeton, NJ, 1996.

\bibitem{alexander02}
D.~E. Alexander.
\newblock {\em Nature's Flyers: Birds, Insects, and the Biomechanics of
  Flight}.
\newblock The Johns Hopkins University Press, Baltimore, MD, 2002.

\bibitem{dudley02}
R.~Dudley.
\newblock {\em The Biomechanics of Insect Flight: Form, Function, Evolution}.
\newblock Princeton University Press, Princeton, NJ, 2002.

\bibitem{vogel03}
S.~Vogel.
\newblock {\em Comparative Biomechanics: {L}ife's Physical World}.
\newblock Princeton University Press, Princeton, NJ, 2003.

\bibitem{ludwig}
W.~Ludwig.
\newblock Zur theorie der {F}limmerbewegung ({D}ynamik, {N}utzeffekt,
  {E}nergiebilanz).
\newblock {\em J. Comp. Physiol. A}, 13:397--504, 1930.

\bibitem{gray68}
J.~Gray.
\newblock {\em Animal locomotion}.
\newblock Norton, London, 1968.

\bibitem{lighthill75}
J.~Lighthill.
\newblock {\em Mathematical Biofluiddynamics}.
\newblock SIAM, Philadelphia, 1975.

\bibitem{lighthill76}
J.~Lighthill.
\newblock Flagellar hydrodynamics---{The John von Neumann} lecture, 1975.
\newblock {\em SIAM Rev.}, 18:161--230, 1976.

\bibitem{brennen77}
C.~Brennen and H.~Winet.
\newblock Fluid mechanics of propulsion by cilia and flagella.
\newblock {\em Ann. Rev. Fluid Mech.}, 9:339--398, 1977.

\bibitem{purcell77}
E.~M. Purcell.
\newblock Life at low {R}eynolds number.
\newblock {\em Am. J. Phys.}, 45:3--11, 1977.

\bibitem{yates86}
G.~T. Yates.
\newblock How microorganisms move through water.
\newblock {\em Am. Sci.}, 74:358--365, 1986.

\bibitem{fauci06}
L.~J. Fauci and R.~Dillon.
\newblock Biofluidmechanics of reproduction.
\newblock {\em Ann. Rev. Fluid Mech.}, 38:371--394, 2006.

\bibitem{holwill66}
M.~E.~J. Holwill.
\newblock Physical aspects of flagellar movement.
\newblock {\em Physiol. Rev.}, 46:696--785, 1966.

\bibitem{jahn72}
T.~L. Jahn and J.~J. Votta.
\newblock Locomotion of protozoa.
\newblock {\em Ann. Rev. Fluid Mech.}, 4:93--116, 1972.

\bibitem{blum79}
J.~J. Blum and M.~Hines.
\newblock Biophysics of flagellar motility.
\newblock {\em Quarterly Rev. Biophys.}, 12:103--180, 1979.

\bibitem{berg00_PT}
H.~C. Berg.
\newblock Motile behavior of bacteria.
\newblock {\em Phys. Today}, 53:24--29, 2000.

\bibitem{braybook}
D.~Bray.
\newblock {\em Cell Movements}.
\newblock Garland Publishing, New York, NY, 2000.

\bibitem{turner00}
L.~Turner, W.~S. Ryu, and H.~C. Berg.
\newblock Real-time imaging of fluorescent flagellar filaments.
\newblock {\em J. Bacteriol.}, 182:2793--2801, 2000.

\bibitem{tadir89}
Y.~Tadir, W.~H. Wright, O.~Vafa, T.~Ord, R.~H. Asch, and M.~W. Berns.
\newblock Micromanipulation of sperm by a laser generated optical trap.
\newblock {\em Fertil. Steri.}, 52:870--873, 1989.

\bibitem{tadir90}
Y.~Tadir, W.~H. Wright, O.~Vafa, T.~Ord, R.~H. Asch, and M.~W. Berns.
\newblock Force generated by human sperm correlated to velocity and determined
  using a laser generated optical trap.
\newblock {\em Fertil. Steri.}, 53:944--947, 1990.

\bibitem{konig96}
K.~Konig, L.~Svaasand, Y.~G. Liu, G.~Sonek, P.~Patrizio, Y.~Tadir, M.~W. Berns,
  and B.~J. Tromberg.
\newblock Determination of motility forces of human spermatozoa using an 800 nm
  optical trap.
\newblock {\em Cell. Mol. Biol.}, 42:501--509, 1996.

\bibitem{mccord05}
R.~P. McCord, J.~N. Yukich, and K.~K. Bernd.
\newblock Analysis of force generation during flagellar assembly through
  optical trapping of free-swimming \textit{{C}hlamydomonas reinhardtii}.
\newblock {\em Cell Mot. Cytos.}, 61:137--144, 2005.

\bibitem{chattopadhyay06}
S.~Chattopadhyay, R.~Moldovan, C.~Yeung, and X.~L. Wu.
\newblock Swimming efficiency of bacterium \textit{{E}scherichia coli}.
\newblock {\em Proc. Natl. Acad. Sci. USA}, 103:13712--13717, 2006.

\bibitem{teff07}
Z.~Teff, Z.~Priel, and L.~A. Gheber.
\newblock Forces applied by cilia measured on explants from mucociliary tissue.
\newblock {\em Biophys. J.}, 92:1813--1823, 2007.

\bibitem{taylor51}
G.~I. Taylor.
\newblock Analysis of the swimming of microscopic organisms.
\newblock {\em Proc. Roy. Soc. A}, 209:447--461, 1951.

\bibitem{hancock53}
G.~J. Hancock.
\newblock The self-propulsion of microscopic organisms through liquids.
\newblock {\em Proc. Roy. Soc. Lond. A}, 217:96--121, 1953.

\bibitem{gray55}
J.~Gray and G.~J. Hancock.
\newblock The propulsion of sea-urchin spermatozoa.
\newblock {\em J. Exp. Biol.}, 32:802--814, 1955.

\bibitem{StevensGalyovStevens2006}
J.~M. Stevens, E.~E. Galyov, and M.~P. Stevens.
\newblock Actin-dependent movement of bacterial pathogens.
\newblock {\em Nature Rev. Microbiol.}, 4:91--101, 2006.

\bibitem{Walsby1994}
A.~E. Walsby.
\newblock Gas vesicles.
\newblock {\em Microbiol. Rev.}, 58:94--144, 1994.

\bibitem{berg00}
H.~C. Berg.
\newblock Motile behavior of bacteria.
\newblock {\em Phys. Today}, 53:24--29, 2000.

\bibitem{berg03}
H.~C. Berg.
\newblock The rotary motor of bacterial flagella.
\newblock {\em Ann. Rev. Biochem.}, 72:19--54, 2003.

\bibitem{berg2004}
H.~C. Berg.
\newblock {\em \textit{E. coli} in Motion}.
\newblock Springer-Verlag, New York, 2004.

\bibitem{LoweMeisterBerg1987}
G.~Low, M.~Meister, and H.~C. Berg.
\newblock Rapid rotation of flagellar bundles in swimming bacteria.
\newblock {\em Nature}, 325:637--640, 1987.

\bibitem{KoyasuShirakihara1984}
S.~Koyasu and Y.~Shirakihara.
\newblock \textit{Caulobacter crescentus} flagellar filament has a right-handed
  helical form.
\newblock {\em J. Mol. Biol.}, 173:125--130, 1984.

\bibitem{ArmitageSchmitt1997}
J.~P. Armitage and R.~Schmitt.
\newblock Bacterial chemotaxis: \textit{{R}hodobacter sphaeroides} and
  \textit{{S}inorhizobium meliloti}---variations on a theme?
\newblock {\em Microbiology}, 143:3671--3682, 1997.

\bibitem{MarwanAlamOesterhelt1984}
W.~Marwan, M.~Alam, and D.~Oesterhelt.
\newblock Rotation and switching of the flagellar motor assembly in
  \textit{Halobacterium halobium}.
\newblock {\em J. Bacteriol.}, 173:1971--1977, 1991.

\bibitem{JarrellMcBride2008}
Ken~F. Jarrell and Mark~J. Mc{B}ride.
\newblock The surprisingly diverse ways that prokaryotes move.
\newblock {\em Nat. Rev. Microbiol.}, 6:466--476, 2008.

\bibitem{Canale-Parola1984}
E.~Canale-Parola.
\newblock The spirochetes.
\newblock In N.~R. Krieg and J.~G. Holt, editors, {\em Bergey's Manual of
  Systematic Bacteriology}, pages 38--70. Williams and Wilkins, Baltimore,
  1984.

\bibitem{Berg1976}
H.~C. Berg.
\newblock How spirochetes may swim.
\newblock {\em J. Theor. Biol.}, 56:269--273, 56.

\bibitem{GoldsteinCharonKreiling1994}
S.~F. Goldstein, N.~W. Charon, and J.~A. Kreiling.
\newblock \textit{Borrelia burgdorferi} swims with a planar waveform similar to
  that of eukaryotic flagella.
\newblock {\em Proc. Natl. Acad. Sci. USA}, 91:3433--3437, 1994.

\bibitem{BergTurner1979}
H.~C. Berg and L.~Turner.
\newblock Movement of microorganisms in viscous environments.
\newblock {\em Nature}, 278:349--351, 1979.

\bibitem{Nakamura_etal2006}
S.~Nakamura, Y.~Adachi, T.~Goto, and Y.~Magariyama.
\newblock Improvement in motion efficiency of the spirochete
  \textit{{B}rachyspira pilosicoli} in viscous environments.
\newblock {\em Biophys. J.}, 90:3019--3026, 2006.

\bibitem{Murphy_etal2008}
G.~E. Murphy, E.~G. Matson, J.~R. Leadbetter, H.~C. Berg, and G.~J. Jensen.
\newblock Novel ultrastructures of \textit{{T}reponema primitia} and their
  implications for motility.
\newblock {\em Mol. Microbiol.}, 67:1184--1195, 2008.

\bibitem{ShaevitzLeeFletcher2005}
J.~W. Shaevitz, J.~Y. Lee, and D.~A. Fletcher.
\newblock \textit{Spiroplasma} swim by a processive change in body helicity.
\newblock {\em Cell}, {\bf 122}:941, 2005.

\bibitem{Berg2002}
H.~C. Berg.
\newblock How spiroplasma might swim.
\newblock {\em J. Bacteriol.}, 184:2063--2064, 2002.

\bibitem{TrachtenbergGilad2001}
S.~Trachtenberg and R.~Gilad.
\newblock A bacterial linear motor: cellular and molecular organization of the
  contractile cytoskeleton of the helical bacterium \textit{Spiroplasma
  melliferum} bc3.
\newblock {\em Mol. Microbiol.}, 41:827--848, 2001.

\bibitem{Wolgemuth2003}
C.~W. Wolgemuth, O.~Igoshin, and G.~Oster.
\newblock The motility of mollicutes.
\newblock {\em Biophys. J.}, 85:828--842, 2003.

\bibitem{WernerSimmons2008}
M.~Werner and L.~W. Simmons.
\newblock Insect sperm motility.
\newblock {\em Biol. Rev.}, 83:191--208, 2008.

\bibitem{Quicke1992}
D.~L.~J. Quicke, S.~N. Ingram, H.~S. Baillie, and P.~V. Gaitens.
\newblock Sperm structure and ultrastructure in {H}ymenoptera ({I}nsecta).
\newblock {\em Zoologica Scripta}, 21:381--402, 1992.

\bibitem{HatsumiWakahami1986}
M.~Hatsumi and K.-I. Wakahama.
\newblock The sperm length and the testis length in \textit{{D}rosophila
  nasuta} subgroup.
\newblock {\em Jpn. J. Genet.}, 61:241--244, 1986.

\bibitem{JolyBressacLachaise1995}
D.~Joly, C.~Bressac, and D.~Lachaise.
\newblock Disentangling giant sperm.
\newblock {\em Nature}, 377:202, 1995.

\bibitem{Mitchell2000}
D.~R. Mitchell.
\newblock \textit{Chlamydomonas} flagella.
\newblock {\em J. Phycol.}, 36:261--273, 2000.

\bibitem{Gibbons1981}
I.~R. Gibbons.
\newblock Cilia and flagella of eukaryotes.
\newblock {\em J. Cell Biol.}, 91:107s--124s, 1981.

\bibitem{sleigh88}
M.~A. Sleigh, J.~R. Blake, and N.~Liron.
\newblock The propulsion of mucus by cilia.
\newblock {\em Am. Rev. Resp. Dis.}, 137:726--741, 1988.

\bibitem{happel}
J.~Happel and H.~Brenner.
\newblock {\em Low {R}eynolds {N}umber {H}ydrodynamics}.
\newblock Prentice Hall, Englewood Cliffs, NJ, 1965.

\bibitem{kimbook}
S.~Kim and J.~S. Karilla.
\newblock {\em Microhydrodynamics: {P}rinciples and {S}elected {A}pplications.}
\newblock Butterworth-Heinemann, Boston, MA, 1991.

\bibitem{leal}
L.~G. Leal.
\newblock {\em Advanced Transport Phenomena: {F}luid Mechanics and Convective
  Transport Processes}.
\newblock Cambridge University Press, Cambridge, UK, 2007.

\bibitem{hinch_lowRe}
E.~J. Hinch.
\newblock Hydrodynamics at low {R}eynolds number: {A} brief and elementary
  introduction.
\newblock In E.~Guyon, Nadal J-P., and Pomeau, editors, {\em Disorder and
  Mixing}, volume~1, pages 43--55. Kluwer, N.A.T.O. A.S.I. E, 1988.

\bibitem{poz}
C.~Pozrikidis.
\newblock {\em Theoretical and Computational Fluid Dynamics}.
\newblock Oxford University Press, Oxford, UK, 2000.

\bibitem{chwang75}
A.~T. Chwang and T.~Y.~T. Wu.
\newblock Hydromechanics of low-{R}eynolds-number flow. 2. {S}ingularity method
  for {S}tokes flows.
\newblock {\em J. Fluid Mech.}, 67:787--815, 1975.

\bibitem{batchelor1967}
B.K. Batchelor.
\newblock {\em An Introduction to Fluid Mechanics}.
\newblock Cambridge University Press, Cambridge, 1967.

\bibitem{pozrikidis_BI}
C.~Pozrikidis.
\newblock {\em Boundary Integral and Singularity Methods for Linearized Viscous
  Flow}.
\newblock Cambridge University Press, Cambridge, UK, 1992.

\bibitem{batchelor70_2}
G.~K. Batchelor.
\newblock The stress system in a suspension of force-free particles.
\newblock {\em J. Fluid Mech.}, 41:545--570, 1970.

\bibitem{pedley92}
T.~J. Pedley and J.~O. Kessler.
\newblock Hydrodynamic phenomena in suspensions of swimming microorganisms.
\newblock {\em Ann. Rev. Fluid Mech.}, 24:313--358, 1992.

\bibitem{stone96}
H.~A. Stone and A.~D.~T. Samuel.
\newblock Propulsion of microorganisms by surface distortions.
\newblock {\em Phys. Rev. Lett.}, 77:4102--4104, 1996.

\bibitem{becker03}
L.~E. Becker, S.~A. Koehler, and H.~A. Stone.
\newblock On self-propulsion of micro-machines at low {Reynolds} number:
  {Purcell's} three-link swimmer.
\newblock {\em J. Fluid Mech.}, 490:15--35, 2003.

\bibitem{shapere87}
A.~Shapere and F.~Wilczek.
\newblock Self-propulsion at low {R}eynolds number.
\newblock {\em Phys. Rev. Lett.}, 58:2051Ð2054, 1987.

\bibitem{shapere89_1}
A.~Shapere and F.~Wilczek.
\newblock Geometry of self-propulsion at low {R}eynolds number.
\newblock {\em J. Fluid Mech.}, 198:557Ð585, 1989.

\bibitem{shapere89_2}
A.~Shapere and F.~Wilczek.
\newblock Efficiencies of self-propulsion at low {R}eynolds number.
\newblock {\em J. Fluid Mech.}, 198:587Ð599, 1989.

\bibitem{yariv06}
E.~Yariv.
\newblock Self-propulsion in a viscous fluid: {A}rbitrary surface deformations.
\newblock {\em J. Fluid Mech.}, 550:139--148, 2006.

\bibitem{avron08}
J.~E. Avron and O.~Raz.
\newblock A geometric theory of swimming: {Purcell's} swimmer and its
  symmetrized cousin.
\newblock {\em New J. Phys.}, 10:063016, 2008.

\bibitem{najafi04}
A.~Najafi and R.~Golestanian.
\newblock Simple swimmer at low {Reynolds} number: {Three} linked spheres.
\newblock {\em Phys. Rev. E}, 69:062901, 2004.

\bibitem{chwang71}
A.~T. Chwang and T.~Y. Wu.
\newblock Helical movement of micro-organisms.
\newblock {\em Proc. Roy. Soc. Lond. B}, 178:327--346, 1971.

\bibitem{childress04}
S.~Childress and R.~Dudley.
\newblock Transition from ciliary to flapping mode in a swimming mollusc:
  {F}lapping flight as a bifurcation in {$Re_\omega$}.
\newblock {\em J. Fluid Mech.}, 498:257--288, 2004.

\bibitem{childress_conf}
S.~Childress.
\newblock Forward flapping flight as a bifurcation in the frequency {R}eynolds
  number.
\newblock In S.-I.~Sohn J.~R.~Kweon, S.-C.~Kim, editor, {\em {Proceedings of
  the 2004 International Workshop on Mathematical Fluid Dynamics and
  Applications}}, pages 9--21. 2004.

\bibitem{vandenberghe04}
N.~Vandenberghe, J.~Zhang, and S.~Childress.
\newblock Symmetry breaking leads to forward flapping flight.
\newblock {\em J. Fluid Mech.}, 506:147--155, 2004.

\bibitem{alben05}
S.~Alben and M.~Shelley.
\newblock Coherent locomotion as an attracting state for a free flapping body.
\newblock {\em Proc. Natl. Acad. Sci. USA}, 102:11163--11166, 2005.

\bibitem{vandenberghe06}
N.~Vandenberghe, S.~Childress, and J.~Zhang.
\newblock On unidirectional flight of a free flapping wing.
\newblock {\em Phys. Fluids}, 18:014102, 2006.

\bibitem{lu06}
X.~Y. Lu and Q.~Liao.
\newblock Dynamic responses of a two-dimensional flapping foil motion.
\newblock {\em Phys. Fluids}, 18:098104, 2006.

\bibitem{lauga_purcell}
E.~Lauga.
\newblock Continuous breakdown of {P}urcell's scallop theorem with inertia.
\newblock {\em Phys. Fluids}, 19:061703, 2007.

\bibitem{camalet99}
S.~Camalet, F.~Julicher, and J.~Prost.
\newblock Self-organized beating and swimming of internally driven filaments.
\newblock {\em Phys. Rev. Lett.}, 82:1590--1593, 1999.

\bibitem{camalet00:njp}
S.~Camalet and F.~Julicher.
\newblock Generic aspects of axonemal beating.
\newblock {\em New J. Phys.}, 2:1--23, 2000.

\bibitem{FuWolgemuthPowers2008}
H.~C. Fu, C.~W. Wolgemuth, and T.~R. Powers.
\newblock Beating patterns of filaments in viscoelastic fluids.
\newblock {\em Phys. Rev. E}, 78:041913--1--12, 2008.

\bibitem{taylor52}
G.~I. Taylor.
\newblock The action of waving cylindrical tails in propelling microscopic
  organisms.
\newblock {\em Proc. Roy. Soc. A}, 211:225--239, 1952.

\bibitem{Lighthill1952}
M.~J. Lighthill.
\newblock On the squirming motion of nearly spherical deformable bodies through
  liquids at very small {R}eynolds numbers.
\newblock {\em Commun. Pure Appl. Math.}, 5:109--118, 1952.

\bibitem{Blake1971a}
J.~R. Blake.
\newblock A spherical envelope approach to ciliary propulsion.
\newblock {\em J. Fluid Mech.}, 46:199--208, 1971.

\bibitem{Blake1971c}
J.~R. Blake.
\newblock Self propulsion due to oscillations on the surface of a cylinder at
  low {R}eynolds number.
\newblock {\em Bull. Austral. Math. Soc.}, 5:255--264, 1971.

\bibitem{landauFM}
L.~D. Landau and E.~M. Lifshitz.
\newblock {\em Fluid mechanics}.
\newblock Butterworth-Heinemann, Ltd., Oxford, 2nd edition, 1987.

\bibitem{Ehlers_etal1996}
K.~M. Ehlers, A.~D.~T. Samuel, H.~C. Berg, and R.~Montgomery.
\newblock Do cyanobacteria swim using traveling surface waves?
\newblock {\em Proc. Natl. Acad. Sci. USA}, 93:8340--8343, 1996.

\bibitem{Reynolds1965}
A.~J. Reynolds.
\newblock The swimming of minute organisms.
\newblock {\em J. Fluid Mech.}, 23:241--260, 1965.

\bibitem{Tuck1968}
E.~O. Tuck.
\newblock A note on a swimming problem.
\newblock {\em J. Fluid Mech.}, 31:305--308, 1968.

\bibitem{yundt75}
A.~P. Yundt, W.~J. Shack, and T.~J. Lardner.
\newblock Applicability of hydrodynamic analyses of spermatozoan motion.
\newblock {\em J. Exp. Biol.}, 62:27--41, 1975.

\bibitem{lauga07_pre}
E.~Lauga.
\newblock Floppy swimming: Viscous locomotion of actuated elastica.
\newblock {\em Phys. Rev. E}, 75:041916, 2007.

\bibitem{Magariyama_etal1995}
Y.~Magariyama, S.~Sugiyama, K.~Muramoto, I.~Kawagishi, Y.~Imae, and S.~Kudo.
\newblock Simultaneous measurement of bacterial flagellar rotation rate and
  swimming speed.
\newblock {\em Biophys. J.}, 69:2154--2162, 1995.

\bibitem{LiTang2006}
G.~Li and J.~X. Tang.
\newblock Low flagellar motor torque and high swimming efficiency of
  \textit{{C}aulobacter crescentus} swarmer cells.
\newblock {\em Biophys. J.}, 91:2726--2734, 2006.

\bibitem{johnson79}
R.~E. Johnson and C.~J. Brokaw.
\newblock Flagellar hydrodynamics - {C}omparison between resistive- force
  theory and slender-body theory.
\newblock {\em Biophys. J.}, 25:113 --127, 1979.

\bibitem{cox70}
R.~G. Cox.
\newblock The motion of long slender bodies in a viscous fluid. {P}art 1.
  {G}eneral theory.
\newblock {\em J. Fluid Mech.}, 44:791--810, 1970.

\bibitem{batchelor70}
G.K. Batchelor.
\newblock Slender body theory for particles of arbitrary cross section in
  {Stokes} flow.
\newblock {\em J. Fluid Mech.}, 44:419 -- 440, 1970.

\bibitem{tillett70}
J.P.K. Tillett.
\newblock Axial and transverse {Stokes} flow past slender axisymmetric bodies.
\newblock {\em J. Fluid Mech.}, 44:401 -- 417, 1970.

\bibitem{keller76-jfm}
J.~B. Keller and S.~I. Rubinow.
\newblock Slender body theory for slow viscous flow.
\newblock {\em J. Fluid Mech.}, 75:705--714, 1976.

\bibitem{geer76}
J.~Geer.
\newblock Stokes flow past a slender body of revolution.
\newblock {\em J. Fluid Mech.}, 78:577 -- 600, 1976.

\bibitem{johnson80}
R.~E. Johnson.
\newblock An improved slender body theory for {S}tokes flow.
\newblock {\em J. Fluid Mech.}, 99:411--431, 1980.

\bibitem{lighthill96_theorem}
J.~Lighthill.
\newblock Reinterpreting the basic theorem of flagellar hydrodynamics.
\newblock {\em J. Eng. Math.}, 30:25--34, 1996.

\bibitem{gueron92}
S.~Gueron and N.~Liron.
\newblock Ciliary motion modeling, and dynamic multicilia interactions.
\newblock {\em Biophys. J.}, 63:1045--1058, 1992.

\bibitem{gueron93}
S.~Gueron and N.~Liron.
\newblock Simulations of three-dimensional ciliary beats and cilia
  interactions.
\newblock {\em Biophys. J.}, 65:499--507, 1993.

\bibitem{gueron97}
S.~Gueron, K.~Levit-Gurevich, N.~Liron, and J.~J. Blum.
\newblock Cilia internal mechanism and metachronal coordination as the result
  of hydrodynamical coupling.
\newblock {\em Proc. Natl. Acad. Sci. USA}, 94:6001--6006, 1997.

\bibitem{gueron99}
S.~Gueron and K.~Levit-Gurevich.
\newblock Energetic considerations of ciliary beating and the advantage of
  metachronal coordination.
\newblock {\em Proc. Natl. Acad. Sci. USA}, 96:12240--12245, 1999.

\bibitem{cortez01}
R.~Cortez.
\newblock The method of regularized stokeslets.
\newblock {\em SIAM J. Sci. Comput.}, 23:1204--1225, 2001.

\bibitem{cortezPC}
R.~Cortez~(Personal communication).

\bibitem{macnab84}
R.~M. Macnab and S.~I. Aizawa.
\newblock Bacterial motility and the bacterial flagellar motor.
\newblock {\em Ann. Rev. Biophys. Bioeng.}, 13:51--83, 1984.

\bibitem{schuster94}
S.~C. Schuster and S.~Khan.
\newblock The bacterial flagellar motor.
\newblock {\em Ann. Rev. Biophys. Biomol. Struct.}, 23:509--539, 1994.

\bibitem{MatsuuraKamiyaAsakura1978}
S.~Matsuura, R.~Kamiya, and S.~Asakura.
\newblock Transformation of straight flagella and recovery of motility in a
  mutant \textit{{E}scherichia coli}.
\newblock {\em J. Mol. Bio.}, 118:431--440, 1978.

\bibitem{wolgemuth00}
C.~W. Wolgemuth, T.~R. Powers, and R.~E. Goldstein.
\newblock Twirling and whirling: {Viscous} dynamics of rotating elastic
  filaments.
\newblock {\em Phys. Rev. Lett.}, 84:1623--1626, 2000.

\bibitem{lim04}
S.~Lim and C.~S. Peskin.
\newblock Simulations of the whirling instability by the immersed boundary
  method.
\newblock {\em Siam J. On Scientific Computing}, 25:2066--2083, 2004.

\bibitem{wada06}
H.~Wada and R.~R. Netz.
\newblock Non-equilibrium hydrodynamics of a rotating filament.
\newblock {\em Europhys. Lett.}, 75:645--651, 2006.

\bibitem{kim_powers2005}
M.~J. Kim and T.~R. Powers.
\newblock Deformation of a helical filament by flow and electric or magnetic
  fields.
\newblock {\em Phys. Rev.}, E \textbf{71}:021914--1--10, 2005.

\bibitem{Bergetal2004private}
H.C. Berg, N.~Darnton, S.~Rojevskaya, and L.~Turner.
\newblock private communication.

\bibitem{takano_et_al2003}
Y.~Takano, K.~Yoshida, S.~Kudo, M.~Nishitoba, and Y.~Magariyama.
\newblock Analysis of small deformation of helical flagellum of swimming
  \textit{{V}ibrio alginolyticus}.
\newblock {\em JSME Int. J.}, C {\bf 46}:1241--1247, 2003.

\bibitem{macnab77_curly}
R.~M. Macnab and M.~K. Ornston.
\newblock Normal to curly flagellar transitions and their role in bacterial
  tumbling---{S}tabilization of an alternative quaternary structure by
  mechanical force.
\newblock {\em J. Mol. Biol.}, 112:1--30, 1977.

\bibitem{hotani1982}
H.~Hotani.
\newblock Micro-video study of moving bacterial flagellar filaments iii. cyclic
  transformation induced by mechanical force.
\newblock {\em J. Mol. Biol.}, {\bf 156}:791, 1982.

\bibitem{machin58}
K.~E. Machin.
\newblock Wave propagation along flagella.
\newblock {\em J. Exp. Biol}, 35:796--806, 1958.

\bibitem{machin63}
K.~E. Machin.
\newblock The control and synchronization of flagellar movement.
\newblock {\em Proc. Roy. Soc. B}, 158:88--104, 1963.

\bibitem{WigginsGoldstein}
C.~H. Wiggins and R.~E. Goldstein.
\newblock Flexive and propulsive dynamics of elastica at low {Reynolds} number.
\newblock {\em Phys. Rev. Lett.}, 80:3879 -- 3882, 1998.

\bibitem{lowe03}
C.~P. Lowe.
\newblock Dynamics of filaments: {M}odelling the dynamics of driven
  microfilaments.
\newblock {\em Philos. Trans. R. Soc. London, Ser. B}, 358:1543--1550, 2003.

\bibitem{lagomarsino03_filament}
M.~C. Lagomarsino, F.~Capuani, and C.~P. Lowe.
\newblock A simulation study of the dynamics of a driven filament in an
  {Aristotelian} fluid.
\newblock {\em J. Theor. Biol.}, 224:215--224, 2003.

\bibitem{yu06}
T.~S. Yu, E.~Lauga, and A.~E. Hosoi.
\newblock Experimental investigations of elastic tail propulsion at low
  {Reynolds} number.
\newblock {\em Phys. Fluids}, 18:091701, 2006.

\bibitem{kosa07}
G.~Kosa, M.~Shoham, and M.~Zaaroor.
\newblock Propulsion method for swimming microrobots.
\newblock {\em {IEEE} Transactions On Robotics}, 23:137--150, 2007.

\bibitem{riveline97}
D.~Riveline, C.~H. Wiggins, R.~E. Goldstein, and A.~Ott.
\newblock Elastohydrodynamic study of actin filaments using fluorescence
  microscopy.
\newblock {\em Phys. Rev. E}, 56:R1330--R1333, 1997.

\bibitem{Wiggins:Biophys}
C.~H. Wiggins, D.~Riveline, A.~Ott, and R.~E. Goldstein.
\newblock Trapping and wiggling: Elastohydrodynamics of driven microfilaments.
\newblock {\em Biophys. J.}, 74:1043--1060, 1998.

\bibitem{landau_lifshitz_elas}
L.~D. Landau and E.~M. Lifshitz.
\newblock {\em Theory of elasticity}.
\newblock Pergamon Press, Oxford, 3rd edition, 1986.

\bibitem{trouilloud08}
R.~Trouilloud, T.~S. Yu, A.~E. Hosoi, and E.~Lauga.
\newblock Soft swimming: Exploiting deformable interfaces for low {R}eynolds
  number locomotion.
\newblock {\em Phys. Rev. Lett.}, 101, 2008.

\bibitem{klapper96}
I.~Klapper.
\newblock Biological applications of the dynamics of twisted elastic rods.
\newblock {\em J. Comp Phys.}, 125:325--337, 1996.

\bibitem{goldstein98}
R.~E. Goldstein, T.~R. Powers, and C.~H. Wiggins.
\newblock Viscous nonlinear dynamics of twist and writhe.
\newblock {\em Phys. Rev. Lett.}, 80:5232--5235, 1998.

\bibitem{manghi06}
M.~Manghi, X.~Schlagberger, and R.~R. Netz.
\newblock Propulsion with a rotating elastic nanorod.
\newblock {\em Phys. Rev. Lett.}, 96:068101, 2006.

\bibitem{qian08}
B.~Qian, T.~R. Powers, and K.~S. Breuer.
\newblock Shape transition and propulsive force of an elastic rod rotating in a
  viscous fluid.
\newblock {\em Phys. Rev. Lett.}, 1:078101, 2008.

\bibitem{coq08}
N.~Coq, O.~du~Roure, J.~Marthelot, D.~Bartolo, and M.~Fermigier.
\newblock Rotational dynamics of a soft filament: {W}rapping transition and
  propulsive forces.
\newblock {\em Phys. Fluids.}, 20:051703, 2008.

\bibitem{Nonaka_etal1998}
S.~Nonaka, Y.~Tanaka, Y.~Okada, S.~Takeda, A.~Harada, Y.~Kanai, M.~Kido, and
  N.~Hirokawa.
\newblock Randomization of left-right asymmetry due to loss of nodal cilia
  generating leftward flow of extraembryonic fluid in mice lacking {KIF3B}
  motor protein.
\newblock {\em Cell}, 95:829--837, 1998.

\bibitem{HilfingerJulicher2008}
A.~Hilfinger and F.~J\"ulicher.
\newblock The chirality of ciliary beats.
\newblock {\em Phys. Biol.}, 5:016003+12, 2008.

\bibitem{satir68}
P.~Satir.
\newblock Studies on cilia. {III}. {F}urther studies on the cilium tip and a
  ``sliding filament'' model of ciliary motility.
\newblock {\em J. Cell Biol.}, 39:77--94, 1968.

\bibitem{summers71}
K.~E. Summers and I.~R. Gibbons.
\newblock Adenosine triphosphate-induced sliding of tubules in trypsin-treated
  flagella of sea-urchin sperm.
\newblock {\em Proc. Natl. Acad. Sci. USA}, 68:3092--3096, 1971.

\bibitem{brokaw72}
C.~J. Brokaw.
\newblock Flagellar movement: {A} sliding filament model.
\newblock {\em Science}, 178:455--462, 1972.

\bibitem{brokaw89}
C.~J. Brokaw.
\newblock Direct measurements of sliding between outer doublet microtubules in
  swimming sperm flagella.
\newblock {\em Science}, 243:1593--1596, 1989.

\bibitem{Riedel-Kruse_etal2007}
I~H. Riedel-Kruse, A.~Hilfinger, J.~Howard, and F.~J\"ulicher.
\newblock How molecular motors shape the flagellar beat.
\newblock {\em HFSP J.}, 1:192--208, 2007.

\bibitem{FuPowersWolgemuth2007}
H.~C. Fu, T.~R. Powers, and H.~C. Wolgemuth.
\newblock Theory of swimming filaments in viscoelastic media.
\newblock {\em Phys. Rev. Lett.}, 99:258101--258105, 2007.

\bibitem{GibbonsGibbons1972}
B.~H. Gibbons and I.~R. Gibbons.
\newblock Flagellar movement and adenosine triphosphatase activity in sea
  urchin sperm extracted with triton {X}-100.
\newblock {\em J. Cell Biol.}, 54:75--97, 1972.

\bibitem{Brokaw1975}
C.~J. Brokaw.
\newblock Molecular mechanism for oscillation in flagella and muscle.
\newblock {\em Proc. Natl. Acad. Sci. U.S.A.}, 72:3102--3106, 1975.

\bibitem{JulicherProst1997}
F.~J\"ulicher and J.~Prost.
\newblock Spontaneous oscillations of collective molecular motors.
\newblock {\em Phys. Rev. Lett.}, 78:4510--4513, 1997.

\bibitem{MuraseHinesBlum1989}
M.~Murase, M.~Hines, and J.~J. Blum.
\newblock Properties of an excitable dynein model for bend propagation in cilia
  and flagella.
\newblock {\em J. Theor. Biol.}, 139:413--430, 1989.

\bibitem{Lindemann1994a}
C.~Lindemann.
\newblock A geometric clutch hypothesis to explain oscillations of the axoneme
  of cilia and flagella.
\newblock {\em J. Theor. Biol.}, 168:175--189, 1994.

\bibitem{Lindemann1994b}
C.~Lindemann.
\newblock A model of flagellar and ciliary functioning which uses the forces
  transverse to the axoneme as the regulator of dynein activation.
\newblock {\em Cell Motil. Cytoskeleton}, 29:141--154, 1994.

\bibitem{Lindemann2002}
C.~Lindemann.
\newblock Geometric clutch model version 3: {T}he role of the inner and outer
  arm dyneins in the ciliary beat.
\newblock {\em Cell Motil. Cytoskeleton}, 52:242--254, 2002.

\bibitem{Brokaw2002}
C.~Brokaw.
\newblock Computer simulation of flagellar movement viii. {C}oordination of
  dynein by local curvature control can generate helical bending waves.
\newblock {\em Cell Motil. Cytoskeleton}, 53:103--124, 2002.

\bibitem{wada07}
H.~Wada and R.~R. Netz.
\newblock Model for self-propulsive helical filaments: {Kink-pair} propagation.
\newblock {\em Phys. Rev. Lett.}, 99(10):108102, 2007.

\bibitem{Charon_etal1984}
N.~W. Charon, G.~R. Daughtry, R.~S. Mc{C}uskey, and G.~N. Franz.
\newblock Microcinematographic analysis of tethered \textit{{L}eptospira
  illini}.
\newblock {\em J. Bacteriol.}, 160:1067--1073, 1984.

\bibitem{GoldsteinCharon1990}
S.~Goldstein and N.~Charon.
\newblock Multiple-exposure photographic analysis of a motile spirochete.
\newblock {\em Proc. Natl. Acad. Sci. U.S.A.}, 87:4895--4899, 1990.

\bibitem{JungMareckFauciShelley2007}
S.~Jung, K.~Mareck, L.~Fauci, and M.~J. Shelley.
\newblock Rotational dynamics of a superhelix towed in a stokes fluid.
\newblock {\em Phys. Fluids}, 19:103105--(1--6), 2007.

\bibitem{mendelson99}
N.~H. Mendelson, A.~Bourque, K.~Wilkening, K.~R. Anderson, and J.~C. Watkins.
\newblock Organized cell swimming motions in \textit{{Bacillus} subtilis}
  colonies: {Patterns} of short-lived whirls and jets.
\newblock {\em J. Bacteriol.}, 181:600--609, 1999.

\bibitem{wu00}
X.~L. Wu and A.~Libchaber.
\newblock Particle diffusion in a quasi-two-dimensional bacterial bath.
\newblock {\em Phys. Rev. Lett.}, 84:3017--3020, 2000.

\bibitem{dombrowski04}
C.~Dombrowski, L.~Cisneros, S.~Chatkaew, R.~E. Goldstein, and J.~O. Kessler.
\newblock Self-concentration and large-scale coherence in bacterial dynamics.
\newblock {\em Phys. Rev. Lett.}, 93:098103, 2004.

\bibitem{tuval05}
I.~Tuval, L.~Cisneros, C.~Dombrowski, C.~W. Wolgemuth, J.~Kessler, and R.~E.
  Goldstein.
\newblock Bacterial swimming and oxygen transport near contact lines.
\newblock {\em Proc. Natl. Acad. Sci. USA}, 102:2277--2282, 2005.

\bibitem{cisneros07}
L.~H. Cisneros, R.~Cortez, C.~Dombrowski, R.~E. Goldstein, and J.~O. Kessler.
\newblock Fluid dynamics of self-propelled micro-organisms, from individuals to
  concentrated populations.
\newblock {\em Exp. Fluids}, 2007.

\bibitem{sokolov07}
A.~Sokolov, I.~S. Aranson, J.~O. Kessler, and R.~E. Goldstein.
\newblock Concentration dependence of the collective dynamics of swimming
  bacteria.
\newblock {\em Phys. Rev. Lett.}, 98:158102, 2007.

\bibitem{Wolgemuth2008}
C.~W. Wolgemuth.
\newblock Collective swimming and the dynamics of bacterial turbulence.
\newblock {\em Biophys. J.}, 95:1564--1574, 2008.

\bibitem{moore02}
Harry Moore, Katerina Dvor‡kov‡, Nicholas Jenkins, and William Breed.
\newblock Exceptional sperm cooperation in the wood mouse.
\newblock {\em Nature}, 418:174--177, 2002.

\bibitem{moore95}
H.~D.~M. Moore and D.~A. Taggart.
\newblock Sperm pairing in the opossum increases the efficiency of sperm
  movement in a viscous environment.
\newblock {\em Biol. Reprod.}, 52:947--953, 1995.

\bibitem{hayashi98}
F.~Hayashi.
\newblock Sperm co-operation in the {Fishfly,} {Parachauliodes} japonicus.
\newblock {\em Funct. Ecol.}, 12:347--350, 1998.

\bibitem{riedel05}
I.~H. Riedel, K.~Kruse, and J.~Howard.
\newblock A self-organized vortex array of hydrodynamically entrained sperm
  cells.
\newblock {\em Science}, 309:300--303, 2005.

\bibitem{dreyfus05}
R.~Dreyfus, J.~Baudry, and H.~A. Stone.
\newblock Purcell's ``rotator": mechanical rotation at low {Reynolds} number.
\newblock {\em Europ. Phys. J. B}, 47:161--164, 2005.

\bibitem{guell88}
D.~C. Guell, H.~Brenner, R.~B. Frankel, and H.~Hartman.
\newblock Hydrodynamic forces and band formation in swimming magnetotactic
  bacteria.
\newblock {\em J. Theor. Biol.}, 135:525--542, 1988.

\bibitem{pooley07_2}
C.~M. Pooley, G.~P. Alexander, and J.~M. Yeomans.
\newblock Hydrodynamic interaction between two swimmers at low {Reynolds}
  number.
\newblock {\em Phys. Rev. Lett.}, 99(22):228103, 2007.

\bibitem{liao07}
Q.~Liao, G.~Subramanian, M.~P. DeLisa, D.~L. Koch, and M.~M. Wu.
\newblock Pair velocity correlations among swimming {Escherichia} coli bacteria
  are determined by force-quadrupole hydrodynamic interactions.
\newblock {\em Phys. Fluids}, 19:061701, 2007.

\bibitem{ishikawa06}
T.~Ishikawa, M.~P. Simmonds, and T.~J. Pedley.
\newblock Hydrodynamic interaction of two swimming model micro-organisms.
\newblock {\em J. Fluid Mech.}, 568:119--160, 2006.

\bibitem{ishikawa_paramecia06}
T.~Ishikawa and M.~Hota.
\newblock Interaction of two swimming \textit{{P}aramecia}.
\newblock {\em J. Exp. Biol.}, 209:4452--4463, 2006.

\bibitem{ishikawa07_bacteria}
T.~Ishikawa, G.~Sekiya, Y.~Imai, and T.~Yamaguchi.
\newblock Hydrodynamic interaction between two swimming bacteria.
\newblock {\em Biophys. J.}, 93:2217--2225, 2007.

\bibitem{simha02}
R.~A. Simha and S.~Ramaswamy.
\newblock Hydrodynamic fluctuations and instabilities in ordered suspensions of
  self-propelled particles.
\newblock {\em Phys. Rev. Lett.}, 89:058101, 2002.

\bibitem{saintillan07}
D.~Saintillan and M.~J. Shelley.
\newblock Orientational order and instabilities in suspensions of
  self-locomoting rods.
\newblock {\em Phys. Rev. Lett.}, 99:058102, 2007.

\bibitem{saintillan08}
D.~Saintillan and M.~J. Shelley.
\newblock Instabilities and pattern formation in active particle suspensions:
  {Kinetic} theory and continuum simulations.
\newblock {\em Phys. Rev. Lett.}, 100:178103, 2008.

\bibitem{ishikawa08}
T.~Ishikawa and T.~J. Pedley.
\newblock Coherent structures in monolayers of swimming particles.
\newblock {\em Phys. Rev. Lett.}, 1:088103, 2008.

\bibitem{hernandez-ortiz05}
J.~P. Hernandez-Ortiz, C.~G. Stoltz, and M.~D. Graham.
\newblock Transport and collective dynamics in suspensions of confined swimming
  particles.
\newblock {\em Phys. Rev. Lett.}, 95:204501, 2005.

\bibitem{ishikawa_pedley_diffusion07}
T.~Ishikawa and T.~J. Pedley.
\newblock Diffusion of swimming model micro-organisms in a semi-dilute
  suspension.
\newblock {\em J. Fluid Mech.}, 588:437--462, 2007.

\bibitem{mehandia08}
V.~Mehandia and P.~R. Nott.
\newblock The collective dynamics of self-propelled particles.
\newblock {\em J. Fluid Mechanics}, 595:239--264, 2008.

\bibitem{underhill08}
P.~T. Underhill, J.~P. Hernandez-Ortiz, and M.~D. Graham.
\newblock Diffusion and spatial correlations in suspensions of swimming
  particles.
\newblock {\em Phys. Rev. Lett.}, 100(24):248101, 2008.

\bibitem{hatwalne04}
Y.~Hatwalne, S.~Ramaswamy, M.~Rao, and R.~A. Simha.
\newblock Rheology of active-particle suspensions.
\newblock {\em Phys. Rev. Lett.}, 92:118101, 2004.

\bibitem{ishikawa_pedley_rheology07}
T.~Ishikawa and T.~J. Pedley.
\newblock The rheology of a semi-dilute suspension of swimming model
  micro-organisms.
\newblock {\em J. Fluid Mech.}, 588:399--435, 2007.

\bibitem{berke08}
A.~P. Berke, L.~Turner, H.~C. Berg, and E.~Lauga.
\newblock Hydrodynamic attraction of swimming microorganisms by surfaces.
\newblock {\em Phys. Rev. Lett.}, 101:038102, 2008.

\bibitem{oldroyd1950}
J.~G. Oldroyd.
\newblock On the formulation of rheological equations of state.
\newblock {\em Proc. Roy. Soc. A}, 200:523--541, 1950.

\bibitem{bird76}
R.~B. Bird.
\newblock Useful non-{N}ewtonian models.
\newblock {\em Ann. Rev. Fluid Mech.}, 8:13--34, 1976.

\bibitem{birdvol1}
R.~B. Bird, R.~C. Armstrong, and O.~Hassager.
\newblock {\em Dynamics of Polymeric Liquids. Second Edition. Vol. 1: {F}luid
  Mechanics.}
\newblock Wiley-Interscience, New York, NY, 1987.

\bibitem{birdvol2}
R.~B. Bird, C.~F. Curtiss, R.~C. Armstrong, and O.~Hassager.
\newblock {\em Dynamics of Polymeric Liquids. Second Edition. Vol. 2: {K}inetic
  Theory.}
\newblock Wiley-Interscience, New York, NY, 1987.

\bibitem{tanner88}
R.~I. Tanner.
\newblock {\em Engineering Rheology, Second Edition}.
\newblock Clarendon Press, Oxford, U.K., 1988.

\bibitem{larson88}
R.~G. Larson.
\newblock {\em Constitutive Equations for Polymer Melts and Solutions}.
\newblock Butterworth-Heinemann, Boston, MA, 1988.

\bibitem{doi88}
M.~Doi and S.~F. Edwards.
\newblock {\em The Theory of Polymer Dynamics}.
\newblock Oxford University Press, Oxford, U.K., 1988.

\bibitem{bird95}
R.~B. Bird and J.~M. Wiest.
\newblock Constitutive equations for polymeric liquids.
\newblock {\em Ann. Rev. Fluid Mech.}, 27:169--193, 1995.

\bibitem{larson99}
R.~G. Larson.
\newblock {\em The Structure and Rheology of Complex Fluids}.
\newblock Oxford University Press, Oxford, U.K., 1999.

\bibitem{fletcher82}
M.~Fletcher and K.~C. Marshall.
\newblock Are solid surfaces of ecological significance to aquatic bacteria?
\newblock {\em Adv. Microb. Ecol.}, 6:199--230, 1982.

\bibitem{cooper88}
G.~L. Cooper, A.~L. Schiller, and C.~C. Hopkins.
\newblock Possible role of capillary action in pathogenesis of experimental
  catheter-associated dermal tunnel infections.
\newblock {\em J. Clin. Microbiol.}, 26:8--12, 1988.

\bibitem{harkes92b}
G.~Harkes, J.~Dankert, and J.~Feijen.
\newblock Bacterial migration along solid surfaces.
\newblock {\em Appl. Environ. Microbiol.}, 58:1500--1505, 1992.

\bibitem{costerton95}
J.~W. Costerton, Z.~Lewandowski, D.~E. Caldwell, D.~R. Korber, and H.~M.
  Lappinscott.
\newblock Microbial biofilms.
\newblock {\em Ann. Rev. Microbiol.}, 49:711--745, 1995.

\bibitem{vanloosdrecht90}
M.~C.~M. Vanloosdrecht, J.~Lyklema, W.~Norde, and A.~J.~B. Zehnder.
\newblock Influence of interfaces on microbial activity.
\newblock {\em Microbiol. Rev.}, 54:75--87, 1990.

\bibitem{harshey03}
R.~M. Harshey.
\newblock Bacterial motility on a surface: {Many} ways to a common goal.
\newblock {\em Ann. Rev. Microbiol.}, 57:249--273, 2003.

\bibitem{katz74}
D.~F. Katz.
\newblock Propulsion of microorganisms near solid boundaries.
\newblock {\em J. Fluid Mech.}, 64:33--49, 1974.

\bibitem{katz75}
D.~F. Katz, J.~R. Blake, and S.~L. Paverifontana.
\newblock Movement of slender bodies near plane boundaries at low {R}eynolds
  number.
\newblock {\em J. Fluid Mech.}, 72:529--540, 1975.

\bibitem{katzblake}
D.F. Katz and J.R. Blake.
\newblock Flagellar motions near walls.
\newblock In T.Y. Wu, C.J. Brokaw, and C~Brennen, editors, {\em Swimming and
  {F}lying in {N}ature}, volume~1, pages 173--184. Plenum, New-York, 1975.

\bibitem{fauci95}
L.~J. Fauci and A.~Mcdonald.
\newblock Sperm motility in the presence of boundaries.
\newblock {\em Bull. Math. Biol.}, 57:679--699, 1995.

\bibitem{ramia93}
M.~Ramia, D.~L. Tullock, and N.~Phan-Thien.
\newblock The role of hydrodynamic interaction in the locomotion of
  microorganisms.
\newblock {\em Biophys. J.}, 65:755--778, 1993.

\bibitem{frymier95}
P.~D. Frymier, R.~M. Ford, H.~C. Berg, and P.~T. Cummings.
\newblock Three-dimensional tracking of motile bacteria near a solid planar
  surface.
\newblock {\em Proc. Natl. Acad. Sci. USA}, 92:6195--6199, 1995.

\bibitem{frymier97}
P.~D. Frymier and R.~M. Ford.
\newblock Analysis of bacterial swimming speed approaching a solid-liquid
  interface.
\newblock {\em {AIChE} J.}, 43:1341--1347, 1997.

\bibitem{vigeant97}
M.~A.~S. Vigeant and R.~M. Ford.
\newblock Interactions between motile {{\it Escherichia coli}} and glass in
  media with various ionic strengths, as observed with a three-dimensional
  tracking microscope.
\newblock {\em Appl. Environ. Microbiol.}, 63:3474--3479, 1997.

\bibitem{diluzio05}
W.~R. DiLuzio, L.~Turner, M.~Mayer, P.~Garstecki, D.~B. Weibel, H.~C. Berg, and
  G.~M. Whitesides.
\newblock Escherichia coli swim on the right-hand side.
\newblock {\em Nature}, 435:1271--1274, 2005.

\bibitem{lauga06}
E.~Lauga, W.~R. DiLuzio, G.~M. Whitesides, and H.~A. Stone.
\newblock Swimming in circles: {Motion} of bacteria near solid boundaries.
\newblock {\em Biophys. J.}, 90:400--412, 2006.

\bibitem{blake71_image}
J.~R. Blake.
\newblock A note on the image system for a {Stokeslet} in a no-slip boundary.
\newblock {\em Proc. Camb. Phil. Soc.}, 70:303--310, 1971.

\bibitem{blake74_image}
J.~R. Blake and A.~T. Chwang.
\newblock Fundamental singularities of viscous-flow. {P}art 1. {I}mage systems
  in vicinity of a stationary no-slip boundary.
\newblock {\em J. Eng. Math.}, 8:23--29, 1974.

\bibitem{rothschild63}
L.~Rothschild.
\newblock Non-random distribution of bull spermatozoa in a drop of sperm
  suspension.
\newblock {\em Nature}, 198:1221--1222, 1963.

\bibitem{winet84}
H.~Winet, G.~S. Bernstein, and J.~Head.
\newblock Spermatozoon tendency to accumulate at walls is strongest mechanical
  response.
\newblock {\em J. Androl.}, 5:19, 1984.

\bibitem{winet84a}
H.~Winet, G.~S. Bernstein, and J.~Head.
\newblock Observations on the response of human spermatozoa to gravity,
  boundaries and fluid shear.
\newblock {\em J. Reprod. Fert.}, 70:511--523, 1984.

\bibitem{cosson03}
J.~Cosson, P.~Huitorel, and C.~Gagnon.
\newblock How spermatozoa come to be confined to surfaces.
\newblock {\em Cell Motil. Cytoskel.}, 54:56--63, 2003.

\bibitem{woolley03}
D.~M. Woolley.
\newblock Motility of spermatozoa at surfaces.
\newblock {\em Reproduction}, 126:259--270, 2003.

\bibitem{fauci90}
L.~J. Fauci.
\newblock Interaction of oscillating filaments---{A} computational study.
\newblock {\em J. Comp. Phys.}, 86:294--313, 1990.

\bibitem{strogatz00}
S.~Strogatz.
\newblock From {K}uramoto to {C}rawford: {E}xploring the onset of
  synchronization in populations of coupled oscillators.
\newblock {\em Physica D}, 143:1--20, 2000.

\bibitem{blake74}
J.~R. Blake and M.~A. Sleigh.
\newblock Mechanics of ciliary locomotion.
\newblock {\em Biol. Rev. Camb. Phil. Soc.}, 49:85--125, 1974.

\bibitem{salathe07}
M.~Salathe.
\newblock Regulation of mammalian ciliary beating.
\newblock {\em Ann. Rev. Physiol.}, 69:401--422, 2007.

\bibitem{smith08}
D.~J. Smith, J.~R. Blake, and E.~A. Gaffney.
\newblock Fluid mechanics of nodal flow due to embryonic primary cilia.
\newblock {\em J. Roy. Soc. Interface}, 5:567--573, 2008.

\bibitem{lenz06}
P.~Lenz and A.~Ryskin.
\newblock Collective effects in ciliar arrays.
\newblock {\em Phys. Biol.}, 3:285--294, 2006.

\bibitem{vilfan06}
A.~Vilfan and F.~Julicher.
\newblock Hydrodynamic flow patterns and synchronization of beating cilia.
\newblock {\em Phys. Rev. Lett.}, 96:058102, 2006.

\bibitem{guirao07}
B.~Guirao and J.~F. Joanny.
\newblock Spontaneous creation of macroscopic flow and metachronal waves in an
  array of cilia.
\newblock {\em Biophys. J.}, 92:1900--1917, 2007.

\bibitem{niedermayer08}
T.~Niedermayer, B.~Eckhardt, and P.~Lenz.
\newblock Synchronization, phase locking, and metachronal wave formation in
  ciliary chains.
\newblock {\em Chaos}, 18:037128, 2008.

\bibitem{anderson75}
R.~A. Anderson.
\newblock Formation of the bacterial flagellar bundle.
\newblock In T.~Y.-T. Wu, C.~J. Brokaw, and C.~Brenner, editors, {\em Swimming
  and Flying in Nature, Vol. 1.} Plenum Press, New York, NY, 1975.

\bibitem{macnab77}
R.~M. Macnab.
\newblock Bacterial flagella rotating in bundles---{S}tudy in helical geometry.
\newblock {\em Proc. Natl. Acad. Sci. USA}, 74:221--225, 1977.

\bibitem{powers02}
T.~R. Powers.
\newblock Role of body rotation in bacterial flagellar bundling.
\newblock {\em Phys. Rev. E}, 65:040903, 2002.

\bibitem{macnab1977}
R.M. Macnab.
\newblock Bacterial flagella rotating in bundles: a study in helical geometry.
\newblock {\em Proc. Natl. Acad. Sci. USA}, {\bf 74}:221, 1977.

\bibitem{kim03}
M.~Kim, J.~C. Bird, A.~J.~Van Parys, K.~S. Breuer, and T.~R. Powers.
\newblock A macroscopic scale model of bacterial flagellar bundling.
\newblock {\em Proc. Natl. Acad. Sci. USA}, 100:15481--15485, 2003.

\bibitem{kim04_PIV}
M.~J. Kim, M.~M.~J. Kim, J.~C. Bird, J.~Park, T.~R. Powers, and K.~S. Breuer.
\newblock Particle image velocimetry experiments on a macro-scale model for
  bacterial flagellar bundling.
\newblock {\em Exp. Fluids}, 37(6):782--788, 2004.

\bibitem{flores05}
H.~Flores, E.~Lobaton, S.~Mendez-Diez, S.~Tlupova, and R.~Cortez.
\newblock A study of bacterial flagellar bundling.
\newblock {\em Bulletin Math. Biol.}, 67:137--168, 2005.

\bibitem{kim04}
M.~Kim and T.~R. Powers.
\newblock Hydrodynamic interactions between rotating helices.
\newblock {\em Phys. Rev. E}, 69:061910, 2004.

\bibitem{reiehert05}
M.~Reiehert and H.~Stark.
\newblock Synchronization of rotating helices by hydrodynamic interactions.
\newblock {\em Europ. Phys. J. E}, 17:493--500, 2005.

\bibitem{block89}
S.~M. Block, D.~F. Blair, and H.~C. Berg.
\newblock Compliance of bacterial flagella measured with optical tweezers.
\newblock {\em Nature}, 338:514--518, 1989.

\bibitem{block91}
S.~M. Block, D.~F. Blair, and H.~C. Berg.
\newblock Compliance of bacterial polyhooks measured with optical tweezers.
\newblock {\em Cytometry}, 12:492--496, 1991.

\bibitem{Bigelow_etal2004}
J.~L. Bigelow, D.~B. Dunson, J.~B. Stanford, R.~Ecochard, C.~Gnoth, and
  B.~Colombo.
\newblock Mucus observations in the fertile window: a better predictor of
  conception then timing of intercourse.
\newblock {\em Hum. Reprod.}, 19:889--892, 2004.

\bibitem{wolf77_1}
D.~P. Wolf, L.~Blasco, M.~A. Khan, and M.~Litt.
\newblock Human cervical mucus. {I}. {R}heologic characteristics.
\newblock {\em Fertility Sterility}, 28:41--46, 1977.

\bibitem{wolf77_2}
D.~P. Wolf, L.~Blasco, M.~A. Khan, and M.~Litt.
\newblock Human cervical mucus. {II}. {C}hanges in viscoelasticity during
  ovulatory menstrual cycle.
\newblock {\em Fertility Sterility}, 28:47--52, 1977.

\bibitem{wolf77_3}
D.~P. Wolf, J.~Sokoloski, M.~A. Khan, and M.~Litt.
\newblock Human cervical mucus. {III}. {I}solation and characterization of
  rheologically active mucin.
\newblock {\em Fertility Sterility}, 28:53--58, 1977.

\bibitem{wolf78}
D.~P. Wolf, L.~Blasco, M.~A. Khan, and M.~Litt.
\newblock Human cervical mucus. {IV}. {V}iscoelasticity and sperm penetrability
  during ovulatory menstrual-cycle.
\newblock {\em Fertility Sterility}, 30:163--169, 1978.

\bibitem{katz78}
D.~F. Katz, R.~N. Mills, and T.~R. Pritchett.
\newblock Movement of human spermatozoa in cervical mucus.
\newblock {\em J. Reprod. Fert.}, 53:259--265, 1978.

\bibitem{dunn76}
P.~F. Dunn and B.~F. Picologlou.
\newblock Viscoelastic properties of cumulus oophorus.
\newblock {\em Biorheol.}, 13:379--384, 1976.

\bibitem{suarez91}
S.~S. Suarez, D.~F. Katz, D.~H. Owen, J.~B. Andrew, and R.~L. Powell.
\newblock Evidence for the function of hyperactivated motility in sperm.
\newblock {\em Biol. Reprod.}, 44:375--381, 1991.

\bibitem{suarez92}
S.~S. Suarez and X.~B. Dai.
\newblock Hyperactivation enhances mouse sperm capacity for penetrating
  viscoelastic media.
\newblock {\em Biol. Reprod.}, 46:686--691, 1992.

\bibitem{suarez96}
S.~S. Suarez.
\newblock Hyperactivated motility in sperm.
\newblock {\em J. Androl.}, 17:331--335, 1996.

\bibitem{ho01}
H.~C. Ho and S.~S. Suarez.
\newblock Hyperactivation of mammalian spermatozoa: {F}unction and regulation.
\newblock {\em Reprod.}, 122:519--526, 2001.

\bibitem{suarez03}
S.~S. Suarez and H.~C. Ho.
\newblock Hyperactivated motility in sperm.
\newblock {\em Reprod. Dom. Anim.}, 38:119--124, 2003.

\bibitem{hwang69}
S.~H. Hwang, M.~Litt, and W.~C. Forsman.
\newblock Rheological properties of mucus.
\newblock {\em Rheol. Acta}, 8:438--448, 1969.

\bibitem{tam80}
P.~Y. Tam, D.~F. Katz, and S.~A. Berger.
\newblock Nonlinear viscoelastic properties of cervical mucus.
\newblock {\em Biorheol.}, 17:465--478, 1980.

\bibitem{eliezer74}
N.~Eliezer.
\newblock Viscoelastic properties of mucus.
\newblock {\em Biorheol.}, 11:61--68, 1974.

\bibitem{litt76}
M.~Litt, M.~A. Khan, and D.~P. Wolf.
\newblock Mucus rheology--{R}elation to structure and function.
\newblock {\em Biorheol.}, 13:37--48, 1976.

\bibitem{lauga07}
E.~Lauga.
\newblock Propulsion in a viscoelastic fluid.
\newblock {\em Phys. Fluids}, 19:083104, 2007.

\bibitem{MontecuccoRappuoli2001}
C.~Montecucco and R.~Rappuoli.
\newblock Living dangerously: how helicobacter pylori survives in the human
  stomach.
\newblock {\em Nature Reviews Mol. Cell Bio.}, {\bf 2}:457, 2001.

\bibitem{WolgemuthCharonGoldstein2006}
C.~W. Wolgemuth, N.~W. Charon, S.~F. Goldstein, and R.~E. Goldstein.
\newblock The flagellar cytoskeleton of the spirochetes.
\newblock {\em J. Mol Microbiol. Biotechno.}, {\bf 11}:221, 2006.

\bibitem{chaudhury79}
T.~K. Chaudhury.
\newblock Swimming in a viscoelastic liquid.
\newblock {\em J. Fluid Mech.}, 95:189--197, 1979.

\bibitem{fulford98}
G.~R. Fulford, D.~F. Katz, and R.~L. Powell.
\newblock Swimming of spermatozoa in a linear viscoelastic fluid.
\newblock {\em Biorheol.}, 35:295--309, 1998.

\bibitem{ross74}
S.~M. Ross and S.~Corrsin.
\newblock Results of an analytical model of mucociliary pumping.
\newblock {\em J. Appl. Physiol.}, 37:333--340, 1974.

\bibitem{FuWolgemuthPowers2009}
H.~C. Fu, C.~W. Wolgemuth, and T.~R. Powers.
\newblock Swimming speeds of filaments in nonlinearly viscoelastic fluids.
\newblock {\em submitted to Phys. Fluids}, 2009.

\bibitem{Ishijima1986}
S.~Ishijima, S.~Oshio, and H.~Mohri.
\newblock Flagellar movement of human spermatozoa.
\newblock {\em Gamete Res.}, {\bf 13}:185--197, 1986.

\bibitem{TeranFauciShelley2008}
J.~Teran, L.~Fauci, and M.~Shelley.
\newblock Peristaltic pumping and irreversibility of a stokesian viscoelastic
  fluid.
\newblock {\em Phys. Fluids}, 20:073101, 2008.

\bibitem{leoni08}
M.~Leoni, J.~Kotar, B.~Bassetti, P.~Cicuta, and M.~Cosentino Lagomarsino.
\newblock A basic swimmer at low reynolds number.
\newblock {\em Preprint}, 2008.

\bibitem{tam07}
D.~Tam and A.~E. Hosoi.
\newblock Optimal stroke patterns for {Purcell's} three-link swimmer.
\newblock {\em Phys. Rev. Lett.}, 98(6):068105, 2007.

\bibitem{najafi05}
A.~Najafi and R.~Golestanian.
\newblock Propulsion at low {Reynolds} number.
\newblock {\em J. Phys. - Cond. Mat.}, 17:S1203--S1208, 2005.

\bibitem{golestanian08}
R.~Golestanian and A.~Ajdari.
\newblock Analytic results for the three-sphere swimmer at low {Reynolds}
  number.
\newblock {\em Phys. Rev. E}, 77(3):036308, 2008.

\bibitem{golestanian08_2}
R.~Golestanian.
\newblock Three-sphere {low-Reynolds-number} swimmer with a cargo container.
\newblock {\em European Phys. J. E}, 25(1):1--4, 2008.

\bibitem{golestanian08_3}
R.~Golestanian and A.~Ajdari.
\newblock Mechanical response of a small swimmer driven by conformational
  transitions.
\newblock {\em Phys. Rev. Lett.}, 1:038101, 2008.

\bibitem{felderhof06}
B.~U. Felderhof.
\newblock The swimming of animalcules.
\newblock {\em Phys. Fluids}, 18:063101, 2006.

\bibitem{avron05:pushme}
J.~E. Avron, O.~Kenneth, and D.~H. Oaknin.
\newblock Pushmepullyou: {A}n efficient micro-swimmer.
\newblock {\em New J. Phys.}, 7:234, 2005.

\bibitem{ogrin08}
F.~Y. Ogrin, P.~G. Petrov, and C.~P. Winlove.
\newblock Ferromagnetic microswimmers.
\newblock {\em Phys. Rev. Lett.}, 100(21):218102, 2008.

\bibitem{koiller96}
J.~Koiller, K.~Ehlers, and R.~Montgomery.
\newblock Problems and progress in microswimming.
\newblock {\em J. Nonlinear Sci.}, 6:507--541, 1996.

\bibitem{laugabartolo08}
E.~Lauga and D.~Bartolo.
\newblock No many-scallop theorem: Collective locomotion of reciprocal
  swimmers.
\newblock {\em Phys. Rev. E}, 78:030901, 2008.

\bibitem{alexander08}
G.~P. Alexander and J.~M. Yeomans.
\newblock Dumb-bell swimmers.
\newblock {\em Euro. Phys. Lett.}, 83:34006, 2008.

\bibitem{dreyfus05_nature}
R.~Dreyfus, J.~Baudry, M.~L. Roper, M.~Fermigier, H.~A. Stone, and J.~Bibette.
\newblock Microscopic artificial swimmers.
\newblock {\em Nature}, 437:862--865, 2005.

\bibitem{yu_unpublished}
T.~S. Yu, E.~Lauga, and A.~E. Hosoi.
\newblock Swimming using elastic tail propulsion.
\newblock {\em (Unpublished manuscript)}.

\bibitem{howse07}
J.~R. Howse, R.~A.~L. Jones, A.~J. Ryan, T.~Gough, R.~Vafabakhsh, and
  R.~Golestanian.
\newblock Self-motile colloidal particles: {From} directed propulsion to random
  walk.
\newblock {\em Phys. Rev. Lett.}, 99(4):048102, 2007.

\bibitem{cebers05}
A.~Cebers.
\newblock Flexible magnetic filaments.
\newblock {\em Curr. Opin. Colloid Interface Sci.}, 10:167--175, 2005.

\bibitem{gauger06}
E.~Gauger and H.~Stark.
\newblock Numerical study of a microscopic artificial swimmer.
\newblock {\em Phys. Rev. E}, 74:021907, 2006.

\bibitem{roper06}
M.~Roper, R.~Dreyfus, J.~Baudry, M.~Fermigier, J.~Bibette, and H.~A. Stone.
\newblock On the dynamics of magnetically driven elastic filaments.
\newblock {\em J. Fluid Mech.}, 554:167--190, 2006.

\bibitem{keaveny08}
E.~E. Keaveny and M.~R. Maxey.
\newblock Spiral swimming of an artificial micro-swimmer.
\newblock {\em J. Fluid Mechanics}, 598:293--319, 2008.

\bibitem{roper08}
M.~Roper, R.~Dreyfus, J.~Baudry, M.~Fermigier, J.~Bibette, and H.~A. Stone.
\newblock Do magnetic micro-swimmers move like eukaryotic cells?
\newblock {\em Proc. Roy. Soc. A}, 464(2092):877--904, 2008.

\bibitem{thaokar07}
R.~M. Thaokar, H.~Schiessel, and I.~M. Kulic.
\newblock Hydrodynamics of a rotating torus.
\newblock {\em European Phys. J. B}, 60:325--336, 2007.

\bibitem{thaokara08}
R.~M. Thaokara.
\newblock Hydrodynamic interaction between two rotating tori.
\newblock {\em European Phys. J. B}, 61:47--58, 2008.

\bibitem{leshansky08}
A.~M. Leshansky and O.~Kenneth.
\newblock Surface tank treading: {P}ropulsion of purcell's toroidal swimmer.
\newblock {\em Phys. Fluids}, 20:063104, 2008.

\bibitem{leshansky07}
A.~M. Leshansky, O.~Kenneth, O.~Gat, and J.~E. Avron.
\newblock A frictionless microswimmer.
\newblock {\em New J. Phys.}, 9:126, 2007.

\bibitem{paxton04}
W.~F. Paxton, K.~C. Kistler, C.~C. Olmeda, A.~Sen, S.~K.~St Angelo, Y.~Y. Cao,
  T.~E. Mallouk, P.~E. Lammert, and V.~H. Crespi.
\newblock Catalytic nanomotors: {Autonomous} movement of striped nanorods.
\newblock {\em J. Am. Chem. Soc.}, 126:13424--13431, 2004.

\bibitem{fournier-bidoz05}
S.~Fournier-Bidoz, A.~C. Arsenault, I.~Manners, and G.~A. Ozin.
\newblock Synthetic self-propelled nanorotors.
\newblock {\em Chem. Comm.}, pages 441--443, 2005.

\bibitem{kline05}
T.~R. Kline, W.~F. Paxton, T.~E. Mallouk, and A.~Sen.
\newblock Catalytic nanomotors: {Remote-controlled} autonomous movement of
  striped metallic nanorods.
\newblock {\em Angew. Chem., Int. Ed.}, 44:744--746, 2005.

\bibitem{mano05}
N.~Mano and A.~Heller.
\newblock Bioelectrochemical propulsion.
\newblock {\em J. Am. Chem. Soc.}, 127:11574--11575, 2005.

\bibitem{paxton06}
W.~F. Paxton, S.~Sundararajan, T.~E. Mallouk, and A.~Sen.
\newblock Chemical locomotion.
\newblock {\em Angew. Chem., Int. Ed.}, 45:5420--5429, 2006.

\bibitem{ruckner07}
G.~Ruckner and R.~Kapral.
\newblock Chemically powered nanodimers.
\newblock {\em Phys. Rev. Lett.}, 98:150603, 2007.

\bibitem{tao08}
Y.~G. Tao and R.~Kapral.
\newblock Design of chemically propelled nanodimer motors.
\newblock {\em J. Chem. Phys.}, 128(16):164518, 2008.

\bibitem{golestanian05}
R.~Golestanian, T.~B. Liverpool, and A.~Ajdari.
\newblock Propulsion of a molecular machine by asymmetric distribution of
  reaction products.
\newblock {\em Phys. Rev. Lett.}, 94(22):220801, 2005.

\bibitem{golestanian07}
R.~Golestanian, T.~B. Liverpool, and A.~Ajdari.
\newblock Designing phoretic micro- and nano-swimmers.
\newblock {\em New J. Phys.}, 9, 2007.

\bibitem{brady08}
U.~M. Cordova-Figueroa and J.~F. Brady.
\newblock Osmotic propulsion: The osmotic motor.
\newblock {\em Phys. Rev. Lett.}, 100:158303, 2008.

\bibitem{kim04_diffusion}
M.~J. Kim and K.~S. Breuer.
\newblock Enhanced diffusion due to motile bacteria.
\newblock {\em Phys. Fluids}, 16(9):L78--L81, 2004.

\bibitem{darnton04}
N.~Darnton, L.~Turner, K.~Breuer, and H.~C. Berg.
\newblock Moving fluid with bacterial carpets.
\newblock {\em Biophys. J.}, 86:1863--1870, 2004.

\bibitem{weibel05}
D.~B. Weibel, P.~Garstecki, D.~Ryan, W.~R. Diluzio, M.~Mayer, J.~E. Seto, and
  G.~M. Whitesides.
\newblock Microoxen: {Microorganisms} to move microscale loads.
\newblock {\em Proc. Natl. Acad. Soc., USA}, 102:11963--11967, 2005.

\bibitem{raz08}
O.~Raz and A.~M. Leshansky.
\newblock Efficiency of cargo towing by a microswimmer.
\newblock {\em Phys. Rev. E}, 77(5):055305, 2008.

\bibitem{raz07}
O.~Raz and J.~E. Avron.
\newblock Swimming, pumping and gliding at low {Reynolds} numbers.
\newblock {\em New J. Phys.}, 9:437, 2007.

\bibitem{kim06:pumping}
Y.~W. Kim and R.~R. Netz.
\newblock Pumping fluids with periodically beating grafted elastic filaments.
\newblock {\em Phys. Rev. Lett.}, 96:158101, 2006.

\bibitem{tabata02}
O.~Tabata, H.~Hirasawa, S.~Aoki, R.~Yoshida, and E.~Kokufuta.
\newblock Ciliary motion actuator using self-oscillating gel.
\newblock {\em Sens. Actuat. A}, 95:234--238, 2002.

\bibitem{pironneau74}
O.~Pironneau and D.~F. Katz.
\newblock Optimal swimming of flagellated microorganisms.
\newblock {\em J. Fluid Mech.}, 66:391--415, 1974.

\bibitem{pironneau_inbook}
O.~Pironneau and D.~F. Katz.
\newblock Optimal swimming of motion of flagella.
\newblock In T.Y. Wu, C.J. Brokaw, and C~Brennen, editors, {\em Swimming and
  {F}lying in {N}ature}, volume~1, pages 161--172. Plenum, New-York, 1975.

\bibitem{higdon79a}
J.~J.~L. Higdon.
\newblock Hydrodynamic analysis of flagellar propulsion.
\newblock {\em J. Fluid Mech.}, 90:685--711, 1979.

\bibitem{tamPhD}
D.~S.-W. Tam.
\newblock {\em Motion at low Reynolds number}.
\newblock PhD thesis, Massachusetts Institute of Technology, Cambridge, MA,
  2008.

\bibitem{higdon79b}
J.~J.~L. Higdon.
\newblock Hydrodynamics of flagellar propulsion - {H}elical waves.
\newblock {\em J. Fluid Mech.}, 94:331--351, 1979.

\bibitem{lighthill96_helical}
J.~Lighthill.
\newblock Helical distributions of stokeslets.
\newblock {\em J. Eng. Math.}, 30:35--78, 1996.

\bibitem{fujita01}
T.~Fujita and T.~Kawai.
\newblock Optimum shape of a flagellated microorganism.
\newblock {\em {JSME} Int. J. Ser. C}, 44:952--957, 2001.

\bibitem{higdon79}
J.~J.~L. Higdon.
\newblock The generation of feeding currents by flagellar motions.
\newblock {\em J. Fluid Mech.}, {\bf 94}:305, 1979.

\bibitem{alouges08}
F.~Alouges, A.~DeSimone, and A.~Lefebvre.
\newblock Optimal strokes for low {Reynolds} number swimmers: {An} example.
\newblock {\em J. Nonlinear Science}, 18(3):277--302, 2008.

\bibitem{avron04:opt}
J.~E. Avron, O.~Gat, and O.~Kenneth.
\newblock Optimal swimming at low {Reynolds} numbers.
\newblock {\em Phys. Rev. Lett.}, 93:186001, 2004.

\end{thebibliography}

\end{document}